\documentclass[fleqn,usenatbib]{mnras}

\usepackage{newtxtext,newtxmath}
\usepackage[T1]{fontenc}
\usepackage{mathtools, cuted}
\usepackage{amsmath}
\usepackage{amsfonts}
\usepackage{amsbsy}	
\usepackage{graphicx}
\usepackage{bm}
\usepackage{color}
\usepackage{ulem}
\usepackage{commath}
\usepackage{stmaryrd}

\allowdisplaybreaks

\usepackage{hyperref}
\hypersetup{
  colorlinks=true,        
  linkcolor=blue,         
  citecolor=cyan,         
  urlcolor=cyan           
}
\newcommand{\cf}{cf.~}
\newcommand{\ie}{i.e.,~}
\newcommand{\eg}{e.g.,~}

\title{Realistic models of general-relativistic differentially rotating
  stars}

\author[M. Cassing and L. Rezzolla]{Marie Cassing$^{1}$, Luciano Rezzolla$^{1,2,3}$\\
  $^{1}$Institut f{\"u}r Theoretische Physik, Max-von-Laue-Strasse 1, 60438
  Frankfurt, Germany \\
  $^{2}$Frankfurt Institute for Advanced Studies, Ruth-Moufang-Strasse 1,
  60438 Frankfurt, Germany \\
  $^{3}$School of Mathematics, Trinity College, Dublin 2, Ireland }

\date{Accepted XXX. Received YYY; in original form ZZZ}

\pubyear{2023}

\begin{document}
\label{firstpage}
\pagerange{\pageref{firstpage}--\pageref{lastpage}}
\maketitle

\begin{abstract}
General-relativistic equilibria of differentially rotating stars are
expected in a number of astrophysical scenarios, from core-collapse
supernovae to the remnant of binary neutron-star mergers. The latter, in
particular, have been the subject of extensive studies where they were
modeled with a variety of laws of differential rotation with varying
degree of realism. Starting from accurate and fully general-relativistic
simulations of binary neutron-star mergers with various equations of
state and mass ratios, we establish the time when the merger remnant has
reached a quasi-stationary equilibrium and extract in this way realistic
profiles of differential rotation. This allows us to explore how well
traditional laws reproduce such differential-rotation properties and to
derive new laws of differential rotation that better match the numerical
data in the low-density Keplerian regions of the remnant. In this way, we
have obtained a novel and somewhat surprising result: the dynamical
stability line to quasi-radial oscillations computed from the
turning-point criterion can have a slope that is not necessarily negative
with respect to the central rest-mass density, as previously found with
traditional differential-rotation laws. Indeed, for stellar models
reproducing well the properties of the merger remnants, the slope is
actually positive, thus reflecting remnants with angular momentum at
large distances from the rotation axis, and hence with cores having
higher central rest-mass densities and slower rotation rates.
\end{abstract}


\section{Introduction}
\label{sec:introduction}

Multi-messenger astronomy and the detection of gravitational waves
provide us with the possibility to study matter under extreme conditions
such as supra-nuclear densities or ultra-strong gravity that are expected
to be present in the remnants of a number of catastrophic astrophysical
scenarios such as core-collapse supernovae~\citep[see, \eg][]{Ott06c,
  Burrows07a, Franceschetti2020}, the accretion-induced collapse of white
dwarfs~\citep{Dessart2006, Abdikamalov:2009aq}, the
phase-transition-induced collapse of neutron stars~\citep{Lin2006,
  Abdikamalov2009}, and, of course, the merger of two neutron
stars~\citep[see, \eg][for some initial works]{Shibata:2003ga, Baiotti08}.
A very important and distinctive feature of the rotational properties of
these remnants is the presence of differential rotation, which is
necessary to handle the large amounts of angular momentum that they
inherit from the processes leading to their formation~\citep[see,
  \eg][for some reviews]{Baiotti2016, Paschalidis2016,
  Mueller2024}. These differentially rotating compact stars are also known to be
subject to non-axisymmetric deformations resulting from the development
of dynamical instabilities~\citep{Rezzolla00, Shibata:2002mr, Saijo2003,
  Sa2004, Baiotti06b, Manca07, Krueger2010, Franci2013} and could lead to
an interesting and peculiar gravitational-wave signal~\citep[see,
  \eg][and references therein]{Mueller2024}.

Since performing numerical simulations producing such differentially
rotating remnants is extremely expensive, alternative approaches have
long since been developed to study the properties of these remnants in
terms of self-gravitating configurations in equilibrium. The first
studies in this direction have started in the 90's to study the
equilibria of uniformly rotating relativistic stars~\citep[see, \eg][for
  some of the early and seminal works]{Komatsu89, Cook92b, Bonazzola1993,
  Stergioulas98}, and have revealed a number of properties of these
objects. Among the most important of such properties we recall: the
existence of a Keplerian stability line beyond which mass shedding
occurs, the existence of a turning-point stability line which represents
a sufficient but not necessary condition for the dynamical instability to
gravitational collapse~\citep{Friedman88, Takami2014}, the
proportionality between the maximum mass allowed by uniform rotation and
the maximum mass of the corresponding nonrotating
configuration~\citep{Cook92b, Lasota1996}, which has then been shown to
reflect a universal-relation for all equilibria along the turning-point
stability line~\citep{Breu2016, Musolino2023b}.

All of these studies of uniformly rotating equilibria have represented
the natural background, both in terms of astrophysical scenarios and of
the numerical techniques employed, for the exploration of the more
complex but also richer space of differentially rotating
equilibria~\citep{Komatsu89b, Goussard1998, Baumgarte00b, Lyford2003,
  Ansorg2009, Gondek2016, Studzinska2016, Bozzola2017, Weih2017,
  Szkudlarek2019, Zhou2019, Espino2019, Bozzola2019, Zhou2019,
  Tsokaros2020c, Staykov2023, Cheong2024, Nishad2024} and of their
subsequent dynamics~\citep{Duez:2006qe, Giacomazzo2011, Weih2017,
  Espino2019b, Szewczyk2023}.

Most the initial works exploring differentially rotating models exploited
the simplest law of differential rotation, namely the ``$j$-constant''
(or KEH) law~\citep{Komatsu89b}, which, as by the name, imposes that the
specific angular momentum $j$ is constant across the star, so that the
angular velocity essentially scales like $1/r_e^2$, where $r_e$ is some
measure of the distance from the rotation axis. However, as numerical
simulations of BNS-mergers have started to explore in greater detail the
properties of the merger remnant~\citep[see, \eg][]{Kastaun2014,
  Hanauske2016, Ciolfi2017, DePietri2018, East2019}, it has appeared that
the structure of the hypermassive neutron star (HMNS) produced after the
merger is highly dynamical and non-trivial, but also that it can be
interpreted effectively in terms of a simple mechanical toy
model~\citep{Takami2015}. Furthermore, the most salient rotational
properties of the remnant are represented by: (i) a slowly and uniformly
rotating core; (ii) a rapid increase in the angular velocity in the outer
regions of the merger remnant; (iii) a radial fall-off reflecting a
Keplerian profile.

As a consequence of this improved understanding of the rotational
properties of BNS merger remnants, more realistic descriptions of the law
of differential rotation have been suggested in the
literature~\citep[see, \eg][]{Uryu2017, Passamonti2020, Xie2020,
  Camelio2021}. Among the various suggestions for the
differential-rotation law of BNS remnants, a parameterised family of
angular velocity profiles has been suggested by~\citet{Uryu2017}, which
in the 4-parameter version (U8 law) or in the 3-parameter (U9 law)
prescription, provides a rather good description of the angular-velocity
profiles of BNS remnants. This was shown first by~\citet{Iosif2020} for
simple polytropic equations of state (EOS) and later extended to
tabulated EOSs in the comprehensive and detailed study
of~\citet{Iosif2021} for the 4-parameter version of the law.

These laws have been extensively studied in the literature and applied to
the analysis of differentially rotating neutron stars and quark
stars. While they already represent a good match to the simulations,
  they be further refined and extended. In fact, to increase the level
of realism in the construction of equilibrium models of differentially
rotating neutrons stars, we have here reconsidered this problem with a
bottom-up approach in which we first consider in detail the rotational
information obtained from three, state-of-the-art BNS merger simulations
performed with different and realistic EOSs and mass ratios. From these
simulations we have then extracted the quasi-stationary rotational
properties of the remnants, both in terms of azimuthally- and
time-averaged profiles of the angular velocity and of the specific
angular momentum. In this way, it was possible to study the existence of
universal features in the angular-velocity profile and to point out a
criterion that allows us to simply recognise the transition between the
``core'' of the remnant and the portion of the HMNS whose
angular-velocity profile follows a Keplerian behaviour, \ie the ``disc''.

Using the differential-rotation profiles extracted from the simulations
we have also improved the fidelity of the U9 law after a simple extension
of the functional expression in terms of an additional exponent, leading
to what we have referred to as the U9+ law.  Because in the outer layers
of the merger remnant the rest-mass density and the angular velocity are
particularly low, both the U9 and the U9+ laws provide there rather
inaccurate descriptions of the remnant properties. Hence, to counter
that, we have devised a novel and more complex law of differential
rotation, that we refer to as the CR law, which, allows us to reproduce
with relative differences that are always below $5\%$ the
angular-velocity profiles also well beyond the nominal stellar surface
and in the disc.

As a final and particularly interesting aspect of our analysis, we have
carried out the first systematic study of the properties of turning-point
stability lines in realistic (and non-realistic) laws of differential
rotation. In this way, we have obtained the surprising result that the
slope of the stability line depends on the properties of the law of
differential rotation and is not necessarily negative. In particular,
those differential-rotations laws that provide an accurate representation
of the simulated post-merger remnant lead to turning-point stability lines
with positive slopes.

The structure of the paper is as follows. First we recall in
Sec.~\ref{sec:section2} the theory of axisymmetric spacetimes and
hydrostationary equilibria, as well as the numerical setup needed in the
optimised \texttt{Hydro-RNS} code to compute such
equilibria. Section~\ref{sec:section4} provides a quick review of the
most traditional differential-rotation laws and shows our ability not
only to reproduce published results with the U8 and U9
laws~\citet{Iosif2020}, but also to extend the space of solutions that
can be obtained with them.  Section~\ref{sec:section3} is dedicated to
the analysis of the data of from the BNS-merger simulations and to
illustrate how to determine the existence of a quasi-stationary
equilibrium and to extract reliable profiles that are averaged in
azimuthal angle and time; this section also provides a discussion on a
simple criterion to determine the transition from the HMNS core to its
disc. Section~\ref{sec:section5} proposes the new laws of differential
rotation, providing evidence of their accuracy in reproducing the
numerical data from the simulations and contrasting their performance in
the various regions of the remnants; the section also discusses the
properties of the turning-point stability lines and how their slope can
be suitably modified across different differential-rotation
laws. Finally, Sec.~\ref{sec:section6} provides a summary of our results
and the prospects for future work. Hereafter, we adopt a set of units in
which $c = 1 = G$, with $c$ and $G$ being the speed of light and the
gravitational constant, respectively. Furthermore, we adopt Greek letters
for indices running from $0$ to $3$ and Latin letters for indices running
from $1$ to $3$.

\section{The theory of rotating stars: a brief review}
\label{sec:section2}

A neutron star in equilibrium can be described as a stationary
axisymmetric self-gravitating perfect fluid whose equilibrium, using a
Newtonian language, is governed by the balance between gravitational,
rotational, and pressure-gradients forces. The corresponding stationary
and axisymmetric spacetime can be described by the following metric in
quasi-isotropic spherical coordinates
\begin{eqnarray}
ds^2 = - e^{2\nu} dt^2 + e^{2 \psi}(d \phi - \omega dt)^2 + e^{2 \mu}(
dr^2 + r^2 d\theta^2)\,.
\label{line_element_rot}
\end{eqnarray}
This spacetime is asymptotically flat and stationary, such that from the
Killing vectors -- defined as those along which Lie-derivative vanishes
-- can be found. The corresponding components are then given by
$t^{\alpha}$, $\phi^{\alpha}$, so that the coordinates $t$ and $\phi$ are
cyclic [\ie the metric~\eqref{line_element_rot} does not depend on
  them]. The quantity $ u^{\alpha} = dx^{\alpha} /d \tau$ is the
four-velocity of the fluid with respect to the proper time $\tau$ and, in
the case of a circular motion in such a spacetime has components
\begin{eqnarray}
u^{\alpha} = u^{t} \left( t^{\alpha} + \Omega \phi^{\alpha} \right)\,,
\label{4velo}
\end{eqnarray}
where the angular velocity $\Omega$ is defined as
\begin{eqnarray}
  \Omega := \frac{u^{\phi}}{u^{t}} = \frac{d \phi}{d t} \,.
\end{eqnarray}

We recall that, given the energy-momentum tensor of a perfect fluid with
energy density $e$, pressure $p$ and four-velocity $u^{\mu}$ is given
by~\citep[see, \eg][]{Rezzolla_book:2013}
\begin{eqnarray}
    T^{\mu \nu} = (e + p)u^{\mu}u^{\nu}+p g^{\mu \nu} \,,
\end{eqnarray}
the equations of hydrostationary equilibrium can be derived from the
conservation of such tensor $\nabla_{\mu}T^{\mu \nu} = 0$. In particular,
for a fluid with four-velocity components $u^t$, and $u^{\phi}$, we can
write the equation of hydrostationary equilibrium for a rotating star as
(or ``rotation integral'')
\begin{eqnarray}
 \frac{\nabla_{\alpha}p}{(e + p)} = \nabla_{\alpha} \ln u^t - u^t
 u_{\phi} \nabla_{\alpha} \Omega\,.
  \label{HydrostatEquil}
\end{eqnarray}
It is now convenient to define the quantity 
\begin{eqnarray}
  j := u^{t} u_{\phi}\,,
  \label{Fmom}
\end{eqnarray}
as the gravitationally redshifted specific angular momentum, which will
play an important role in the investigation of rotation profiles of
differentially rotating stars (We note that the quantity $j$ is instead
indicated as $F$ in~\citet{Stergioulas98}, which is an important
reference in the context equilibria of rotating stars). Using definition
of the specific angular momentum~\eqref{Fmom}, the Euler equation
(\ref{HydrostatEquil}) can also be expressed in the form that will be
useful later on
\begin{eqnarray}
  \label{eq:Eul_1}
  \frac{\nabla_{\alpha}p}{(e + p)} = \nabla_{\alpha} \ln u^t - j
\nabla_{\alpha} \Omega\,.
\end{eqnarray}

Another useful quantity valid for barotropic fluids, \ie fluids whose 
energy density depends only on the pressure, is the dimensionless enthalpy 
\begin{eqnarray}
H(p) := \int_0^p \ \frac{dp'}{e(p')+p'}\,,
\end{eqnarray}
which, after using the second law of thermodynamics, can be shown to
satisfy the identity
\begin{eqnarray}
 \nabla H = \nabla \ln h - \frac{T}{h} \nabla s\,,
\end{eqnarray}
where $\rho$ is the rest-mass density, $T$ the temperature, $s$ the
specific entropy density, and $h := (e + p)/\rho$ the specific enthalpy.
As a result, the Euler equation~\eqref{eq:Eul_1} can be rewritten as
\begin{eqnarray}
  \label{eq:Eul_2}
  \nabla \left( H - \ln u^{t}\right) = -j \nabla \Omega \,,
\end{eqnarray}
where $\Omega$ is either a constant in the case of uniformly rotating
stars, or can be a function of position if the star is rotating
differentially. Employing the relativistic version of the
Poincar\'e-Wavre theorem~\citep{Tassoul-1978:theory-of-rotating-stars},
the specific angular momentum is a function of the angular velocity only,
\ie $j=j(\Omega)$, so that we can rewrite the Euler
equation~\eqref{eq:Eul_2} in an integral form and obtain the first
integral of hydrostationary equilibrium
\begin{eqnarray}
  \label{eq:Eul_3}
  H - \ln u^{t} + \int_{\Omega_0}^{\Omega} j(\Omega')d \Omega' = {\rm const.} \,.
\end{eqnarray}
In the case of uniform rotation and taking $\Omega_0 = \Omega_{\rm pol}$,
Eq.~\eqref{eq:Eul_2} reduces to $H - \ln u^{t} = \nu|_{\rm pol}$, where
$\nu|_{\rm pol}$ is the value of the metric function $\nu$ at the pole.
If $\Omega$ is not uniform, Eq.~\eqref{eq:Eul_3} does not hold and
the resulting rotation integral~\eqref{HydrostatEquil} will be a function
of $\Omega$. Depending on the chosen rotation law $j=j(\Omega)$, the
integral in~\eqref{eq:Eul_3} can either be computed analytically or, in
more general conditions, it will have to be calculated numerically.

\subsection{Numerical setup}

The numerical calculations reported here have been performed making use
of a modified version of the \texttt{Hydro-RNS} thorn, which is part of
the Einstein-toolkit~\citep{EinsteinToolkit_etal:2020_11}.
\texttt{Hydro-RNS} is based on the original \texttt{RNS}
code~\citep{Stergioulas95, Stergioulas04} computes equilibrium solutions
of rotating stars with analytic and tabulated EOSs that are either in
uniform or differential rotation. In addition, by construction, it can
compute sequences of stellar models having either constant central
rest-mass densities $\rho_c$, or constant ratio between the polar and
equatorial radii $r_p/r_e$. In practice, the code integrates the
Einstein equations for stars described as perfect fluids employing the
so-called Komatsu-Eriguchi-Hachisu (KEH) formulation in which the
Einstein equations are rewritten as a set of elliptic equations and are
solved using Green functions~\citep{Komatsu89}. In particular, the
rotating-star models are assumed to be described by the generic
line-element
\begin{eqnarray}
ds^2 &=& - e^{\gamma + \varrho}dt^2 + e^{\gamma -\varrho} r^2 \sin^2 \theta
\left(d\phi - \omega dt\right)^2 \nonumber \\
&& + e^{2\alpha}\left(d r^2 + r^2
\sin^2{\theta} \right)\,,
\label{metric_RNS}
\end{eqnarray}
where the metric functions $\gamma,\, \varrho,\, \alpha$ and $\omega$
depend only on the coordinates $r$ and $\theta$ ($\omega$ accounts for
the frame-dragging effect) and where the radial quasi-isotropic
coordinate $r$ is related to the proper circumferential radial coordinate
$R$ via the simple expression $R := e^{\gamma -\varrho} r^2 \sin^2
\theta$. We use small "$r$" for quantities related to the quasi-isotropic
radius as, for example, $r_e$ and capital $R$ for the corresponding
proper value, that is the integral of the infinitesimal proper distance.
The code uses the angular variable $\mu := \cos(\theta)$ and compactifies
the radial coordinate by $s := r/(r_e+r)$, such that $s=0.5$ at the
equator $r_e$.

Quite generically, and independently of whether the rotation is uniform
or differential, \texttt{Hydro-RNS} first computes the solution for a
spherically symmetric configuration and uses it as an initial guess while
slightly decreasing the axis ratio $r_p/r_e$. For each value of the
latter, an iteration procedure to obtain an equilibrium stellar model is
carried out, starting with guesses for the metric functions and ending if
a desired accuracy in the quasi-isotropic equatorial radius $r_e$ is
reached.  Given a first guess of the metric functions $\rho$, $\gamma$,
$\mu$, $\omega$, of the energy density $\epsilon$ and of the angular
velocity $\Omega$ some of these quantities are rescaled to dimensionless
units (denoted with a hat) to be used in the code, such as 
\begin{align}
  &&\hat{\gamma} := \gamma / r_e^2\,, \qquad
  &\hat{\rho} := \rho / r_e^2\,, \qquad
  &\hat{\alpha} := \alpha / r_e^2\,,\\
  &&\hat{\omega} := \omega r_e\,, \qquad
  &\hat{\Omega} := \Omega r_e\,. \qquad
  &\phantom{\hat{\alpha} := \alpha / r_e^2\,,}
\end{align}

In a second step, the radius at the equator $r_e$ is calculated from the
first integral of the hydrostationary equilibrium equated at the pole and
at the centre. Next, the equatorial rotation frequency $\Omega_{e}$ is
obtained from equating the Euler equation at the pole and at the equator
and finding the root with a Brent-algorithm.  Since now that $\Omega_e$
is found, the central rotation frequency $\Omega_c$ is calculated
accordingly from the rotation law.  The whole distribution of angular
velocity $\Omega$ on the grid in $s$ and $\mu$ is obtained by finding the
root of $j(\Omega)=u^t u_{\phi}$ at each grid-point. Using the
distribution $\Omega(s,\mu)$, it is possible to calculate the velocity of
the fluid $v(s,\mu)$ and the specific enthalpy at each grid-point
$h(s,\mu)$ through the equation of hydrostationary
equilibrium~(\ref{eq:Eul_3}). At this stage, using the specific enthalpy
$h(s,\mu)$ it is possible to calculate at each grid-point thermodynamic
quantities such as the pressure and energy density. In the last step of
the iteration over $r_e$, the metric functions and the metric potentials
$S_{\gamma}, \, S_{\rho}, \, S_{\omega}$ and $S_{\alpha}$ are
calculated. After that the procedure is repeated until convergence is
reached.

Once a solution for a rotating stellar model with the desired accuracy is
found, the corresponding integral quantities, such as the gravitational
mass $M$ or the angular momentum $J:=\int j dM_{0}$ (with baryonic mass
$M_{0}$), can be used to construct sequences. Particularly useful are
sequences of stellar models having the same angular momentum $J$, as
these sequences are normally characterised by a maximum value of the
gravitational mass, that is a mass satisfying the condition
\begin{eqnarray}
  \left. \frac{\partial M}{\partial \rho_c} \right \vert_{J} = 0 \,.
  \label{turningPoint_crit}
\end{eqnarray}

A well-known criterion, the ``turning-point''
criterion~\citep{Friedman88} then states that the
condition~\eqref{turningPoint_crit} is a sufficient condition for the
quasi-radial dynamical instability leading to a gravitational collapse to
black holes. A less well-known result is that the ``turning-point
stability line'', \ie the line marked by all stellar models satisfying
Eq.~\eqref{turningPoint_crit}, is distinct (for $J > 0$) from the
``neutral-stability line'', which is instead marked by all stellar models
whose fundamental $f$-mode frequency of quasi-radial oscillations
vanishes~\citep{Takami:2011}. By construction, and independently of the
type of rotation law, such a line represents a necessary condition for
the appearance of a dynamical instability leading to a gravitational
collapse, hence, truly distinguishing stellar models that are unstable
from those that are instead stable. Unfortunately, the neutral-stability
line can be determined either via complex perturbative
studies~\citep{Dimmelmeier06, Zink:2010a, Takami:2011} or through direct
numerical-relativity calculations. More specifically, time-dependent
  simulations in full general relativity can be carried out by perturbing
  rotating and nonrotating stellar models and by measuring the
  corresponding $f$-mode frequencies, which become increasingly smaller
  as the stellar models approach the neutral-stability line. A
  multi-dimensional fitting procedure is then applied to extrapolate such
  frequencies to vanishing values, hence obtaining the neutral-stability
  line in the $(\rho_c,M$) plane.

calculations Because of its simplicity, and bearing in mind that the
differences between the two lines in terms of masses and central
rest-mass densities are of a few percent at most, we will hereafter make
use only of the turning-point stability line.

\section{Traditional differential-rotation laws}
\label{sec:section4}

Several laws of differential rotation that are meant to mimic those
encountered in the HMNS produced in a BNS merger have been proposed over
the years and a particularly rich family has been proposed
in~\citet{Uryu2017}, and feature the presence of a peak in the rotation
profiles together with a (Keplerian) fall-off. Within this family of
rotation laws, two are particularly interesting and have been employed in
previous works, namely, [see Eq. (8) of~\citet{Uryu2017}]
\begin{eqnarray}
   \Omega(j,A,B,p,q) = \Omega_c \frac{1+(j/(B^2 \Omega_c))^p}{1+(j/(A^2
     \Omega_c))^{(q+p)}}
   \label{EQ8}
\end{eqnarray}
and [see Eq. (9) of~\citet{Uryu2017}]
\begin{eqnarray}
   \Omega(j,A,B,p) = \Omega_c \left(1+\left( \frac{j}{B^2 \Omega_c}
   \right)^p \right) \left( 1 - \left( \frac{j}{A^2 \Omega_c} \right)
   \right) \,,
   \label{EQ9}
\end{eqnarray}
and hereafter we will refer them as the ``Uryu-8'' (U8) and ``Uryu-9''
(U9) laws, respectively~\citep{Uryu2017}. Note that the first law
(\ref{EQ8}) has four parameters, $p,q,A,B$, while the second one
(\ref{EQ9}) has only three, $p,A,B$. Furthermore, because these laws are
expressed as functions $\Omega=\Omega(j)$ and not in the inverse form
$j=j(\Omega)$ needed to fulfil the integrability condition, it is
necessary to perform a variable transformation
\begin{eqnarray}
  \int j d\Omega = \int j \frac{d\Omega}{dj} dj \,,
  \label{intcon}
\end{eqnarray}
to compute the integral of the Euler equation~\eqref{eq:Eul_3}.

\subsection{The Uryu-8 rotation law}
\label{sec:section4.1}

In this subsection we focus on the properties of the U8 law (\ref{EQ8})
first concentrating on solutions obtained when varying two relevant
parameters ($\lambda_1$ and $\lambda_2$; see Sec.~\ref{sec:section4.1.1}
for a definition) and then when considering variations of all parameters
(Sec.~\ref{sec:section4.1.2}).

\subsubsection{Varying $\lambda_1$ and $\lambda_2$}
\label{sec:section4.1.1}

We start by recalling that the shape of the rotation profile
$\Omega=\Omega(R)$ is mainly changed by the parameter $A$, which
essentially sets the outer radius of the star, and by the parameter $B$,
which essentially sets the maximum of the angular velocity. At the same
time, rather than working directly with the parameters $A$ and $B$, it is
more convenient to work through suitable combinations of them. More
specifically, following the parameterisation first suggested
by~\citet{Iosif2020}, we introduce the new parameters $\lambda_1$ and
$\lambda_2$ that are functions of $A$ and $B$ and essentially
parameterisations of the maximum angular velocity $\Omega_{\rm max}$ and
of angular velocity at the equator $\Omega_e:=\Omega(R=R_{e})$ in terms
of the central angular velocity $\Omega_c:=\Omega(R=0)$
\begin{eqnarray}
  \label{eq:lambda1}
&&\lambda_1 := \frac{\Omega_{\rm max}}{\Omega_c} = \frac{1+(j_{\rm
    max}/(B^2 \Omega_c))^p}{1+(j_{\rm max}/(A^2 \Omega_c))^{(q+p)}} \,,
  \\
  \label{eq:lambda2}
&&\lambda_2 := \frac{\Omega_{e}}{\Omega_c} =
\frac{1+(j_{e}/(B^2 \Omega_c))^p}{1+(j_{e}/(A^2
  \Omega_c))^{(q+p)}}\,.
\end{eqnarray}

\begin{figure*}
  \centering
  \includegraphics[width=0.8\textwidth]{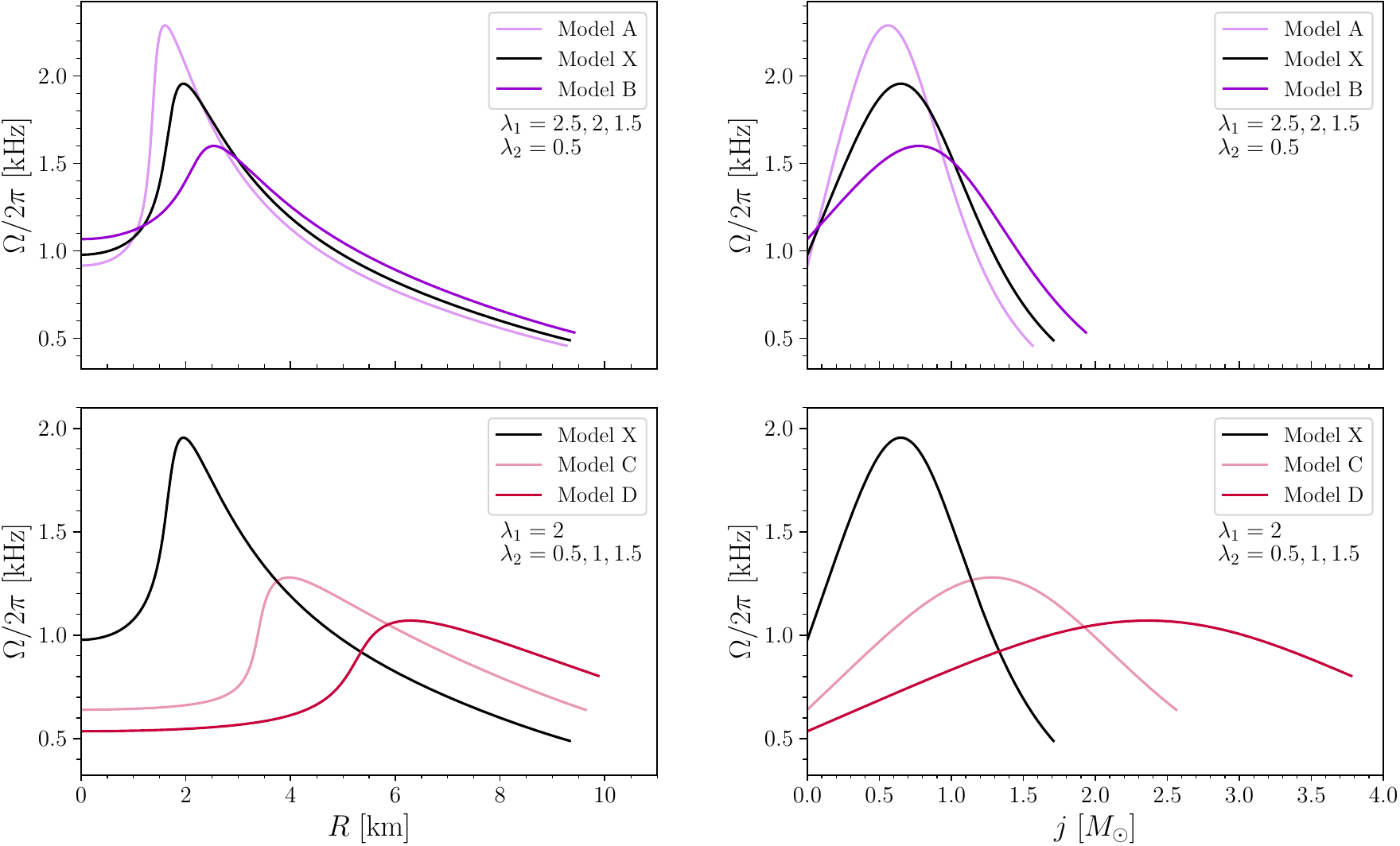}
  \caption{Angular velocity as a function of the proper circumferential
    radial coordinate $R$ (left panel) and angular velocity as a function
    of the specific angular momentum $j$ (right panel) for a stellar
    model obeying the DD2 EOS, having a central rest-mass density $\rho_c
    = 8.024 \times 10^{14}\, {\rm g~cm^{-3}}$, and an axis ratio
    $r_p/r_e = 0.8$. The underlying differential rotation law is U8
    with $p=1, \, q=3$ and varying parameters $\lambda_1,
    \lambda_2$.}
  \label{U8_L1_L2_var}
\end{figure*}

A maximal rotation is achieved at an angular momentum $j_{\rm max}$. In
practice, because \texttt{Hydro-RNS} employs dimensionless quantities, we
use the equatorial radius $r_e$ to compute the ``hatted'' quantities
\begin{align}
&&\hat{j}_{\rm max} &:= j_{\rm max}/r_e\,, &
  \qquad
\hat{j}_{e} &:=  j_{e}/r_e\,, &\\
&&\hat{\Omega}_e &:=  \Omega_e r_e\,, &
\qquad
\hat{\Omega}_{\rm max} &:= \Omega_{\rm max} r_e\,. &
\end{align}
so that we can compute the angular velocity at the equator by equating
the Euler equation at the pole with its value at the equator
\begin{eqnarray}
0 &=& \ln \left(\frac{u_{\rm pol}^t}{u_{\rm eq}^t}\right) + \int_0^{\hat{F}_e} j
\frac{d\Omega}{dj} dj \,. \label{EulerOmE}
\end{eqnarray}

Using in Eq. (\ref{EulerOmE}) the metric \eqref{metric_RNS} and the
four-velocity component $u^t$,
\begin{eqnarray}
  u^t = \frac{e^{-(\gamma + \rho)/2}}{\sqrt{1-v^2}} \,, \qquad \qquad
  v = \left( \Omega - \omega \right)r \sin \theta e^{-\rho}\,,
\end{eqnarray}
we obtain an expression in dimensionless units, which we  solve
by a Brent-method
\begin{eqnarray}
\label{eq:re}
  0 = r_e^2 (\hat{\gamma}_{\rm e} + \hat{\rho}_e - \hat{\gamma}_p -
\hat{\rho}_p) + \ln \left[ 1 - \left( \hat{\Omega}_e -
  \hat{\omega}_e\right)^2 e^{-2 r_e^2 \hat{\rho}_e} \right] \nonumber
\\ + 2\int_0^{\hat{F}_e} j \frac{d\Omega}{dj} dj \,.
\end{eqnarray}
where
\begin{eqnarray}
    \hat{j}_{e} = \frac{ \left( \hat{\Omega}_e -
      \hat{\omega}_e\right)^2 e^{-2 r_e^2 \hat{\rho}_e}}{1 - \left(
      \hat{\Omega}_e - \hat{\omega}_e\right)^2 e^{-2 r_e^2 \hat{\rho}_e}}
    \,, \label{j_e}
\end{eqnarray}
represents the dimensionless specific angular momentum at the equator. 

Since $r_e$ is a function of the axis ratio and exhibits a maximum $r_{e,
  {\rm max}}$ at some value of the axis ratio $(r_e/r_p)_{\rm max}$,
experience has shown that in order to solve Eq.~(\ref{eq:re}), the
guesses of the iterated quantity $r_e$ for the U8 law should be close to
the values given by the function
\begin{eqnarray}
\label{eq:re_max}
  r_e = r_{e, {\rm max}}- k_1 \left[(r_e/r_p)_{\rm max} \right]^{k_2}\,,
\end{eqnarray}
which depends on the chosen rotation parameters or on the central
rest-mass density; we found that $k_2=2.62$ approximates well guesses for
different central rest-mass densities. The parameter $k_1$ in
Eq.~\eqref{eq:re_max} can be calculated as
\begin{equation}
k_1 = ((r_{e,{\rm max}}-r_{e,{\rm curr}})/((r_e/r_p)_{\rm
  max}-(r_e/r_p)_{\rm curr})^{k_2}\,,
\end{equation}
where $(r_e/r_p)_{\rm curr}$ and $r_{e,{\rm curr}}$ are respectively the
axis ratio at the current iteration and the corresponding equatorial
(coordinate) radius and if the iteration scheme has reached some axis
ratio beyond the maximum $(r_e/r_p)_{\rm max}$. The value of the maximum
dimensionless specific angular momentum $\hat{j}_{\rm max}$ is obtained
numerically from the profile $\Omega(j)$ where the parameters $\hat{A}$
and $\hat{B}$ are obtained from Eqs. (A5) and (A6)
in~\citet{Iosif2020}. Once $\Omega_e$ and $\lambda_2$ are known, we use
Eq. (\ref{eq:lambda2}) to obtain the angular velocity at the center, \ie
$\Omega_c = \Omega_e / \lambda_2$.

As discussed by~\citet{Iosif2020}, the rotational integral in
(\ref{EulerOmE}) has an analytical solution for the specific set of
parameters $p=1$ and $q=3$, in which case it reads
\begin{eqnarray}
 \int_0^{j} j' \frac{d\Omega}{dj'} dj'\!\! &=& \!\!j \Omega - \frac{A^4
    \Omega_c^2 }{2B^2} \arctan\left( \frac{j^2}{A^4 \Omega_c^2}\right)
  \nonumber \\ && \!\!- \frac{\sqrt{2}A^2 \Omega_c^2}{4}\Biggl[ \arctan\left(
    1- \frac{j\sqrt{2}}{A^2 \Omega_c} \right)
    \nonumber \\
    && \!\! - \arctan\left(1 + \frac{j\sqrt{2}}{A^2 \Omega_c} \right)
    + \tanh^{-1}\left( \frac{A^2\Omega_c j
      \sqrt{2}}{j^2+A^4 \Omega_c^2} \right)\Biggr] \,, \nonumber \\
  \label{U8_int}
\end{eqnarray}
where we have used the identity: $\tanh^{-1}(x) = 0.5 \ln((1+x)/(1-x))$.

How the parameters $\lambda_1$ and $\lambda_2$ change the shape of the
rotation profiles can be appreciated from Fig. \ref{U8_L1_L2_var}, where
we report models computed for the DD2 EOS~\citep{Hempel2009} at a fixed
central rest-mass density of $\rho_c = 8.024 \times 10^{14}\, {\rm
  g~cm^{-3}}$, an axis ratio $r_e/r_p = 0.8$, and with $p = 1,\,q =
3$. Note how the parameter $\lambda_1$ mostly affects the height of the
maximum of the angular velocity (top row), while its position is only
minimally affected; in all cases, the equatorial angular velocity is
smaller than the central one. Small are also the changes in the profiles
of $\Omega(j)$, which essentially reflect those in $\Omega(R)$. On the
other hand, the parameter $\lambda_2$ controls both the value of the
equatorial angular velocity $\Omega_e$ and the position of the
angular-velocity maximum (bottom row). Note how $\lambda_2$ also controls
the slope of $\Omega(j)$, leading to much smaller gradients and
increasing the maximum value of the specific angular momentum at the
equator. The equilibrium properties as the total mass or equatorial
radius of all Models can be found in Table
\ref{tab:Uryu8_physical_quantities} in Appendix \ref{sec:appendixA}.

\begin{figure}
  \centering
  \includegraphics[width=0.9\columnwidth]{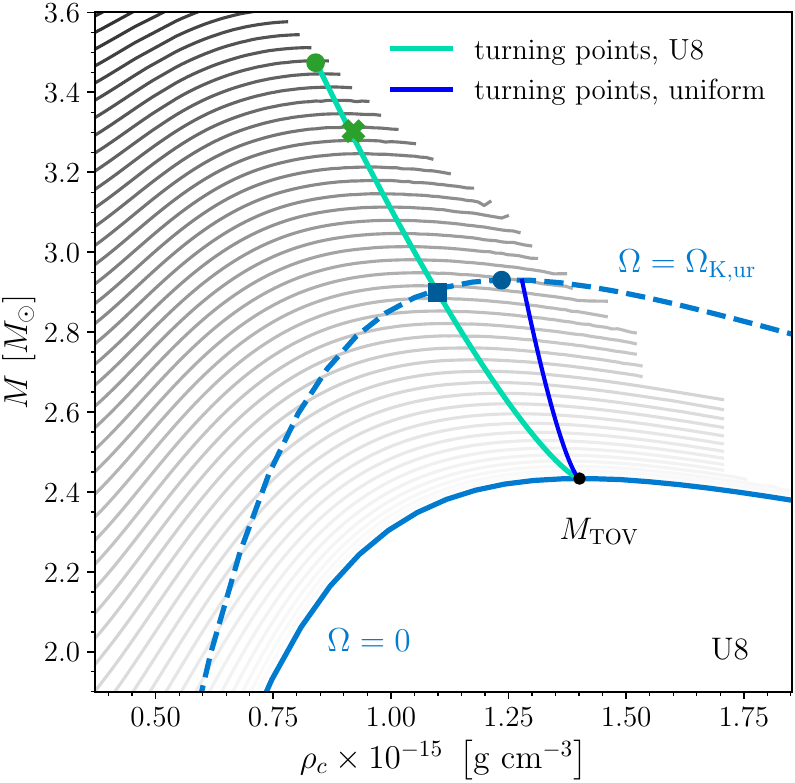}
  \caption{Total mass $M$ as function of the central rest-mass density
    $\rho_c$ for stellar models obeying the DD2 EOS and following the U8
    differential-rotation law. Shown with grey shadings are sequences of
    constant total angular momentum $J \in [0.25,9.5]$ and the
    corresponding turning-point stability line (light-blue), which starts
    at $M_{\rm TOV}$ (black circle). The green cross marks the maximum
    mass reached previously by~\citet{Iosif2021} on the stability line
    for this differential-rotation law and the green circle the
    corresponding maximum mass reached in this work. Also marked are the
    nonrotating sequence ($\Omega=0$) and the Keplerian stability line
    $\Omega=\Omega_{\rm K, ur}$ of uniformly rotating models with the
    same EOS; on such a line we mark the maximum mass (filled blue
    circle) and the maximum mass along the turning-point stability line
    (filled blue square). Finally, shown as a reference with a blue solid
    line is the turning-point stability line for the uniformly rotating
    models.}
  \label{U8_DD2_stability}
\end{figure}

The full space of solutions of differentially rotating stars obtained
with the DD2 EOS and the parameters choice $p=1\,, q=3\,, \lambda_1=1.5$,
and $\lambda_2=0.5$ is shown in Fig.~\ref{U8_DD2_stability} in the $(M,
\rho_c$) plane and matches the equivalent Fig.~3 of \citet{Iosif2021},
where the central energy density is used instead. Shown with grey solid
lines of different transparency we report $J={\rm const.}$ sequences,
with $J\in [0.25, 9.5]$ ([light-grey, dark-grey]). As discussed in the
Introduction, the maxima of these sequences of constant $J$ mark
turning-point stability line (light-blue solid line). The stability line
starts at the maximal mass $M_{\rm TOV}$ (black filled circle) of the
static configurations, \ie with $\Omega=0$ (blue solid line), crosses the
Keplerian limit of uniformly rotating configurations, \ie with
$\Omega=\Omega_{\rm K,ur}$ (blue dashed line) at a mass $M_{\rm tp; K,
  ur}$ (blue filled square) and terminates with a model having a mass of
$M=3.474\,M_{\odot}$ (green filled circle); this value exceeds the
maximum mass along the turning-point stability line reported
by~\citet{Iosif2021} (green cross) and has been achieved thanks to the
additional iteration control in the extended version of our code. Note
that for this EOS, the maximum mass of the Keplerian configurations with
uniform rotation $M_{\rm max; K,ur}$ (blue filled circle) is different
and actually larger than $M_{\rm tp; K, ur}$ so that the corresponding
stellar model is very likely to be dynamically unstable~\citep[see
  also][]{Takami:2011}. As we will comment later on, the fact that
$M_{\rm tp; K, ur} \leq M_{\rm max; K,ur}$ appears to be a robust result
quite independently of the EOS and law of differential rotation (see
Figs.~\ref{U9_DD2_stability} and~\ref{U9Q_DD2_stability}). Note that
despite the improvements to the code and the solution strategy, our
turning-point stability line does not cross the Keplerian sequence for
this EOS and the U8 law. This will be possible only for the $j$-constant
law and $\hat{A}\gtrsim 3.33$ (see Fig.~\ref{Models_stability_lines} but
also~\citet{Weih2017}).

\subsubsection{Varying $A, B, p$ and $q$}
\label{sec:section4.1.2}

We next consider the impact on the differential-rotation law when varying
the parameters $A$ and $B$ rather than $\lambda_1$ and $\lambda_2$ (the
values of $p$ and $q$ will be set to be constant as in
Sec.~\ref{sec:section4.1.1}). The advantage of this choice is that via the
parameter $B$ it is possible to scale the height of the values of central
and maximum angular velocity, while via the parameter $A$ it is possible
to set the position of the maximum in radial direction. Since we are using
$A,\, B$ instead of $\lambda_1, \, \lambda_2$, we need to compute the
latter from our choices for $\hat{A} = A/r_e$ and $\hat{B} = B/r_e$. In
particular, the parameter $\lambda_2$, \ie the ratio $\Omega_e/\Omega_c$
can be calculated as a root of the following nonlinear equation
\begin{eqnarray}
    0 = \frac{1 + \left( j_e \lambda_2/(B^2 \Omega_e)\right)^p}{1 +
      \left( j_e \lambda_2/(A^2 \Omega_e)\right)^{q+p}} - \lambda_2\,,
\end{eqnarray}
while $\lambda_1$ can be computed through Eq.~\eqref{eq:lambda1} once
$\Omega_c$ is found.

\begin{figure*}
  \centering
  \includegraphics[width=0.8\textwidth]{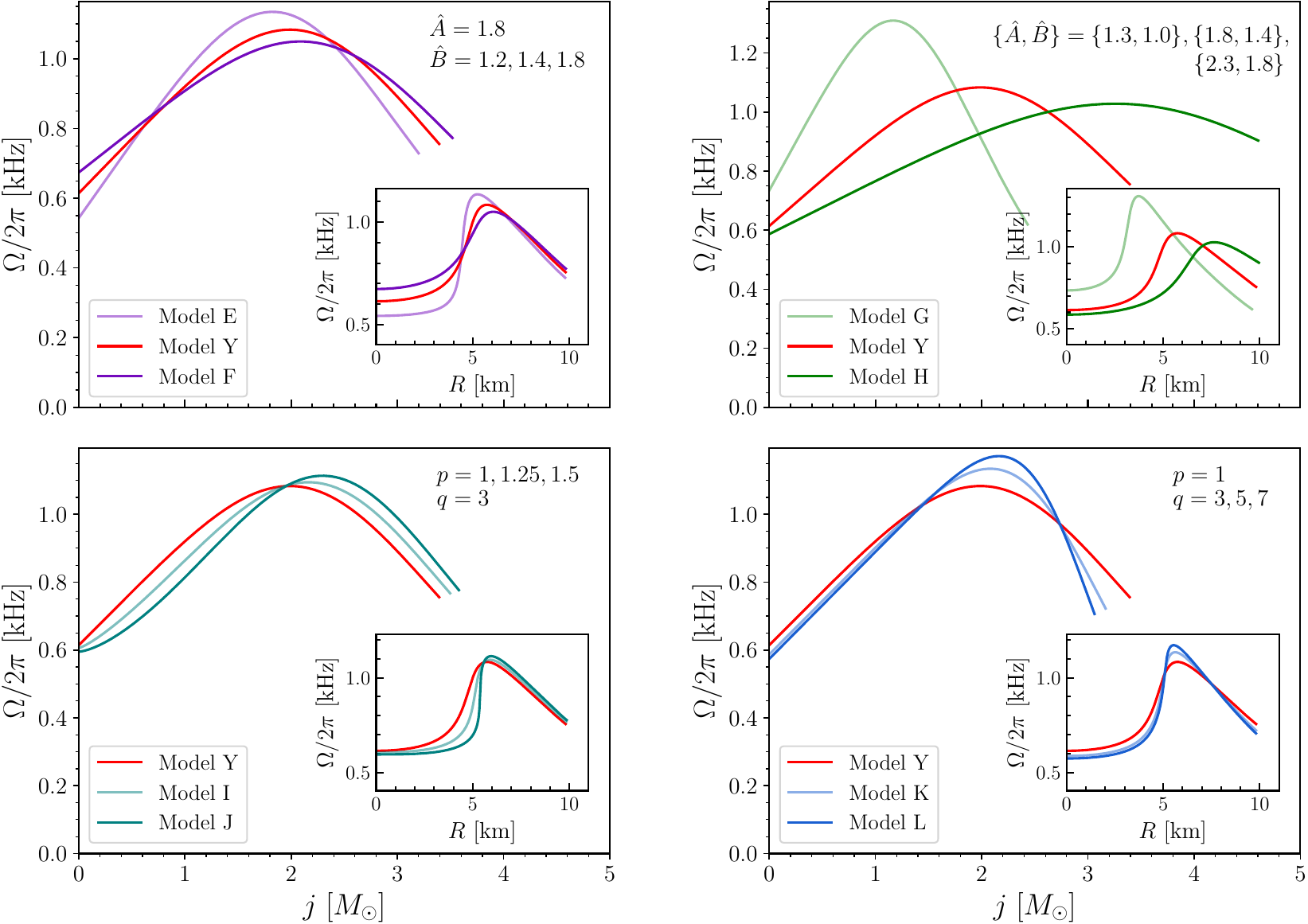}
  \caption{Angular velocity as a function of the specific angular
    momentum $j$ for a stellar model obeying the DD2 EOS, having a
    central rest-mass density $\rho_c = 8.024 \times 10^{14}\, {\rm
      g~cm^{-3}}$, and an axis ratio $r_p/r_e = 0.8$. The underlying
    differential rotation law is U8 and the different panels show how the
    functional form changes when varying the relevant parameters
    $\hat{A}, \hat{B}, p, q$ indicated in the top-right part of each
    panel. Note that each panel contains an inset reporting, for the same
    models, the angular velocity as a function of the proper
    circumferential radial coordinate $R$. In addition, in the top
      row the parameters $\hat{A}$ and $\hat{B}$ are varied but such that
      their ratio remains constant, thus leading to three different
      curves only.}
  \label{U8_P_Q_var}
\end{figure*}

To illustrate the impact of the variation of the parameters $\hat{A},
\hat{B}, p$ and $q$ we consider the DD2 EOS and set as reference ``Model
${\rm Y}$'', that is, a star with $\rho_c = 8.024 \times 10^{14}\, {\rm
  g~cm^{-3}}$, $\hat{A}=1.8$, $\hat{B} = 1.4$, $p=1$, and $q=3$. Such a
model, which has a mass $M_{\rm Y}=2.29\,M_{\odot}$ a circumferential
equatorial radius of $R_{e,{\rm Y}} = 9.84~{\rm km}$ and an axis ratio
$r_p/r_e =0.8$ is reported with solid red lines in all panels of
Fig.~\ref{U8_P_Q_var}.

More specifically, the top-left panel of Fig.~\ref{U8_P_Q_var} highlights
the impact of varying $\hat{B}$ while leaving all the other parameters
fixed (\ie $\hat{A}=1.8,\, p=1,\, q=3$) shows that it impacts the
position and value of $\Omega_{\rm max}$, as well as the value of
$\Omega_c$, while leaving mostly unaffected the radial slope between
$\Omega_{\rm max}$ and $\Omega_e$ (see inset). Similarly, the top-right
panel of Fig.~\ref{U8_P_Q_var} shows the changes introduced when varying
$\hat{A}$ and $\hat{B}$, but maintaining a constant ratio
$\hat{B}/\hat{A}$ to exclude large changes in $\lambda_1$ and keeping the
other parameters as in the top-left panel (\ie $p=1,\, q=3$). Clearly, as
the values of $\hat{A}$ and $\hat{B}$ increase (\ie going from model
${\rm G}$ to model ${\rm H}$) the rotation profile changes considerably,
with the maximal angular velocity, moving to larger radii and larger
angular momenta. Indeed, model ${\rm H}$ is the one with the largest
angular momenta and with the outermost location of the angular-velocity
maximum. 

Because our version of \texttt{Hydro-RNS} can handle values of the
exponents $p$ and $q$ in a rather wide range, a large space of parameters
can be explored in this way and a summary of this exploration is shown in
the bottom panels of Fig.~\ref{U8_P_Q_var}. In particular, the bottom
left reports different models obtained when varying $p$ while keeping the
other parameters fixed, \ie $\hat{A}=1.8$, $\hat{B}=2$, and $q=3$; in a
similar fashion, the bottom right panel models are obtained varying $q$ while
keeping $\hat{A}=1.8$, $\hat{B}=2$, and $p=1$. Considering model ${\rm
  Y}$, (red solid line) as the reference as it appears in all panels, it
is possible to appreciate that an increase in $p$ leads to more extended
stars, spanning larger angular momenta and having an angular-velocity
maximum farther away from the center.  By contrast, an increase in $q$
leads to more compact stars, spanning smaller angular momenta and having
an angular-velocity maximum that is both larger and closer to the center
Also in this case, additional details on the various stellar models can
be found in Table \ref{tab:Uryu8_physical_quantities} in
Appendix~\ref{sec:appendixA}.

\subsection{The Uryu-9 rotation law}
\label{sec:section4.2}

We next concentrate on the rotation profiles and hydrostatic equilibria
that can be computed with the U9 rotation law (\ref{EQ9}). This law has
a number of advantages. First, the derivatives and rotation integral can
be calculated analytically for all possible parameters and not only for a
specific parameter choice as for the U8 law with $p=1, \, q=3$. Second,
the law is fully described by a set of only three free parameters
${p,A,B}$ instead of four as in U8. Finally, its functional form is such
that it can produce solutions in the angular-velocity profile exhibiting
an additional minimum close to the center of the star. We recall that the
U9 rotation law reads as~\citep{Uryu2017}
\begin{eqnarray}
  \label{eq:U9}
  \Omega (j) = \Omega_{c}\left( 1 + \left( \frac{j}{B^2
      \Omega_c}\right)^p\right)\left( 1 - \left( \frac{j}{A^2\Omega}
    \right)\right) \,,
\end{eqnarray}
and it is possible to calculate analytically the first and second
derivative of the angular velocity as
\begin{eqnarray}
  && \hspace{-0.5cm}
  \frac{d \Omega}{d j} = \frac{p}{j} (A^2 \Omega_c-j)\left(
    \frac{j}{B^2 \Omega_c}\right)^p - \frac{1}{A^2}\left[ 1+ \left(
      \frac{j}{B^2 \Omega_c}\right)^p \right]\,,\\
    &&\hspace{-0.5cm}
    \frac{d^2 \Omega}{d j^2} = - \frac{p}{A^2 j^2}\left( -A^2 \Omega_c p
    + A^2 \Omega_c + pj + j\right)\left(
    \frac{j}{B^2\Omega_c}\right)^p\,,
\end{eqnarray}
while the normalised maximum and equatorial angular velocities are
expressed as
\begin{eqnarray}
\label{eq:l1_u9}
  &&\lambda_1 := \frac{\Omega_{\rm max}}{\Omega_c} = \left( 1+ \left(
\frac{j_{\rm max}}{B^2 \Omega_c} \right)^p\right) \left(1 - \frac{j_{\rm
    max}}{A^2 \Omega_c} \right)\,,\\
\label{eq:l2_u9}
&&\lambda_2 := \frac{\Omega_{e}}{\Omega_c} = \left( 1+ \left(
\frac{j_{e}}{B^2 \Omega_c} \right)^p\right) \left(1 - \frac{j_{e}}{A^2
  \Omega_c} \right)\,.
\end{eqnarray}

\begin{figure*}
  \centering
  \includegraphics[width=0.45\textwidth]{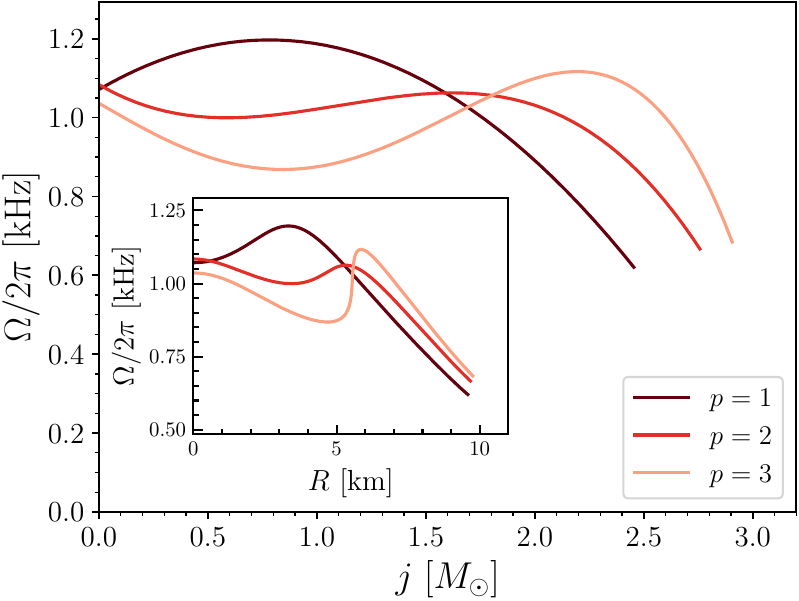}
  \hspace{0.5cm}
  \includegraphics[width=0.45\textwidth]{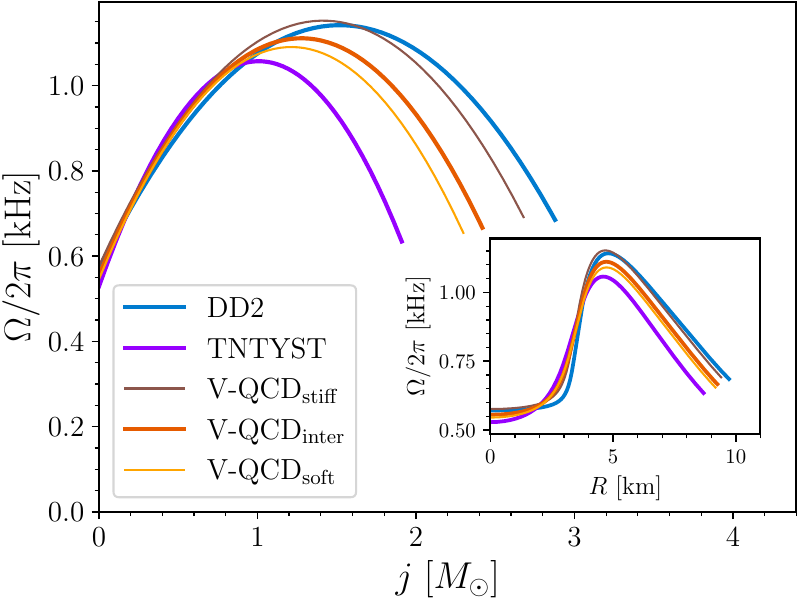}
  \caption{\textit{Left panel:} the same as in Fig.~\ref{U8_P_Q_var} but
    with the U9 differential-rotation law when keeping fixed
    $\hat{A}=1.4$ and $\hat{B}=1$ and varying the exponent $p= 1,
    2,3$. \textit{Right panel:} the same as in the left panel for $p=1$,
    $\lambda_1 = 2$, $\lambda_1 = 1.2$. The various lines refer to the
    different EOSs: DD2 (blue), TNTYST (purple) and V-QCD with different
    degrees of stiffness (shades of orange). Also in this case, the
    insets in the panels offer a view of the angular velocity as a
    function of $R$.}
  \label{U9_p123}
\end{figure*}

After a bit of algebra, it is possible to show that the expressions for
$A$ and $B$ appearing in \eqref{eq:l1_u9} and \eqref{eq:l2_u9} follow
from the solution of the quadratic equation in $B^2$
\begin{eqnarray}
 &&\hspace{-0.5cm} 0 = B^{4p}\Omega_{c}^2 \left[ j_{\rm max} (1-
    \lambda_2)-j_{e}(1-\lambda_1)\right] + j_{\rm max}^{p+1} j_{e}^p
  - j_{e}^{p+1} j_{\rm max}^p \nonumber \\
  &&\hspace{-0.5cm} \phantom{0 =} +B^{2p}\Omega_{c} \left[ j_{\rm max} j_{e}^{p+1}
    (1-\lambda_2) - j_{e} j_{\rm max}^p - j_{e}^{p+1}(1-\lambda_1)\right] \,.\nonumber \\
\end{eqnarray}
so that, after solving analytically for $B$, we can write an expression
for $A$ as
\begin{eqnarray}
    A^2 = \frac{1}{\Omega_{c}} \left[\frac{j_{\rm max}(B^2\Omega_{c})^p
      + j_{\rm max}^{p+1}}{(B^2\Omega_{c})^p(1-\lambda_1) + j_{\rm
        max}^p} \right]= \hat{A}^2 r_e^2 \,.
\end{eqnarray}

As anticipated, the rotation integral~\eqref{HydrostatEquil} for the U9
law can be found analytically for arbitrary parameters $p, \,q$.
\begin{eqnarray}
        \int_0^j j' \frac{d\Omega}{dj'} dj' &=& \Omega_{c} \Biggl[
          \frac{p}{p+1}\frac{j^{p+1} }{(B^2 \Omega_{c})^p }-
          \frac{p}{p+2} \frac{j^{p+2}}{(B^2 \Omega_{c})^p (A^2
            \Omega_{c})}  \nonumber \\ && 
             - \frac{j^2}{2 A^2 \Omega_{c}} -
          \frac{j^{p+2}}{p+2}\frac{1}{(B^2 \Omega_{c})^p (A^2
            \Omega_{c})} \Biggr] \,,
\end{eqnarray}
while, in order to obtain the rotation frequency on the equatorial plane,
we have to solve the Euler equation
\begin{eqnarray}
    && 0 = r_e^2 (\hat{\gamma}_{\rm e} + \hat{\rho}_e - \hat{\gamma}_p -
  \hat{\rho_p}) + \ln \left[ 1 - \left( \hat{\Omega}_e -
    \hat{\omega}_e\right)^2 e^{-2 r_e^2 \hat{\rho}_e} \right] \nonumber
  \\
    && + 2 \frac{\hat{\Omega}_{e}}{\lambda_2}\Biggl[
    \frac{p}{p+1}\frac{\hat{j}_e^{p+1} }{(\hat{B}^2 \hat{\Omega}_{e}/
      \lambda_2)^p }- \frac{p}{p+2} \frac{\hat{j}_e^{p+2}}{(\hat{B}^2
      \hat{\Omega}_{e}/ \lambda_2)^p (A^2 \hat{\Omega}_{\rm
        e}/\lambda_2)} \nonumber \\
    && - \frac{\hat{j}_e^2}{2 \hat{A}^2 \hat{\Omega}_{e}/\lambda_2} -
    \frac{\hat{j}_e^{p+2}}{p+2}\frac{1}{(\hat{B}^2 \hat{\Omega}_{e}/
      \lambda_2)^p (\hat{A}^2 \hat{\Omega}_{e}/\lambda_2)} \Biggr]\,,
\end{eqnarray}
with $j_{e}$ as in Eq. (\ref{j_e}), and then obtain the central
rotational frequency $\Omega_{c} = \Omega_{e}/\lambda_2$. 

We note that the numerical calculation of equilibrium models with the U9
law can become problematic and convergence be lost, at least with
the \texttt{Hydro-RNS} code. In particular, convergence often fails at
large values of $\lambda_2$, where $\Omega_e > \Omega_c$, as the iterated
equatorial radius $r_e$ alternates between two different solutions with
different masses and radii for the same value of the axis ratio (these
two branches merge again at small axis ratios). Considering that the
branch yielding lower masses also exhibits an unphysical relation between
the mass $M$ and the specific angular momentum $j$ (\ie $M(j)$ is no
longer injective with $j$) and non-smooth profiles in $\Omega(r)$ as well
as $M(j)$, we consider such branch as ``unphysical'' and discuss
hereafter only the results relative to the ``physical'' branch, that is,
the branch yielding larger masses for the same axis ratio.

Examples of solutions with the U9 law are presented in the left panel of
Fig.~\ref{U9_p123}, which shows how the rotation profile changes with the
parameter $p$ at a fixed axis ratio of $r_p/r_e = 0.8$ and a central
rest-mass density of $\rho_c = 8.024 \times 10^{14}\, {\rm g~cm^{-3}}$.
Note how the variation of the index $p$ allows for considerably different
functional behaviours of the rotation law and the appearance of an
additional minimum which is seen for $p=2,3$ (red and salmon lines). In
essence, such values of the parameter $p$ give rise to a region in the
star which is rotating slower than both the center and the surface on the
equatorial plane; while such a profile has not yet been observed in
binary mergers, it may appear in the presence of phase
transitions~\citep{Weih:2019xvw, Hanauske2021}. Also seen in
Fig.~\ref{U9_p123} is that the maximum in the angular velocity
$\Omega_{\rm max}$ shifts to larger radii and larger angular momenta. On
the other hand, the impact of variations in the parameters $\hat{A}$,
$\hat{B}$ or $\lambda_1$, $\lambda_2$ when using the U9 law is analogous
to that discussed above for the U8 law and reported in
Fig.~\ref{U8_P_Q_var}.

The right panel of Fig.~\ref{U9_p123} offers some representative rotation
profiles obtained with the U9 law for a fixed set of parameters, \ie
$p=1$, $q=1$, $\lambda_1 = 2$, $\lambda_2=1.2$, $\rho_c = 8.024 \times
10^{14}\,{\rm g~cm^{-3}}$, $r_p/r_e = 0.8$, and different EOSs. Note
that the angular-velocity profiles extends to larger values of angular
momentum for stiffer EOSs, with the DD2 being the stiffest and the
TNTYST~\citep{Togashi2017} the softest of the EOSs for such a value of
the central rest-mass density [a similar behaviour is seen also for the
  V-QCD EOS~\citep{Jarvinen:2011qe, Ishii:2019gta, Ecker2019} in the
  three different cases considered].

\begin{figure}
  \centering
  \includegraphics[width=0.9\columnwidth]{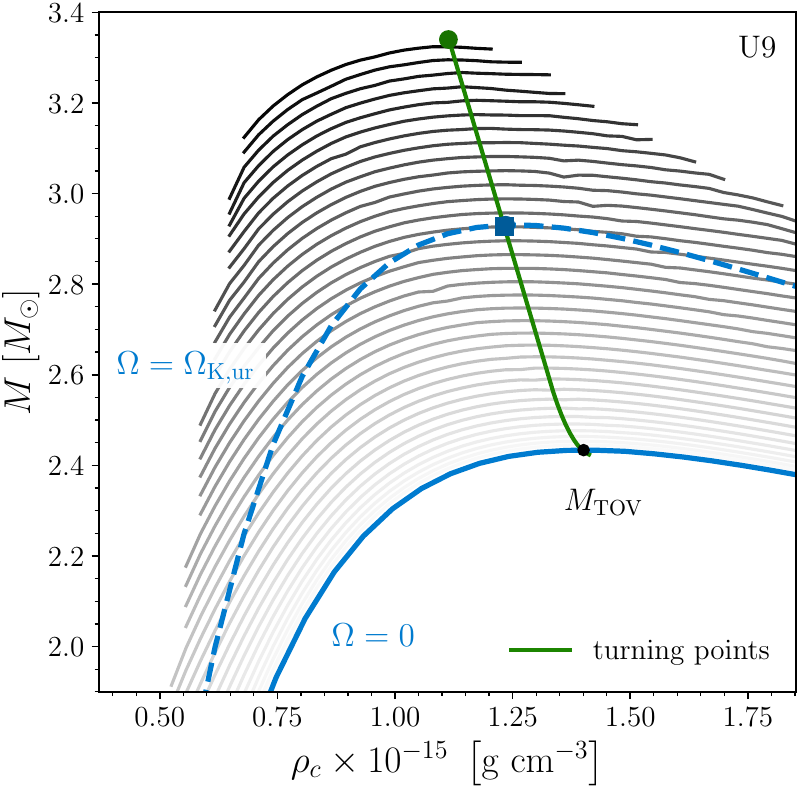}
  \caption{The same as in Fig.~\ref{U8_DD2_stability} but for the U9
    differential-rotation law.  Note that in this case the turning-point
    stability line (green solid line) intersects the Keplerian stability
    line at the maximum mass of Keplerian uniform-rotation models (blue
    filled square).}
  \label{U9_DD2_stability}
\end{figure}

We conclude the analysis of the U9 rotation law by considering the full
space of solutions after setting $p=1$, $q=1$, $\lambda_1 = 2$,
$\lambda_2=1.2$ (as in Fig.~\ref{U9_DD2_stability}) and when considering
the DD2 EOS. In analogy with Fig.~\ref{U8_DD2_stability}, sequences of
constant angular momentum are presented with a grey shading for $J=0.25,
\, 0.5, \,..., 9.25$, while the thick solid light-blue (blue) line shows
the static case with $\Omega=0$ (the Keplerian limit of uniform rotation
$\Omega_{\rm K,ur}$) and the green solid line indicates the stability
line obtained with the turning point criterion.

Interestingly, the turning-point stability line crosses the Keplerian
limit of uniform rotation exactly at its maximum $M_{\rm tp; K}=M_{\rm
  max; K,ur}$, which is marked by a light blue square and terminates at
the mass $M=3.34\,M_{\odot}$. Note also that the turning-point
stability line has a negative slope, \ie higher masses on the turning
point line are at smaller central densities, in analogy with what we
have seen in Fig.~\ref{U8_DD2_stability} for the U8 law and the DD2
EOS. As we will comment later on (see Sec.~\ref{sec:section5.2}), this
behaviour of the turning-point stability line is common but not the only
one possible.

\section{Differential-rotation laws from BNS merger simulations}
\label{sec:section3}

\begin{figure*}
  \centering
  \includegraphics[width=0.9\columnwidth]{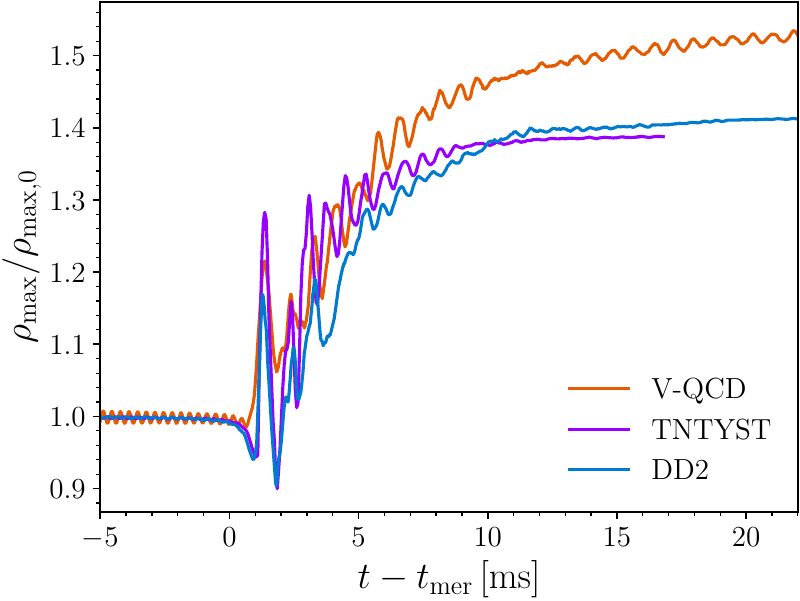}	
  \hspace{1.0cm}
  \includegraphics[width=0.9\columnwidth]{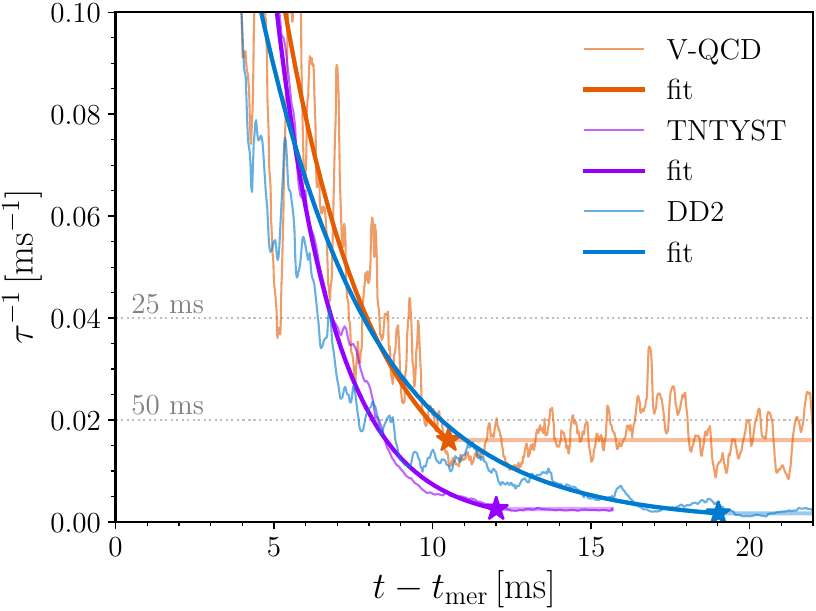}
  \caption{\textit{Left panel:} evolution of the central rest-mass
    density normalised to its initial value for the DD2 (blue line),
    TNTYST (violet line) and V-QCD EOS (orange line). \textit{Right
      panel:} evolution of the inverse of the equilibration timescale for
    the same EOSs as in the left panel. The bold lines represent the fit
    to the exponentially decaying part at early times and to the
    approximately constant part at later times. Grey dotted lines mark
    representative timescales.}
  \label{Tau_all}
\end{figure*}

When building differentially rotating equilibria that are meant to model
the HMNS produced in the merger of a binary system of neutron stars, it
is clear that the only way to obtain \textit{realistic}
differential-rotation profiles it to extract this information from
accurate and general-relativistic simulations of binary mergers using a
state-of-the-art microphysical description. Here, we consider the results
of BNS-merger simulations with three different temperature-dependent
EOSs: DD2~\citep{Hempel2009}, TNTYST~\citep{Togashi2017}, and the V-QCD
of intermediate stiffness~\citep{Demircik2020, Demircik:2021zll}.

More precisely, the first remnant we consider is the product of an
equal-mass BNS with component masses $M_{1,2} = 1.4\,M_{\odot}$ evolved
with the DD2 EOS by~\citet{Ng2023b}; since it survives for times
$t-t_{\rm mer} \gtrsim 100 \,{\rm ms}$ it represents a very good
candidate to study how the properties and rotation profiles change with
time and reach a quasi-stationary equilibrium. The second simulation
considers the merger of a magnetized equal-mass binary with component
masses $M_{1,2} = 1.275\,M_{\odot}$ and the TNTYST
EOS~\citep{Togashi2017}. In this case, a dipolar magnetic field is
present inside the star with maximum magnetic-field strength $B \sim
10^{14}\,{\rm G}$~\citep{Chabanov2022} and the merger remnant is an HMNS
that does not collapse promptly to a black hole. In practice, even
  the presence of an initial magnetic field of $10^{14}\,{\rm G}$ does
  not modify the dynamics of the remnant significantly and we observe a
  behaviour in terms of differential rotation that is very similar to the
  case of zero magnetic fields. It may be necessary to consider much
  larger magnetic-field strengths, \eg $10^{14}-10^{17}\,{\rm G}$ to
  appreciate differences in the dynamics and equilibria of the remnant.
Finally, the third merger remnant refers to an unequal-mass binary with
mass ratio $q:= M_2/M_1 = 0.7$, where $M_1 = 1.64\, M_{\odot}$ and $M_2 =
1.15\,M_{\odot}$. The binary is evolved with the V-QCD EOS of
intermediate stiffness, where the temperature dependence is introduced at
low densities via the HS-DD2 EOS (including relativistic mean-field
interactions)~\citep{Hempel2010, Typel:2009sy} and in the dense
nuclear-matter phase via a van der Waals model~\citep{Demircik:2021zll}.
In this case, the merger remnant has a lifetime of $t-t_{\rm mer} \gtrsim
37 \,{\rm ms}$, which, as we will show below, is sufficient to extract
robust rotation profiles. The basic properties of the binaries considered
here and some of the corresponding dynamical properties have been
summarised in Table~\ref{tab:merger_core}.

\begin{table*}
  \centering
  \caption{Properties of the BNS merger configurations. Reported are the
    initial masses $M_1$, $M_2$, the EOS, and the collapse time $t_{\rm
      coll}$. Also reported are the coefficients of the linear fit
    \eqref{eq:Om_rho}, the time $t-t_{\rm mer}$ at which is performed,
    and the corresponding values for the core radius $R_{\rm core}$, the
    rest-mass density $\rho_{\rm core-disc}$ and angular velocity
    $\Omega_{\rm core}$ of the core.}
  \label{tab:merger_core}
  \begin{tabular}{lccccccccc}
    \hline
    EOS & $M_1$ & $M_2$ & $t_{\rm coll}$ & $t-t_{\rm mer}$ & $a\times 10^{-2}$ & $b$ & $\rho_{\rm core-disc}\times 10^{-13}$ & $\Omega_{\rm core}/2\pi$ & $R_{\rm core}$ \\
    & $[M_{\odot}]$ & $[M_{\odot}]$ &$[\rm ms]$  &$[\rm ms]$ & $[{\rm kHz}]$ &  $[{\rm kHz}]$ & $[{\rm g~cm^{-3}}]$ & $[{\rm kHz}]$ & $[{\rm km}]$ \\
    \hline 
    \texttt{DD2}    & $1.400$ & $1.400$ & $>100$ & $20$ & $1.480$ & $2.456$ & $5.963$ & $1.088$ & $11.982$ \\
    \texttt{TNTYST} & $1.275$ & $1.275$ & $>20$  & $13$ & $1.957$ & $3.086$ & $3.355$ & $1.157$ & $11.129$ \\
    \texttt{V-QCD}  & $1.640$ & $1.150$ & $>37$  & $12$ & $2.038$ & $3.084$ & $2.911$ & $1.060$ & $12.023$\\
    \hline
  \end{tabular}
\end{table*}

\subsection{Expressions in the 3+1 decomposition}
\label{3+1_rotation}

The simulations of BNS mergers are performed with the \texttt{Einstein
  Toolkit}~\citep{loeffler_2011_et, EinsteinToolkit_etal:2020_11} within
a 3+1 decomposition of spacetime, where the hydrodynamical variables and
the metric coefficients are stored on a Cartesian grid on spacelike
hypersurfaces~\citep{Rezzolla_book:2013}. To extract the necessary
information from the simulations output, we make use of the open-source
post-processing tool \texttt{Kuibit}~\citep{kuibit21}, which is able to
analyze different quantities from the \texttt{Einstein Toolkit}. However,
since the simulation data is produced on a Cartesian grid while the
equilibria are computed by \texttt{RNS} assuming a quasi-isotropic
spherical coordinates system, a coordinate transformation is necessary to
compute the angular velocity, the three-velocities and angular momentum
in Cartesian coordinates. 

More specifically, in the standard $3+1$ decomposition of spacetime where
$\alpha$ is the lapse function, $\boldsymbol{\beta}$ the shift vector,
$\boldsymbol{\gamma}$ the spatial part of the metric, and
$\boldsymbol{u}$ the fluid four-velocity, we can compute the Lorentz
factor as
\begin{eqnarray}
  W := \alpha u^t = \frac{1}{\sqrt{1 - v^{i} v_{i}}} \,.
  \label{lorentz_3+1}
\end{eqnarray}
The expression for the angular velocity and for the angular momentum can
be derived recalling that $u_{i} = v_{i} \alpha u^{t}$, so that the
angular velocity is given by
\begin{eqnarray}
  \Omega := \frac{u^{\phi}}{u^{t}} = \alpha v^{\phi} - \beta^{\phi}  \,,
\end{eqnarray}
while the specific angular momentum is
\begin{eqnarray}
j := u^t u_{\phi} = u^t v_{i} \alpha u^t = \frac{W^2}{\alpha} v_{\phi} \,.
\end{eqnarray}
The corresponding quantities in Cartesian coordinates are instead given
by 
\begin{eqnarray}
\Omega = \alpha \left( \frac{x v^{y} - y v^{x}}{x^2 + y^2} \right) -
\frac{x \beta^{y} - y \beta^{x}}{\sqrt{x^2 + y^2}} \label{Om3p1} \,,
\end{eqnarray}
and 
\begin{eqnarray}
 j &=& \frac{W^2}{\alpha}\Biggl[ \left( \frac{x v^{y} - y v^{x}}{x^2 + y^2}
 \right) \left( y^2 \gamma_{xx} - 2 xy \gamma_{xy} + x^2 \gamma_{yy}
 \right) +
 \nonumber \\ &&
 \phantom{\frac{W^2}{\alpha}\Biggl[ }
 \left( \frac{x v^x + y
   v^y}{\sqrt{x^2 + y^2}} \right) \left(\frac{yx}{\sqrt{x^2 + y^2}}
 \gamma_{xx} + \frac{y^2}{\sqrt{x^2 + y^2}} \gamma_{xy} \right.- 
 \nonumber  \\ &&
 \phantom{\left( \frac{x v^x + y v^y}{\sqrt{x^2 + y^2}}\right)\biggl(}
 \left. \frac{x^2}{\sqrt{x^2 + y^2}} \gamma_{yx} -
 \frac{xy}{\sqrt{x^2 + y^2}} \gamma_{yy} \right) +
 \nonumber \\ && 
 \phantom{\frac{W^2}{\alpha}\Biggl[ }
 v^z \left( y \gamma_{xz} - x \gamma_{yz} \right)\Biggr]
 \,. 
 \label{jmom_3p1}
 \end{eqnarray}
 
\subsection{Quasi-stationarity and rotation profiles}
\label{sec:qs_rp}

\begin{figure*}
  \centering
  \includegraphics[width=0.31\textwidth]{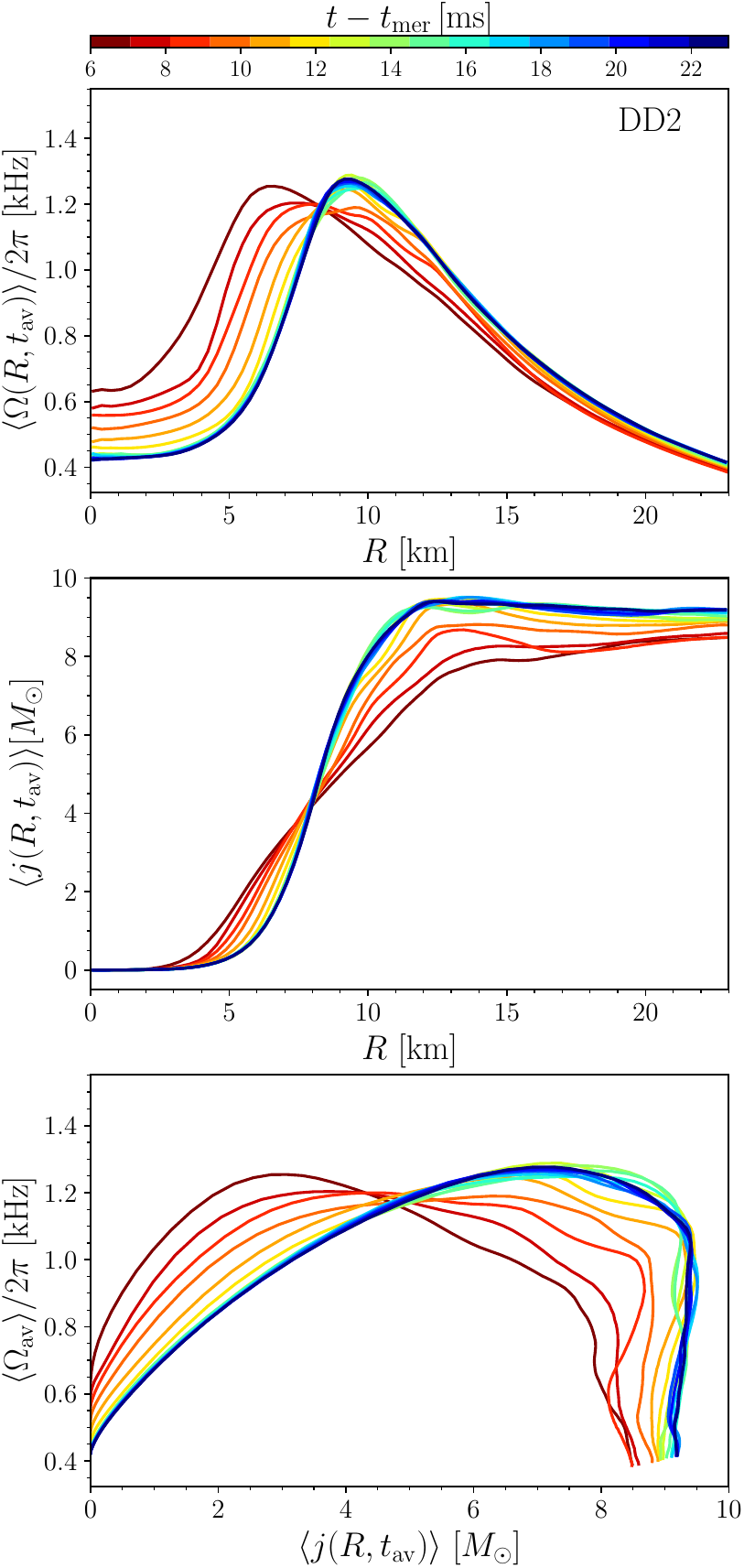}
  \includegraphics[width=0.31\textwidth]{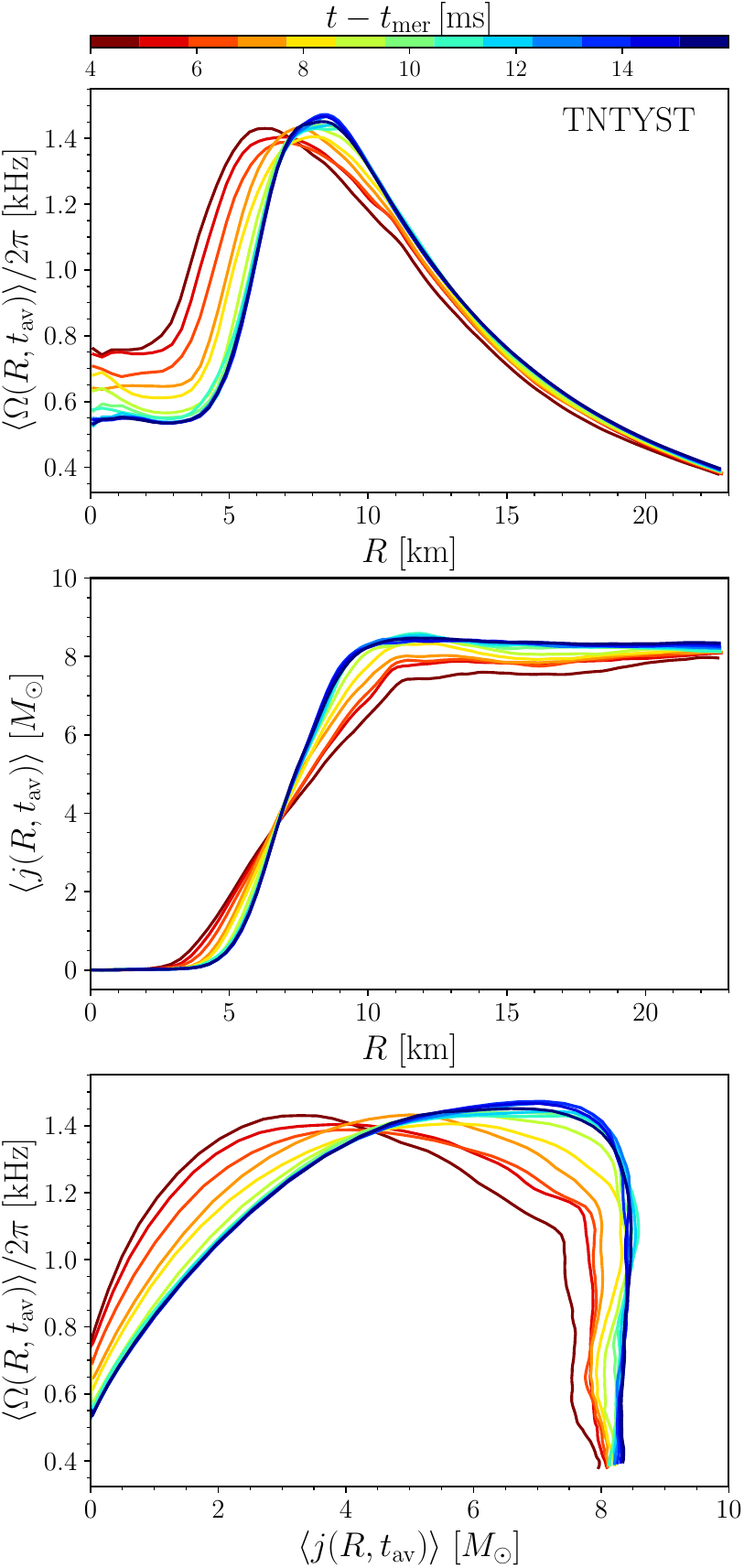}
  \includegraphics[width=0.31\textwidth]{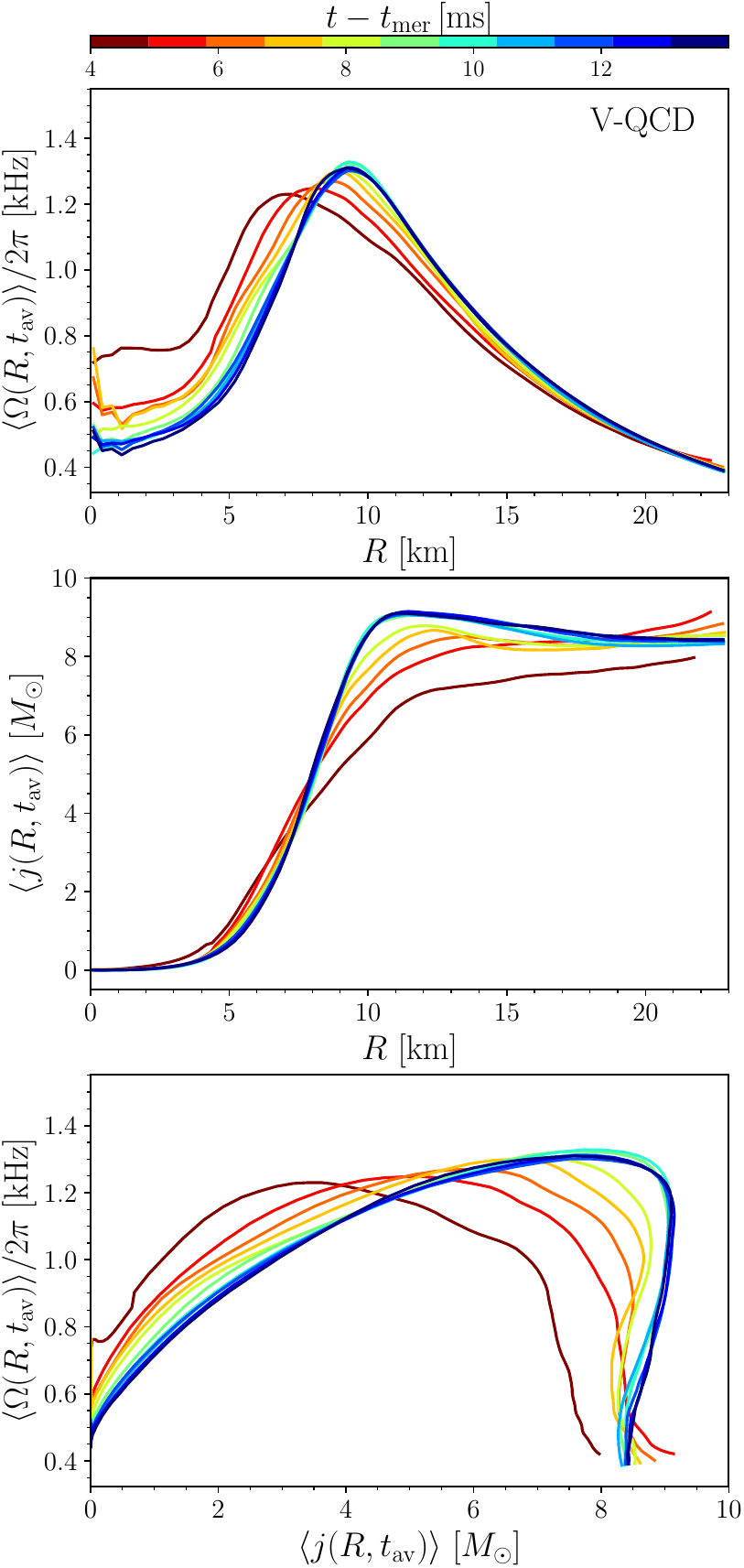}
  \caption{Evolution of the rotational properties of the BNS merger
    remnant obtained with the DD2 EOS (left column), the TNTYST EOS
    (middle column) and the V-QCD EOS (right column). In each panel,
    coloured lines following the colourmap at the top show the different
    profiles as measured at different average times $t_{\rm av}$ after
    the merger with a cadence of $1\,{\rm ms}$. From top to bottom, the
    different rows show the angular velocities and the specific angular
    momentum as a function of $R$, as well as the angular velocity as a
    function of $j$ for the three EOSs. All quantities show are averaged
    in azimuthal angle and time (see also Fig.~\ref{DD2_outAll} for a
    higher cadence).}
  \label{fig:Data_1ms_Av}
\end{figure*}

For each of the simulations discussed above, we analyse the post-merger
rotation profile of the long-living remnants, extracting rotation
quantities such as the angular velocity $\Omega$ and the specific angular
momentum $j$. Furthermore, since the merger remnant has an equilibrium
that varies rapidly right after the merger, attention needs to be paid to
make sure that the rotational properties are measured when the remnant
has reached a quasi-stationary equilibrium. Thus, we define a time
window after the merger with extremes $t-t_{\rm mer} = t_1$ and $t-t_{\rm
  mer} = t_2$ (where $0< t_1 < t_2$) during which we monitor the
evolution of properties of the merger remnant. In particular, we use
$t_1$ as a reference time from when the system has started to reach a
quasi axisymmetric equilibrium and $t_2$ to mark the time when this
equilibrium has been reached to a given tolerance.

More specifically, while any hydrodynamical quantity could be employed to
this scope, we take the evolution of the maximum rest-mass density of the
merger remnant normalised to the initial value as a good proxy for the
measurement of $t_1$ and $t_2$. The left panel of Fig.~\ref{Tau_all}
reports the evolution of this quantity for the three EOSs considered here
and represented with a colorcode that we will employ hereafter, and
clearly shows that while the inspiral is characterised by an
almost-constant behaviour in time, the merger is marked by a large
increase accompanied by large-amplitudes oscillations that are gradually
damped in time, thus leading to a new, almost-constant behaviour. Using
this information, we can define an ``equilibration timescale'' $\tau :=
\rho_{\rm max}(t)/\dot{\rho}_{\rm max}(t)$, with a dot indicating a time
derivative, and whose evolution is shown in the right panel of
Fig.~\ref{Tau_all}. While this timescale has high-frequency oscillations
(thin solid lines), it also exhibits a clear exponential-decay behaviour
from $t-t_{\rm mer}=t_1$ that we capture with a fit of the type
$\ln(\tau)= m (t-t_{\rm mer}) + b$ with slope $m<0$ (thick solid
lines). The exponential decay is followed by an evolution in which $\tau
\simeq {\rm const}.$ and we use the transition between the exponential
decay and the constant-in-time evolution to mark $t-t_{\rm mer}=t_2$
(these times are shown with filled stars in Fig.~\ref{Tau_all}). As a
result, the corresponding values for the three remnants are 
\begin{align}
  \quad\bullet~ &{\rm DD2:}&        &t_1=6 \,{\rm ms}\,, &t_{2}=19.0 \,{\rm ms}\,,
  \nonumber \\
  \quad\bullet~ &{\rm TNTYST:}&     &t_1=4 \,{\rm ms}\,, &t_{2}=12.0 \,{\rm ms}\,,
  \nonumber \\
  \quad\bullet~ &{\rm V\!-\!QCD:}&  &t_1=4 \,{\rm ms}\,, &t_{2}=10.5 \,{\rm ms}\,.
  \nonumber
\end{align}
Note in the right panel of Fig.~\ref{Tau_all} that the equilibration
timescales are different for the various EOSs and that $\tau$ reaches a
constant value rather rapidly for the V-QCD EOS, but also that this value
is larger than for the other two (dotted grey lines report the
corresponding values of the equilibration timescale in
milliseconds). This is is also reflected in the left panel of
Fig.~\ref{Tau_all}, where the corresponding normalised rest-mass density
continues to maintain small oscillations as a result of the original mass
asymmetry in the binary. Similarly, note that even though the remnant of
the DD2 merger has the largest equilibration timescale, it reaches it
only at a comparatively later time.

Having defined the window in time where it is interesting to study the
evolution of the rotational properties of the merger remnant, we
introduce azimuthal and time-averaged representation of the angular
velocity and of the specific angular momentum, namely, 
\begin{eqnarray}
\label{eq:Omega_avrg}
  \langle \Omega(R,t_{\rm av}) \rangle &:=& \frac{1}{2\pi\Delta
  t}\int^{2\pi}_{0} \int^{t_{\rm av}
  + \Delta t/2}_{t_{\rm av} - \Delta t/2}\Omega(R,\phi,t)d\phi\,dt\,, \\
\label{eq:j_avrg}
\langle j(R,t_{\rm av}) \rangle &:=& \frac{1}{2\pi\Delta t}
\int^{2\pi}_{0} \int^{t_{\rm av} + \Delta t/2}_{t_{\rm av} - \Delta t/2}
j(R,\phi,t)d\phi\,dt \,.
\end{eqnarray}
where we set $\Delta t=1 \, {\rm ms}$ and the averaging window is taken
to vary in the interval $t_1 < t_{\rm av} < t_2$. Using these averaged
quantities \eqref{eq:Omega_avrg} and \eqref{eq:j_avrg}, it is possible to
study the behaviour of the rotation profiles $\Omega(j)$ and how they
change from soon after the merger (\ie $t_{\rm av} \simeq t_1$) to the
time when the HMNS has reached a quasi-stationary equilibrium (\ie
$t_{\rm av} \simeq t_2$).

Figure~\ref{fig:Data_1ms_Av} presents the evolution of $\langle \Omega
\rangle$ (top row) and $\langle j \rangle$ (bottom row) shown as a
function of the proper radius $R$ for the remnants with three different
EOSs (different columns) and with the averaging time $t_{\rm av}$ having
a cadence of $1\,{\rm ms}$ as being indicated by the colormap going from
$t_1$ (red) to $t_2$ (blue; Fig.~\ref{DD2_outAll} in
Appendix~\ref{sec:appendixA} offers a similar evolution for the DD2 EOS
where $t_{\rm av}$ is chosen across the interval $5 \, {\rm ms} <
t-t_{\rm mer} < 23 \, {\rm ms}$ but with a finer time interval, \ie
$\Delta t = 0.1 \, {\rm ms}$.)

Clearly, a very similar behaviour is shown by all remnants in
Fig.~\ref{fig:Data_1ms_Av}, with only minor differences related to the
EOS and the mass ratio. In particular, at early times $t_{\rm av}
\gtrsim t_1$ (red lines) the angular velocities have maxima $\Omega_{\rm
  max}$ which are larger than at later times and are positioned closer to
the center (top row); the angular profile is essentially Keplerian
outside the maximum and can be associated to the low rest-mass density
``disc'' around the HMNS~\citep{Kastaun2014, Hanauske2016}; a similar
behaviour has been seen also in the actual discs produced when the HMNS
collapses to a BH~\citep[see, \eg][]{Rezzolla:2010}. Hence, the position
of the maximum of the angular velocity can be taken as to mark the
``outer edge'' of the HMNS and the ``inner edge'' of the disc surrounding
it. Between $t_1$ and $t_2$, the maxima move to smaller values of the
angular velocity and outwards from the center (yellow and green
lines). At late times, $t_{\rm av} \lesssim t_2$ (blue lines), the
changes in time are much smaller and the maxima in the angular velocities
attain their equilibrium values and positions. Importantly, note that the
``core'' of the HMNS is rotating almost uniformly and it slows down and
grows in size between $t_1$ and $t_2$ for all the binaries considered,
reaching a size of $\simeq 5\,{\rm km}$~\citep[see also][]{Kastaun2014,
  Hanauske2016}.

A similar behaviour is shown also by the specific angular momentum
(middle row), which is however monotonic and joins two almost-constant
states in the core of the HMNS and in its disc. Unsurprisingly, the
presence of a uniformly-rotating core in the HMNS is clear also in terms
of the specific angular momentum, which is very constant and close to zero,
and whose size increases with time reaching the quasi-stationary size of
of $\simeq 5\,{\rm km}$. Note also that as time progresses the specific
angular momentum in the HMNS interior (not only in the core) decreases,
while it increases outside and in the HMNS disc; this happens at $R\simeq
10\,{\rm km}$. A slight deviation from the behaviour described above is
seen for the V-QCD EOS but we attribute this mostly to the fact that the
data refers to a binary with a considerable difference in mass and for
which the disc may reach a constant state on longer timescales than those
considered here.

\begin{figure}
  \centering
  \includegraphics[width=0.9\columnwidth]{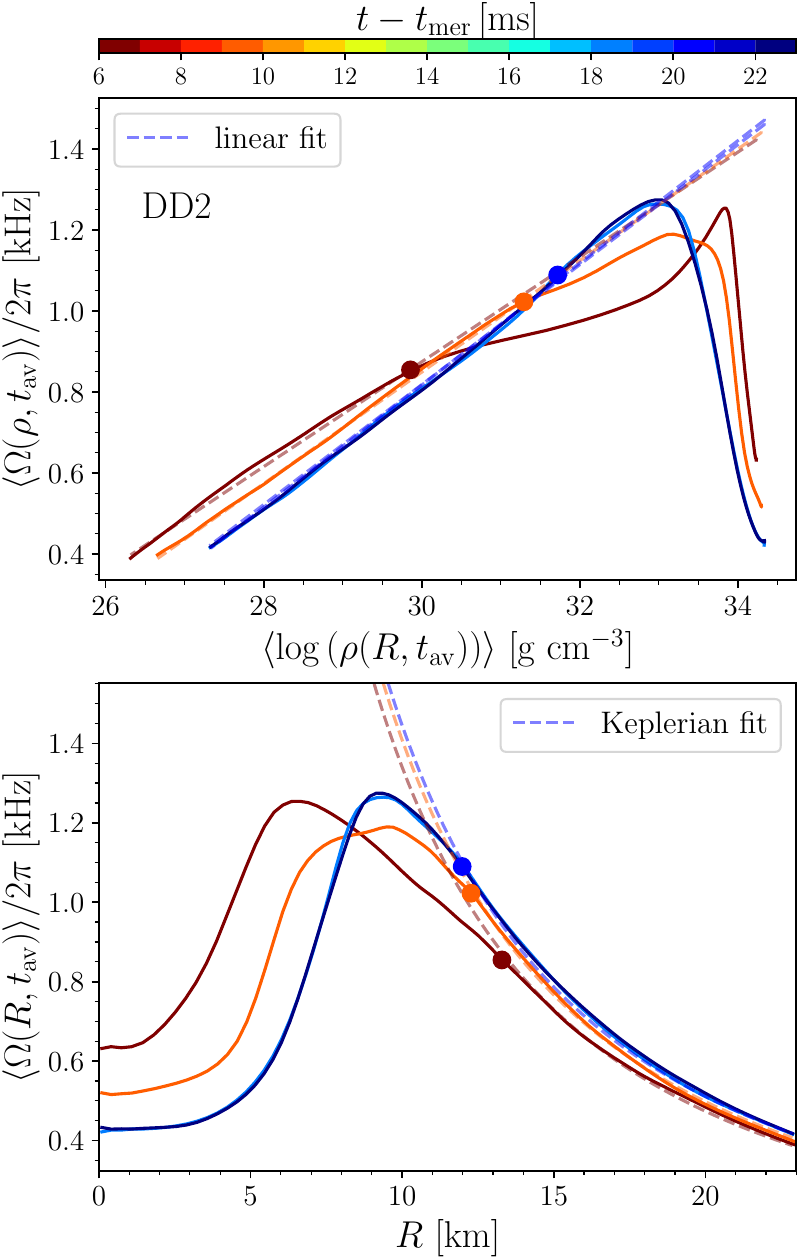}
  \caption{\textit{Top panel:} angular velocity as a function of the
    logarithm of the rest-mass density for the BNS remnant modeled with
    the DD2 EOS at three different averaging times after the merger
    $t_{\rm av}=6,9,18,22\,{\rm ms}$ (see colourmap on the
    top). Indicated with dashed lines are the linear fits and marked with
    coloured filled circles are the locations of the departures from a
    linear behaviour. \textit{Bottom panel:} the same as in the top panel
    but when the angular velocity is shown as a function of $R$. The
    dashed lines reproduce the Keplerian fall-off of $R^{-3/2}$ and it
    should be noted that the fit is starts to be valid from the position
    of the three filled circles, which can therefore be taken to
    distinguish the core from the disc in the HMNS.}
  \label{disc_core}
\end{figure}

Finally, the bottom row of Fig.~\ref{fig:Data_1ms_Av} presents the evolution
of $\langle \Omega \rangle$ as a function of $\langle j \rangle$, \ie
$\langle \Omega \rangle(\langle j \rangle)$, which is instructive, since
this is the quantity that needs to be integrated to compute the first
integral of the hydrostationary equilibrium [see
  Eq.~\eqref{eq:Eul_3}]. Obviously, this quantity combines the behaviours
discussed in the top and middle row and reflects how, from $t_1$ to
$t_2$, the decrease of specific angular momentum leads to a decrease of
$\langle \Omega \rangle(\langle j \rangle)$ in the HMNS interior and to a
corresponding increase in the exterior, as expected from a rough
conservation of angular-momentum (we recall that angular momentum within
the HMNS is not perfectly conserved as it is radiated via GWs and lost in
part via shocks). In the following sections we will use the numerical
profiles of $\langle \Omega \rangle(\langle j \rangle)$ at $t_2$ to
construct realistic equilibrium models of differentially rotating stars.

\subsection{Distinguishing core and disc in the remnant}
\label{sec:section3.1}
A problem frequently arising when modelling an HMNS is to distinguish the
actual ``core'' from what is normally referred as ``disc''. This
distinction is often made in terms of cut-off densities, \eg
$10^{13}\,{\rm g/cm}^3$. While this choice is reasonable, it is
fundamentally arbitrary and any other cut-off density could be chosen in
principle, hence leading to different results.

\begin{figure*}
  \centering
  \includegraphics[width=0.95\textwidth]{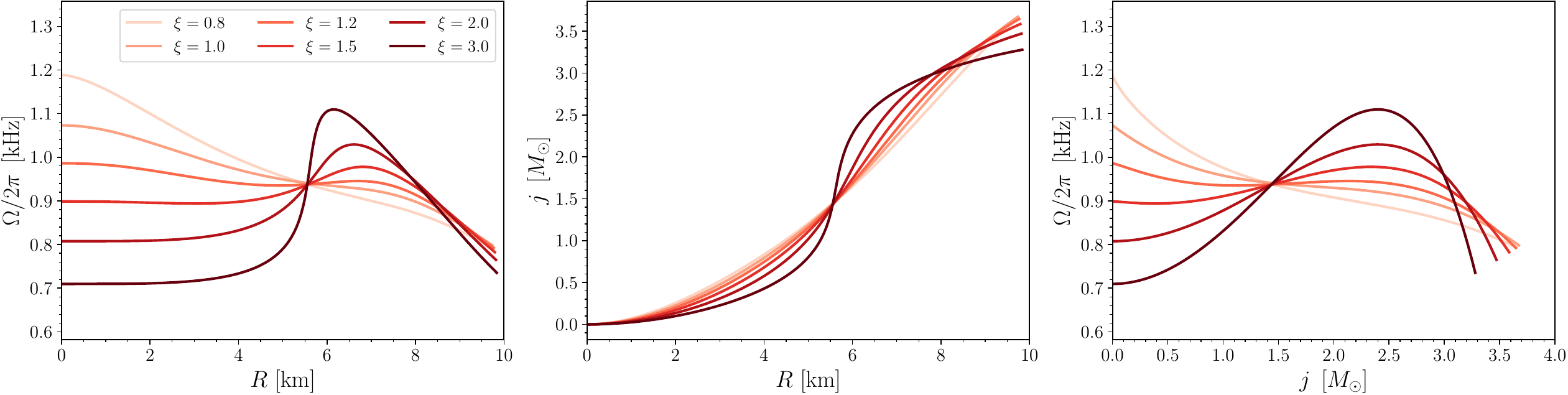}
  \caption{Rotational properties of stellar models obeying the U9+
    differential-rotation law. From left to right we show the radial
    profiles of the angular velocity (left panel), the radial profile of
    the specific angular momentum (middle panel) and the angular velocity
    as a function of the specific angular momentum (right panel). The
    different curves are obtained when keeping fixed the parameters $p=2,
    \hat{A}=1.8, \hat{B}=1.4$ and varying the exponent $\xi \in
        [0.6,3]$; also in this case, the stellar model obeys the DD2 EOS,
        has a central rest-mass density $\rho_c = 8.024 \times 10^{14}\,
        {\rm g~cm^{-3}}$, and an axis ratio $r_p/r_e = 0.8$.}
  \label{U9q_Q_var}
\end{figure*}

To counter this arbitrariness and define a criterion that is grounded on
the physical properties of the HMNS once a quasi-stationary configuration
has been reached, we have investigated the functional dependence of the
remnant's angular velocity as a function of the rest-mass density at
different times after the merger. This is shown in the top panel
Fig.~\ref{disc_core}, where we report the azimuthally averaged and over a
$1\,{\rm ms}$-window averaged angular-velocity profiles as a function of
the logarithm of the averaged (in time and azimuth) rest-mass density for
the DD2 EOS and at four different times, \ie $t-t_{\rm mer} = 6, 9, 18$
and $22\,{\rm ms}$ (brown, orange, light-blue, and dark-blue,
respectively; note that the last lines are essentially overlapping because
of the attained quasi-stationarity). Note the interesting linear
behaviour of $\Omega=\Omega(\log(\rho))$, which can be effectively fitted
with a simple function of the type
\begin{equation}
  \label{eq:Om_rho}
  \Omega(\rho)=a \log(\rho) + b\,,
\end{equation}
with
parameters $a$ and $b$ that are functions of time and are reported in
Table~\ref{tab:merger_core}. Since the linear (in log) fitting works well
for the low-density regions of the disc but fails as the densities
increase and the core is approached, we exploit the change of behaviour
to mark the transition between the disc and the core. This is shown with
filled circles, whose positions vary in time and highlight that the core
in the HMNS actually ``shrinks'' over time but soon settles to a
quasi-stationary value. Obviously, as the core shrinks, the inner
position of the disc also decreases, as can be appreciated from the
bottom panel of Fig.~\ref{disc_core}, which reports the angular-velocity
profile as a (linear) function of the radius and where the filled circles
are the same as in the top panel.

A few remarks are worth making. First, note that the point of departure
from a linear-in-log behaviour of the angular velocity with the rest-mass
density coincides with the corresponding departure of the angular
velocity from a Keplerian profile with radius. This is not surprising but
underlines that the point of departure can be interpreted either as the
``beginning'' of the Keplerian disc, \ie $R_{\rm core}$ or as the ``end''
location of where the density in the HMNS stops growing as a power-law
with radius, \ie $\rho_{\rm core-disc}$. Second, note that the
``core-disc'' transition does not coincide with the maximum of the
angular velocity, which systematically lies at smaller radii; this point,
which was already remarked by~\citet{Hanauske2016}, warns against using
the angular velocity maximum as a particularly significant location in
the HMNS. Third, note that at least in the limited sample considered here
$\rho_{\rm core-disc}$ varies by a factor two at most, \ie between
$\simeq 3 \times 10^{13}\,{\rm g/cm}^3$ and $\simeq 3 \times
10^{13}\,{\rm g/cm}^3$ and $\simeq 6 \times 10^{13}\,{\rm g/cm}^3$ (see
Table \ref{tab:merger_core}). Finally, note that $R_{\rm core}$ has a
much smaller variance (\ie $\lesssim 7\%$) and while a more systematic
analysis may constrain this value more precisely, it is interesting that
very different binaries lead to essentially the same core radius.

All things considered, the suggestion of determining the start of the
disc as the location where the angular velocity does no longer follow a
power-law with the rest-mass density appears to be easy to employ and
physically well-grounded; furthermore it points out that values
$\rho_{\rm core-disc} \simeq 4 \times 10^{13}\,{\rm g/cm}^3$ and
$R_{\rm core} \simeq 12\,{\rm km}$ can be taken as a reasonable
references for the core density and radius, respectively.

\section{New and realistic differential-rotation laws}
\label{sec:section5}

Having discussed in detail in the previous section the laws of differential
rotation that are actually measured from BNS merger simulations, we note
that the traditional prescriptions to model differential rotation laws
presented in Sec.~\ref{sec:section4} generically offer difficulties in
generating stellar models that have, at the same time, large radii and
large angular-velocity maxima. This motivates us to seek new differential
rotation laws that are either extensions of the Uryu laws
(Sec.~\ref{sec:section5.1}) or novel formulations (Sec.~\ref{sec:section5.2}),
and that provide a very accurate description of what is measured from fully
nonlinear simulations. In what follows, we discuss such new
differential-rotation laws and their properties.

\subsection{The extended Uryu-9 rotation law}
\label{sec:section5.1}

The simplest way to produce a differential-rotation law that provides a
better match with the results of the simulations is to ``extend'' the U9
law via the introduction of a new exponent $\xi$ whose role is to
increase the magnitude of the angular-velocity maximum, so that
expression~\eqref{eq:U9} can now be written as
\begin{eqnarray}
    \Omega (j) = \Omega_{c}\left( 1 + \left( \frac{j}{B^2
      \Omega_c}\right)^p\right)\left( 1 - \left( \frac{j}{A^2\Omega}
    \right)^{\xi}\right) \,.
\end{eqnarray}
such that $\xi=1$ yields the standard U9 law and $\xi \neq 1$ leads to the
extended U9 law, or ``U9+'' hereafter. Also in this case, it is possible
to compute analytically the first and second-order derivatives of
$\Omega(j)$ with respect to the specific angular momentum and obtain
respectively
\begin{eqnarray}
  &&\frac{d \Omega}{dj} = \Omega_{c} \Biggl[ p \frac{j^{p-1}}{(B^2
      \Omega_c)^p} \left( 1- \left( \frac{j}{A^2 \Omega_c}
    \right)^{\xi}\right) \nonumber \\
    && \phantom{\Omega_{c}}- \xi \frac{j^{\xi -1}}{(A^2\Omega_c)^{\xi}}
       \left( 1+ \left( \frac{j}{B^2 \Omega_c} \right)^{\xi}\right)
       \Biggr]\,, 
\end{eqnarray}
and 
\begin{eqnarray}
    && \frac{d^2 \Omega}{d j^2} = \Omega_c \Biggl[ p(p-1)
    \frac{j^{p-2}}{(B^2\Omega_c)^p}\left( 1 - \left(\frac{j}{A^2
      \Omega_c}\right)^{\xi}\right) \nonumber \\
    && - p \frac{j^{p-1}}{(B^2 \Omega_c)^p}\xi
    \frac{j^{\xi -1}}{(j/(A^2 \Omega_c))^{\xi}}
    - p \frac{j^{p-1}}{(B^2\Omega_c)^p} \left( \xi
    \frac{j^{\xi -1}}{(A^2\Omega_c)^{\xi}}\right) \nonumber \\
    && - \left( 1+ \left(
    \frac{j}{B^2 \Omega_c} \right)^p\right) \frac{\xi(\xi -1)
      j^{\xi-2}}{(j/(A^2\Omega_c))^{\xi}} \Biggr] \,. \nonumber
\end{eqnarray}

In this case, the rotation integral is given by 
\begin{eqnarray}
  && \int j' \frac{d\Omega}{dj'} dj' = \nonumber \\
  && \Omega_{c}\Biggl[
    \frac{p}{p+1}\frac{j^{p+1} }{(B^2 \Omega_{c})^p }-
    \frac{p}{p+\xi +1} \frac{j^{p+\xi +1}}{(B^2 \Omega_{c})^p (A^2
      \Omega_{c})^{\xi}} \nonumber \\
  && - \frac{\xi}{\xi +1}\frac{j^{\xi +1}}{( A^2
      \Omega_{c})^{\xi}} - \frac{\xi}{p+\xi +1}\frac{j^{p+\xi
        +1}}{(B^2 \Omega_{c})^p (A^2 \Omega_{c})^{\xi}} \Biggr]
  \,.
\end{eqnarray}
and, as for the U9 law, we can compute $B$ from the solution of the
quadratic equation 
\begin{eqnarray}
 &&\hspace{-0.5cm} 0 = B^{2p}\Omega_{c}^{p} \biggl\{ B^{2p}
  \Omega_{c}^{p}\left[ j_{\rm max}^{\xi} (1-
    \lambda_2)-j_{e}^{\xi}(1-\lambda_1)\right] + j_{\rm max}^{p+\xi}
  (1-\lambda_2)\nonumber \\ && - j_{e}^{\xi} j_{\rm max}^p + j_{e}^{p}
  j_{\rm max}^{\xi} - j_{e}^{p+\xi}(1-\lambda_1) \biggr\} - j_{\rm
    max}^{p+\xi} j_{e}^p - j_{e}^{p+\xi} j_{\rm max}^p\,.\nonumber \\
\end{eqnarray}
so that, again, after solving analytically for $B$, we can write an
expression for $A$ as
\begin{eqnarray}
   A^2 = \frac{1}{\Omega_c} \left( \frac{-(B^2 \Omega_c)^p F_{\rm
        max}^{\xi} - j_{\rm max}^{p+\xi}}{(\lambda_1 - 1)(B^2\Omega_c)^p
      - j_{\rm max}^p} \right)^{\frac{1}{\xi}} \,.
\end{eqnarray}

In Fig.~\ref{U9q_Q_var} we report some representative U9+ solutions
obtained for stars modeled with the DD2 EOS with a fixed central
rest-mass density of $\rho_c = 8.024 \times 10^{14}\, {\rm g~cm^{-3}}$
and an axis ratio of $r_p/r_e = 0.8$. The different curves refer to
different values of the exponent $\xi$, while the other parameters in the
U9+ law, namely, $\hat{A}=2.5, \, \hat{B}=1.8, p=2$, are kept fixed. As
seen from the left and right panels, values $\xi<3$ tend to change the
values of the angular velocity at the center and beginning of the
disc. In particular, for $\xi < 1.2$ the angular-momentum maximum is
reached at the center of the star, and for $\xi < 1.0$ the
angular-velocity essentially decreases monotonically with radius, thus
behaving in a way that is similar to the $j$-constant law.

\begin{figure}
  \centering
   \includegraphics[width=0.9\columnwidth]{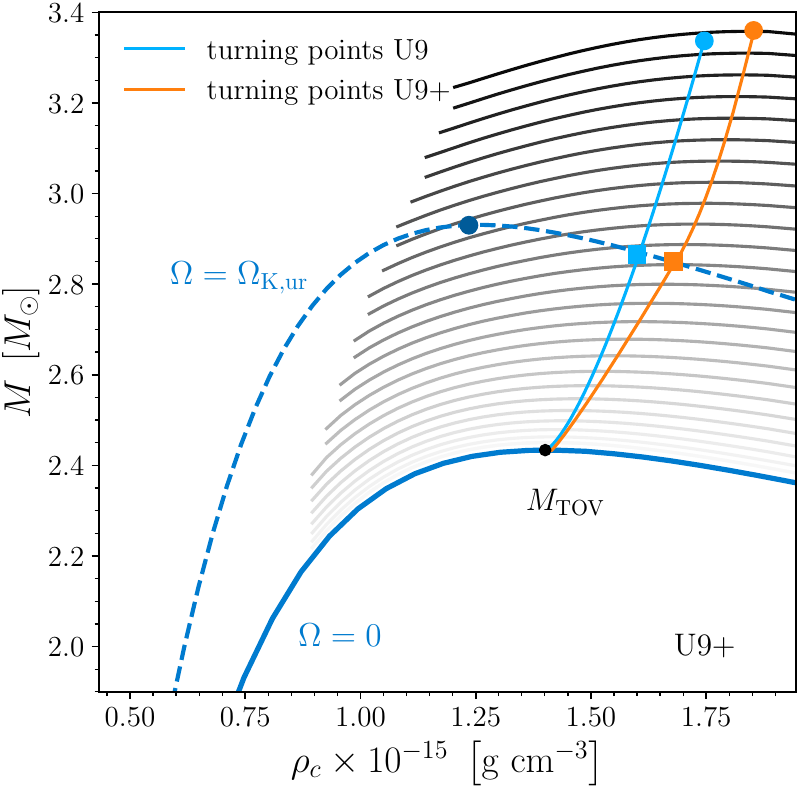}
   \caption{The same as in Fig.~\ref{U8_DD2_stability}, but for the new
     U9+ differential-rotation law (orange line) and for a different
     parameterisation of the U9 law (light-blue line; see also
     Fig.~\ref{U9_DD2_stability}). Note that for these rotation laws, the
     turning-point stability line has a positive slope (see also
     Fig.~\ref{U9Q_DD2_stability})}
  \label{U9Q_DD2_stability}
\end{figure}

An interesting feature that emerges from the use of the U9+ law, although
it is not peculiar of this law, can be appreciated from
Fig.~\ref{U9Q_DD2_stability}, which reports the usual sequences of
constant angular momentum $J$ for models with the DD2 EOS and the
following set of parameters: $p=0.8$, $\xi=3$, $\lambda_1=3$,
$\lambda_2=2.8$. Also shown in Fig.~\ref{U9Q_DD2_stability} is the
standard stability line obtained by joining the various turning
points. When comparing with the corresponding
Figs.~\ref{U8_DD2_stability} and \ref{U9_DD2_stability} is it quite
apparent that the turning-point stability line (orange solid line) has a
positive slope, \ie that $\partial M/\partial \rho_c|_{\rm tp} >0$, thus
in stark contrast with what is seen for the space of solutions with the U8
and U9 differential-rotation laws. The most important consequence of this
behaviour is that it is in principle possible to construct two stellar
models located on the Keplerian stability line that have the same
gravitational mass but different central rest-mass densities, and
obviously different internal structure. We postpone a more detailed
discussion of what these differences in the stellar structure amount to
until Sec.~\ref{sec:comp_U9-CR}, where we will be able to discuss them
from a more general point of view and when considering also other
rotation laws. However, it can suffice here to say that this novel
behaviour of the turning-point critical line is not unique nor an
intrinsic property of the U9+ rotation law. Indeed, it is possible to
reproduce a very similar behaviour for the same DD2 EOS also for the U9
law once the following parameters are carefully chosen: $p=1$,
$\lambda_1=3$, $\lambda_2 =2.8$. The corresponding turning-point
stability line is shown in Fig.~\ref{U9Q_DD2_stability} (light blue solid
line)\footnote{Note that for clarity we do not show in
Fig.~\ref{U9Q_DD2_stability} the constant-$J$ sequences for the U9 law,
which are obviously different from those of the U9+ law.} and thus
highlights that constructing differentially rotating models whose maximum
mass at constant angular momentum increases with central density is more
common than one may naively assume.

\subsection{A new differential-rotation law}
\label{sec:section5.2}

As we will discuss in more detail in Sec.~\ref{sec:comp_U9-CR}, the added
flexibility introduced in the U9+ law allows it to provide a rather
accurate description of realistic differential-rotation profiles measured
from BNS mergers. At the same time, it can be further improved via a
novel differential-rotation law that is directly informed from the
simulations and there the angular velocity can be modelled as
\begin{eqnarray}
  \label{eq:CR_law}
  \Omega (j) = \left\{
  \begin{array}{ll} \Omega_c \left[ \lambda_1
      -(\lambda_1-1) \left( 1- j/j_{\rm max} \right)^{2 \beta} \right], &
    j< j_{\rm max} \\
    \nonumber \\
    \Omega_c \lambda_1 -\Omega_c \lambda_1 \left( j/j_{\rm max} -1
    \right)^{2\gamma}A, & j\geq j_{\rm max} \end{array}\right. \,,
\end{eqnarray}
Because the first derivative is simply given by 
\begin{eqnarray}
  \frac{d\Omega (j)}{dj} = \left\{
  \begin{array}{ll} 2 \beta \Omega_c
    (\lambda_1-1)/j_{\rm max} \left( 1- j/j_{\rm max} \right)^{2
      \beta-1}\,, & j< j_{\rm max} \\
    \nonumber \\
    2 \gamma \Omega_c \lambda_1 /j_{\rm max} \left( j/j_{\rm max} -1
    \right)^{2\gamma-1} A, & j\geq j_{\rm
      max} \end{array}\right. \nonumber \\
\end{eqnarray}
the associated rotation integral has an analytical expression for the
rotation integral and varies whether $j \leq j_{\rm max}$ or $j > j_{\rm
  max}$. In the former case, the integral is given by 
\begin{eqnarray}
  && \hspace{-0.7cm} j_{\rm max} \Omega_c \left( \lambda_1 \! - \! 1
  \right) \Biggl[ 1 \! - \! \left( 1 \! - \! \frac{j}{j_{\rm max}}
    \right)^{2 \beta} \! - \!  \frac{2 \beta}{2 \beta+1} \left( 1 \! - \!
    \left( 1 \! - \! \frac{j}{j_{\rm max}}\right)^{2 \beta +1} \right)
    \Biggr] \,, \nonumber \\
\end{eqnarray}
while in the latter case it has the more complex expression
\begin{eqnarray}
  && \hspace{-0.9cm}j_{\rm max} \Omega_c (\lambda_1 -1 )\biggl\{\left( 1 - \frac{2
    \beta}{2\beta +1} \right) - \nonumber \\
  && \hspace{-0.2cm} 2 A \gamma \Biggl[
    \frac{1}{2\gamma +1} \left( 1 - \frac{j}{j_{\rm max}}\right)^{2\gamma
      +1} - \frac{1}{2\gamma}\left( 1 - \frac{j}{j_{\rm
        max}}\right)^{2\gamma} \Biggr]\biggr\}\,.
\end{eqnarray}
Note that when implementing the rotation law~\eqref{eq:CR_law}
-- that hereafter will refer to as the ``CR'' law -- in the
\texttt{Hydro-RNS} code we need to introduce the standard (rescaled)
parameters $\hat{j}_{\rm max}, \hat{j}_{e}, \lambda_1, \lambda_2$, and
$A$, but also the new parameters $\beta$ and $\gamma$. 

\begin{figure*}
  \centering
  \includegraphics[width=0.9\textwidth]{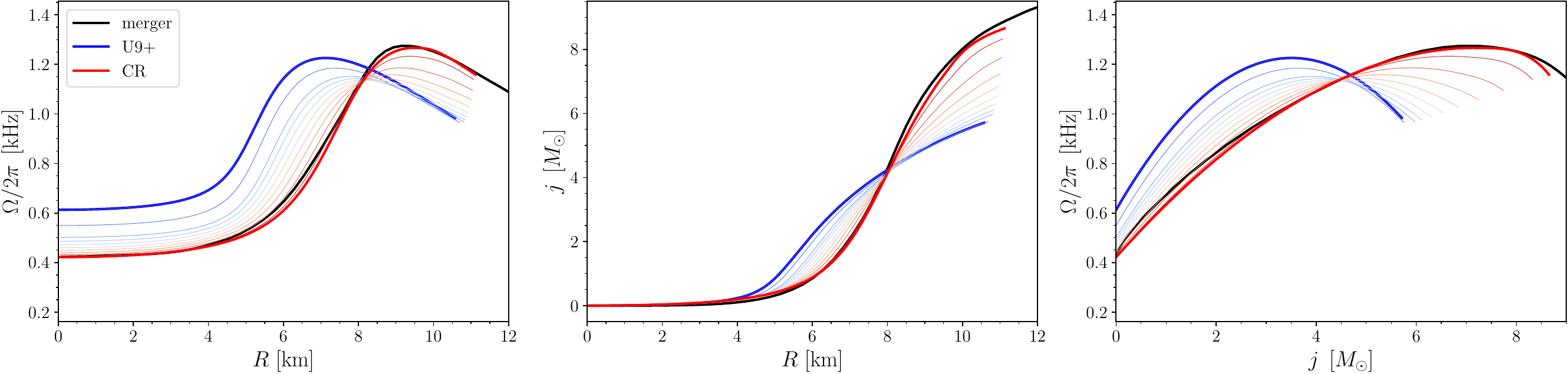}
  \caption{``Blending'' of the rotational properties of stellar models
    with the U9+ and and CR differential-rotation laws to transit from a
    pure-U9+ model (thick blue line) to a pure-CR model (thick red line)
    that provides an accurate description of the DD2 merger remnant
    (thick black line). All models obey the DD2 EOS, have a central
    rest-mass density of $\rho_c=8.024 \times 10^{14}\, {\rm g~cm^{-3}}$
    and an axis ratio $r_p/r_e=0.7$.}
  \label{MC4_lin_combi}
\end{figure*}

Following the structure of the code with auxiliary $\lambda_2$
we are left with five free parameters $\beta$,
$\gamma, A, \lambda_1$, and $j_{\rm max}/j_{e}$ which, together, control
the shape of the profile. In particular, the parameter $\lambda_1$
regulates the maximum angular velocity and can be read of from the ratios
of the angular velocities from the merger-profiles. The parameter $\beta$
controls instead the slope of the angular velocity for angular momenta
smaller than $j_{\rm max}$, which tends to become constant (linear
behaviour) for smaller values of $\beta$. The parameters $A$ and $\gamma$
are responsible for the part with angular momenta larger than $j_{\rm
  max}$, with $\gamma$ controlling how far from the maximum the decreasing
part is extended, and $A$ regulating the slope of the angular velocity in
the Keplerian region. Note that larger values of the ratio $\gamma/A$
yield profiles that extend to larger angular momenta.

Once a set of parameters of the CR law has been chosen, we can compute
$\lambda_2$ from the definition (\ref{eq:CR_law}) at $j=j_{e}$ and when
$j \geq j_{\rm max}$ part 
\begin{eqnarray}
\lambda_2 = \lambda_1 - \lambda_1 A \left( \frac{j_{e}}{j_{\rm max}}
-1 \right)^{2 \gamma}\,.
\end{eqnarray}
This fixes the value of $\lambda_2$, which can be employed in the Euler
equation (\ref{HydrostatEquil}) to find $\Omega_e$ and hence $\Omega_c =
\Omega_e / \lambda_2$.

\begin{figure}
  \centering
  \includegraphics[width=0.9\columnwidth]{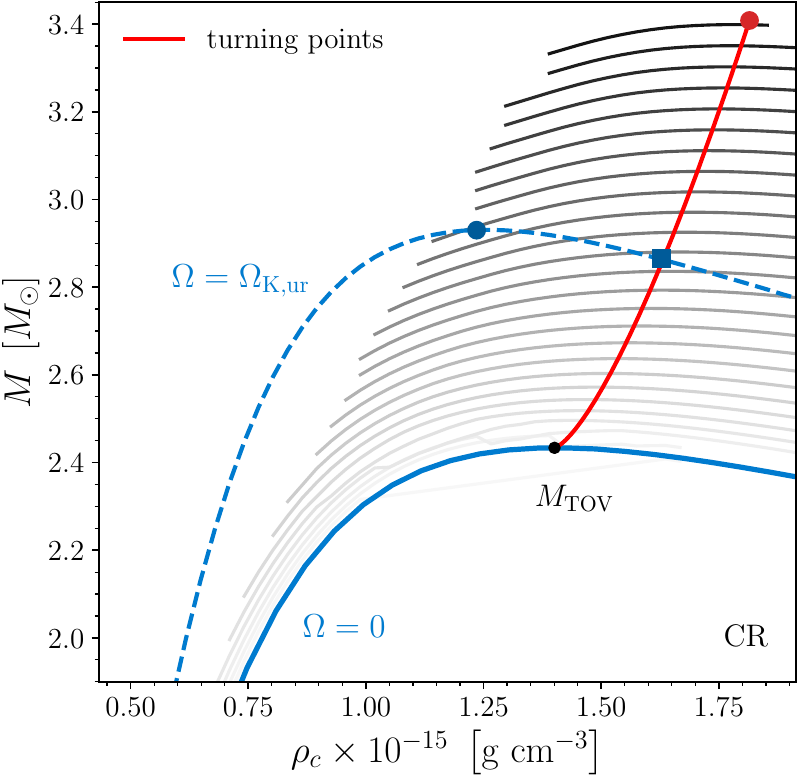}
   \caption{The same as in Fig.~\ref{U8_DD2_stability}, but for the new
     CR differential-rotation law; note that also in this case, the
     turning-point stability line has a positive slope (see also
     Fig.~\ref{U9Q_DD2_stability}).}
  \label{CR_DD2_stability}
\end{figure}

While the CR law provides a more accurate representation of the remnant
rotational properties, it also offers additional challenges during the
numerical calculation and can easily lead to a loss of convergence in the
iterative procedure employed by \texttt{Hydro-RNS} in computing a
solution. In particular, for large values of $\lambda_1 =\Omega_{\rm
  max}/\Omega_c$, the range of possible guesses for $r_e$ and of the
corresponding solutions of $\Omega_e$ grows substantially. In addition,
if the guess for $r_e$ is rather different from the one for which a
solution exists, the code fails to converge because no solution with the
guessed $\Omega_e$ can be found with the Euler equation. It is in
principle possible to manually control the iterations of $r_e$ and to extend
the range of search of $\Omega_e$ so that a solution is found, but this
is tedious and not robust.
However, an approach that tackles these problems consists ``blending''
the solution with the new CR rotation law with the solution obtained with
the U9 law, which is less accurate but less challenging in the regions of
large specific angular momentum. Hence, we introduce a smooth transition
from the converging U9 rotation law to the new CR law via a simple linear
combination 
%
\begin{eqnarray}
  \label{eq:blend_omega}
  \Omega(j) = a \ \Omega_{\rm U9+}(j) + b \ \Omega_{\rm CR}(j)\,,
\end{eqnarray}
where $a, b \in [0,1]$ and $a + b = 1$. Hence, starting
with $a=1,\, b=0$ and progressively marching to a situation in
which $a=0,\, b=1$, the \texttt{Hydro-RNS} code can find
equilibrium solutions with axis ratios that are larger than those allowed
by the U9+ rotation law. Clearly, the strategy adopted in the
blending~\eqref{eq:blend_omega} with the U9+ law can be employed also with
other differential-rotation laws.
Figure~\ref{MC4_lin_combi} illustrates the result of this approach by
showing how an equilibrium solution with the CR law with $j_{\rm
  max}/j_e=0.82, \beta=0.95, \gamma=1.8, A=20, \lambda_1=3$ (red thick
line) can be obtained from an initial U9+ law with $p=1, \xi=1,
\lambda_1=2, \lambda_2=1.6$ (blue thick line) via a sequence of
intermediate solutions which represent the various flavours of the
blending ~\eqref{eq:blend_omega} (blue and red thin lines). The final
solution is compared with the angular velocity profile of the DD2 BNS
simulation (black thick line) at $t - t_{\rm mer} = 20\,{\rm ms}$,
showing the superior accuracy of the new CR law. 
We postpone to
Sec.~\ref{sec:comp_U9-CR} the discussion of how the CR-law
 performs when modelling the remnants of BNS mergers, we
here briefly provide evidence that this law allows one to construct
numerical models of differentially rotating stars by presenting in
Fig.~\ref{CR_DD2_stability} a representative space of solutions for
stars modelled with the DD2 EOS and with the following set of fixed
parameters: $\hat{j}_{\rm max}/\hat{j}_{e}=0.82$, $\beta=0.95$,
$\gamma=1.8$, $A=20$, $\lambda_1=3$ and that was obtained after
performing a progressive linear blending with the U9+ law. 
Note that also in this case, the choice of
parameters leads to a turning-point stability line such that
$\partial M/\partial \rho_c|_{\rm tp} >0$. As we comment next, this is
not a coincidence and reflects a general behaviour of
differential-rotation laws that are inspired by BNS mergers.

\subsection{Comparison of various differential-rotation laws}
\label{sec:comp_U9-CR}
\begin{figure*}
  \centering
  \includegraphics[width=0.99\textwidth]{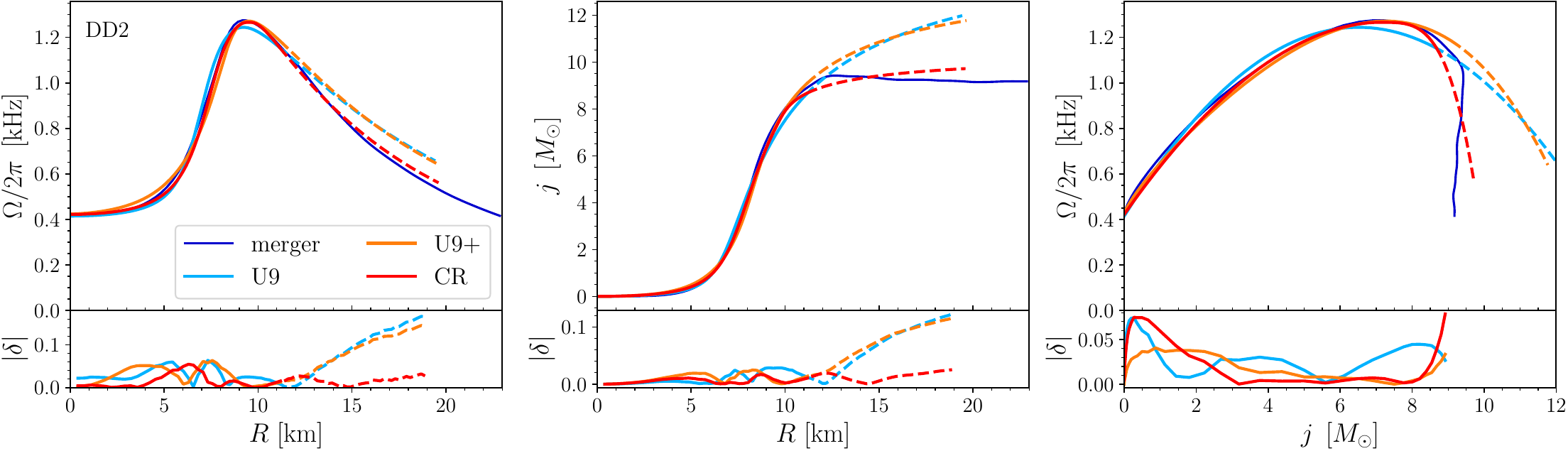}
  \caption{Comparison of the rotational properties of a BNS merger
    remnant with different laws of differential rotation. From left to
    right we show the radial profiles of the angular velocity (left
    panel), the radial profile of the specific angular momentum (middle
    panel) and the angular velocity as a function of the specific angular
    momentum (right panel). Shown with blue lines are the data from the
    numerical simulations at $t-t_{\rm mer}=20 \, {\rm ms}$, while lines
    of other colours show the behaviour of the various laws: light blue
    for U9, orange for U9+, and red for CR. Note that for these laws, the
    transition from solid to dashed lines refers to transition from the
    core to the disc or, equivalently, from a numerical to an analytic
    solution.}
  \label{RNS_20ms_fits}
\end{figure*}

After having reviewed the traditional differential-rotation laws (and
their extensions) and discussed how new laws can be introduced and solved
numerically, it is now useful to actually compare and contrast the
different laws against some representative differential-rotation as
deduced from fully nonlinear simulations. We do this by considering the
BNS merger computed with the DD2 EOS and whose differential-rotation
profile has been extracted at $t-t_{\rm mer} = 20\,{\rm ms}$ (see
Fig.~\ref{DD2_outAll} in Appendix~\ref{sec:appendixA}), which we
reproduce using three different laws: U9, U9+ and CR. In all cases, we
consider models with a central rest-mass density $\rho_c = 8.024 \times
10^{14}\,{\rm g~cm^{-3}}$ and an axis ratio $r_p/r_e=0.7$. In addition,
for the U9 (U9+) laws we use respectively the following set of parameters
$p=1, \lambda_1 = 3, \lambda_2 = 2.8$ ($p=0.8, \xi=3, \lambda_1 =3,
\lambda_2 = 2.8$), which have been chosen so as to optimise the match
with the numerical data. The stellar model following the CR law, on the
other hand, has been obtained after fixing $j_{\rm max}/j_e=0.82 ,
\beta=0.95, \gamma=1.8, A=20$, and $\lambda_1=3$\footnote{Obviously, a
similarly good fit can be obtained also for the TNTYST EOS at $t - t_{\rm
  mer} = 13 \, {\rm ms}$. For completeness, and to aid reproducibility,
we list below the best-fit parameters. For the U9 (U9+) law, the relevant
parameters are $p=1, \lambda_1=2.7, \lambda_2=2.5$ ($p=0.8, \xi=2.5,
\lambda_1=2.7, \lambda_2=2.5$). Similarly, for the CR law they are
$j_{\rm max}/j_e=0.84, \beta=1.12, \gamma=2.1, A=90$, and
$\lambda_1=2.7$. The best-matching models computed with
\texttt{Hydro-RNS} have a central rest-mass density of $\rho_c = 1.074
\times 10^{15}\, {\rm g~cm^{-3}}$ and an axis ratio of $r_p/r_e =
0.68$.}.

A summary of the comparison of the data is offered in
Fig.~\ref{RNS_20ms_fits}, which reports the remnant profiles from the
simulations (dark-blue lines) and the corresponding matches with the U9
(light-blue lines), the U9+ law (orange lines), and the CR law (red
lines). As in previous figures, the different panels offer views of
$\Omega=\Omega(R)$ (left panel), $j=j(R)$ (middle panel), and
$\Omega=\Omega(j)$ (right panel). Note that for all the three different
fitting laws we show with solid (dashed) lines the profiles computed by
\texttt{Hydro-RNS} inside (outside) the stellar model. Also reported at
the bottom of each panel is the relative difference of the given quantity
$\phi$ between the numerical-relativity data and the corresponding
fitting law,\ie $|\delta| := \left|1- \phi_{\rm BNS}/\phi_{\rm fit}
\right|$. Overall, the comparison performed in Fig.~\ref{RNS_20ms_fits}
indicates that all of the rotation laws provide a very good match with
the data, but also that the ability to adjust the $\xi$ exponent in the
U9+ law allows the latter to achieve a closer representation of the BNS
merger data. However, both the U9 and the U9+ laws, provide rather
inaccurate representations of the data in the portion of the solution
that we have considered to be the disc and that is indicated with dashed
light-blue and orange lines, respectively. The CR law, on the other hand,
provides a more accurate representation of the numerical data both in the
central parts of the merger remnant and of the corresponding disc, with a
relative error that remains $|\delta|\lesssim 5\%$ (for $\phi=\Omega(R)$
and $\phi=\Omega(j)$). While this better match is also the result of the
more complex functional structure of the CR law, the fact that its
solution is not computationally more challenging than that of the U9 or
the U9+ laws, makes the newly proposes CR differential-rotation law a
potentially optimal description of the properties of the post-merger
remnants from BNS simulations.

\begin{figure}
  \centering
  \includegraphics[width=0.9\columnwidth]{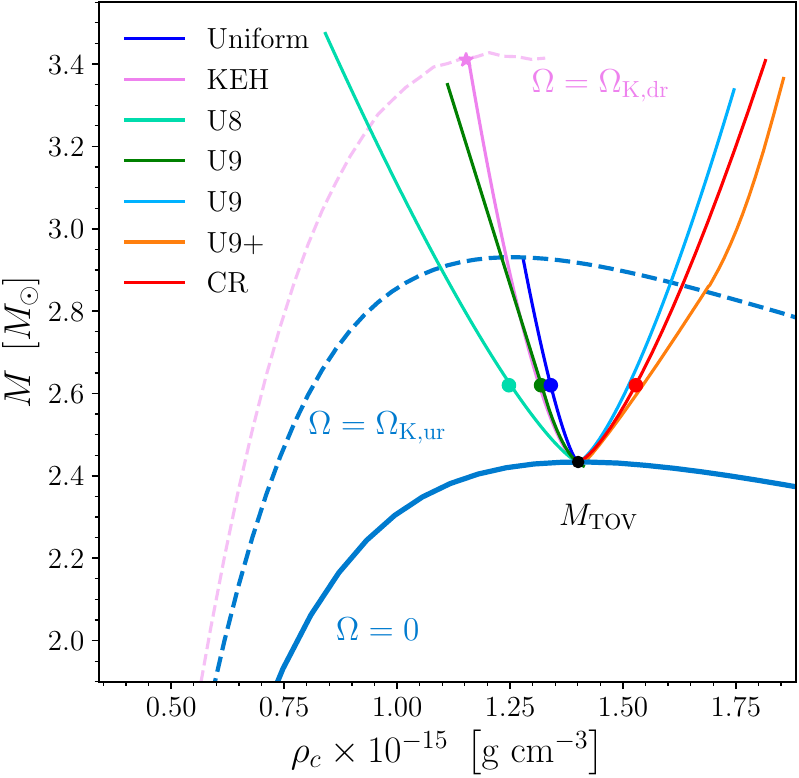}
  \caption{The same as in Fig.~\ref{U8_DD2_stability}, but for all of the
    differential-rotation laws considered here: $j$-constant law (violet
    line), U9 law (green and light-blue lines) the U9+ law (orange line)
    and the new CR law (red line); see also Figs.~\ref{U9_DD2_stability},
    \ref{U9Q_DD2_stability}, and \ref{CR_DD2_stability}. The Keplerian
    limit of differentially rotating models obeying the $j$-constant law
    for $\hat{A}=3.33$, \ie $\Omega=\Omega_{\rm K, dr}$, is shown with a
    pink dashed line; the solid filled circles mark the stellar models
    shown in Fig.~\ref{Models_mass262}.}
  \label{Models_stability_lines}
\end{figure}

\begin{figure}
  \centering
  \includegraphics[width=0.9\columnwidth]{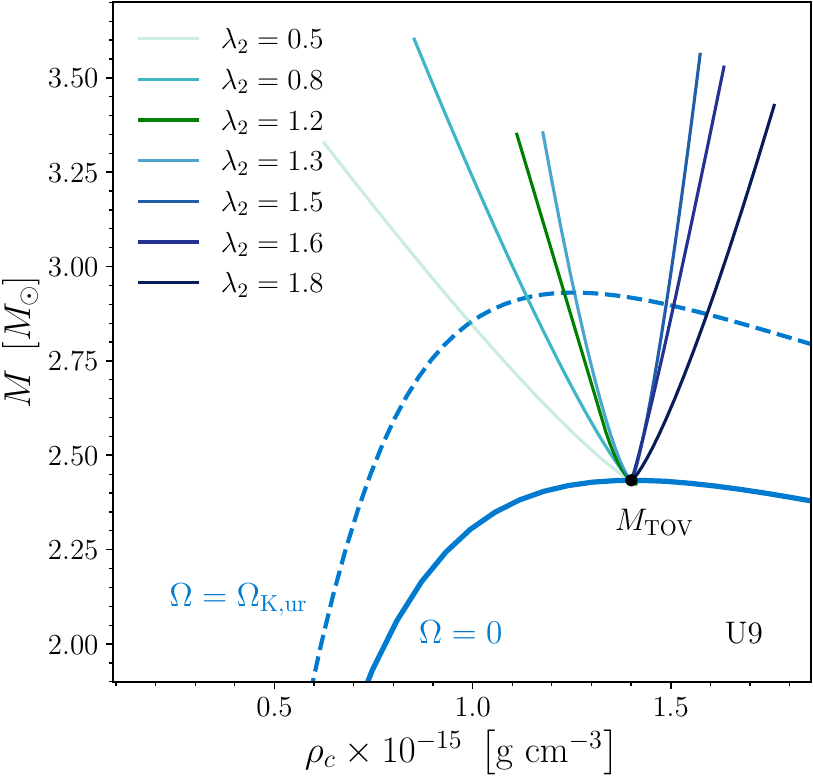}
  \caption{The same as in Fig.~\ref{U9_DD2_stability} for the U9 law but
    with different values of the parameters $\lambda_2$ at fixed
    $\lambda_1=2$. Note how the increase in $\lambda_2$ with constant
    $\lambda_1$ leads to a change in the slope of the turning-point
    stability lines.}
  \label{U9_stability_clock}
\end{figure}

We conclude this section by returning to a topic already covered in the
previous sections, namely that of the turning-point stability line, but
that we here discuss in a more comprehensive way and collecting all the
differential-rotation laws we have considered in this
work. Figure~\ref{Models_stability_lines} offers in a single plot a
complete view of the turning-point stability lines for the various laws
considered so far and referred to stellar models obeying a DD2 EOS. As in
previous versions of this figure, the thick blue line shows the
nonrotating solutions and the blue dashed line the Keplerian limit of
uniform rotation. Besides the turning-point stability line of uniform
rotation (dark blue line), we report the results of the $j$-constant law
for $\hat{A}=3.33$ (violet line)\footnote{We note that this parameter
choice represents the smallest value of $\hat{A}$ for the ``type-A''
models in the classification of~\citet{Ansorg2009}.} and for which the
Keplerian limit of differential rotation is reached at a maximal mass of
$M_{\rm max}=3.41\,M_{\odot}$ (violet star), as already reported
by~\citet{Weih2017}. Unfortunately, this is also the only law of
differential rotation for which a maximum mass on the Keplerian limit was
found as extending the turning-point stability lines beyond the values
reported in the figure (that are below the Keplerian limit) has turned
out not to be possible from a computational point of view.

Other turning-point stability lines reported in
Fig.~\ref{Models_stability_lines} are for the U8 law with $p=1$, $q=3$,
$\lambda_1 = 1.5$, $\lambda_2 = 0.5$ (turquoise line; see also
Fig.~\ref{U8_DD2_stability}), the U9 law with $p=1$, $\lambda_1=2$,
$\lambda_2=1.2$ (light-green line) and $p=1$, $\lambda_1=3$,
$\lambda_2=2.8$ (light-blue line; see also Figs.~\ref{U9_DD2_stability}
and \ref{U9Q_DD2_stability}), the U9+ law with $p=0.8$, $\xi=3$,
$\lambda_1=3$, $\lambda_2=2.8$ (orange line; see also
Fig.~\ref{U9Q_DD2_stability}), and, finally, the CR law with
$\hat{j}_{\rm max}/\hat{j}_{e}=0.82$, $\beta=0.95$, $\gamma=1.8$, $A=20$,
$\lambda_1=3$ (red line; see also Fig.~\ref{CR_DD2_stability}). Beside
the variety of behaviours and slopes in the turning-point stability lines
shown in Fig.~\ref{Models_stability_lines}, what we find particularly
remarkable is that all the differential rotation laws that best match the
angular-velocity profiles deduced from BNS merger simulations (U9, U9+
and CR) actually have turning-point stability lines with positive
slopes. To the best of our knowledge, this is the first time this
universal behaviour has been reported, and highlights how merger remnants
tend to have critical masses that increase with increasing central
rest-mass density.

In order to establish what produces the change of slope in the
turning-point stability lines, we explore stellar models described by the
U9 differential-rotation law -- which shows turning-point stability lines
with both positive and negative slopes (see Figs.~\ref{U9_DD2_stability}
and \ref{U9Q_DD2_stability}) -- at fixed parameters $p=1, \lambda_1 = 2$
and when varying the parameter $\lambda_2$, which changes the value of
the equatorial angular velocity $\Omega_e$ and therefore the
angular-velocity profile.

\begin{figure*}
  \centering
  \includegraphics[width=0.99\textwidth]{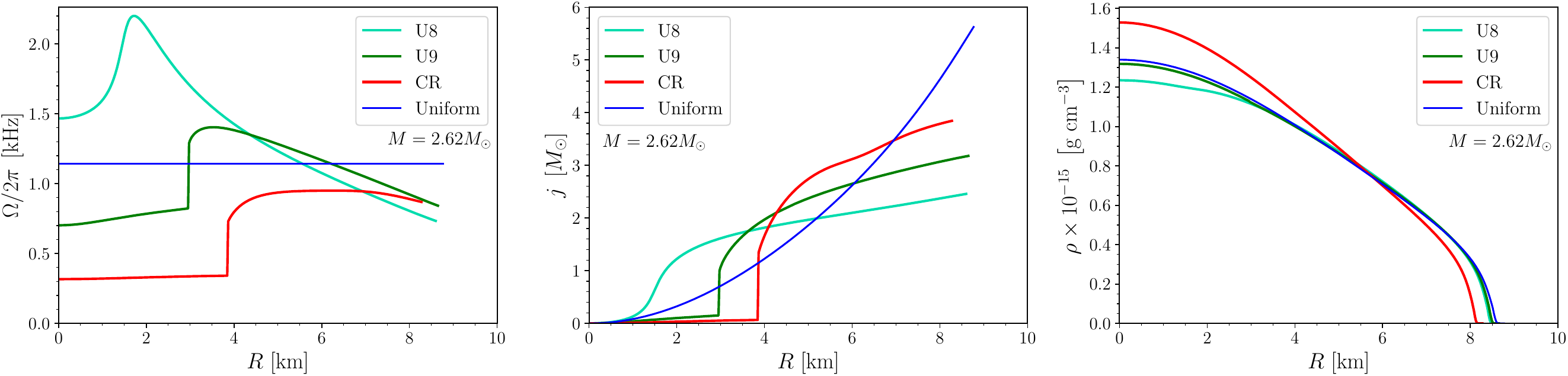}
  \caption{Rotational and structural properties of the four stellar
    models with mass $M=2.62\,M_{\odot}$ and marked as filled circles on
    the stability lines in Fig.~\ref{Models_stability_lines}. Shown are
    the U8 law (light blue; $p=1$, $q=3$, $\lambda_1=1.5$,
    $\lambda_2=0.5$) in turquoise, the U9 law (green;
    $p=1$,$\lambda_1=2$, $\lambda_1 = 1.2$) the CR law (red). From left
    to right we report the radial profiles of the angular velocity (left
    panel), of the specific angular momentum (middle panel) and of the
    rest-mass density (right panel).}
  \label{Models_mass262}
\end{figure*}

In Fig.~\ref{U9_stability_clock} we present the stability lines for the
U9 law for seven different values $\lambda_2 = 0.5, \, 0.8, \, 1.2, \, 1.3,
\, 1.5, \, 1.6, \, 1.8$. Note how the variation of the value of
$\lambda_2$ clearly provides a change in the slope, so that for sequences
with $0.5 \leq \lambda_2 \leq 1.3$ the slope of the turning-point
stability line is negative, while for sequences with $1.3 \leq \lambda_2
\leq 1.8$ the slope is positive. Recalling that $\lambda_2 := \Omega_e
/\Omega_c$ [Eq.~\eqref{eq:lambda2}] and since $\lambda_1 := \Omega_{\rm
  max}/\Omega_c$ [Eq.~\eqref{eq:lambda1}] is kept fixed in
Fig.~\ref{U9_stability_clock}, we can convert the threshold value in
$\lambda_2$ for the change of slope to a threshold value in the ratio
$\Omega_e/\Omega_{\rm max}$. In other words, we conjecture that stellar
models for which the equatorial angular velocity is about 3/4 times
larger than the maximum, \ie $\Omega_{e} \gtrsim 3/4\,\Omega_{\rm max}$
will have turning-point stability lines with positive slopes. In
addition, if the impact of the $\lambda_1$ parameter is weak, it may also
be possible to convert the threshold value in $\lambda_2$ for the change
of slope to a threshold value in the ratio $\Omega_e/\Omega_c$, thus
concluding that stellar models in which the equatorial angular velocity
is about 1.4 times larger than the value at the stellar center, \ie
$\Omega_e \gtrsim 1.4\,\Omega_c$, will have turning-point stability lines
with positive slopes. While these results have been obtained for the U9
law and the DD2 EOS, it is possible that the general criteria deduced
here for the appearance of turning-point stability lines with positive
slopes, \ie $\Omega_{e} \gtrsim 3/4\,\Omega_{\rm max}$ and $\Omega_{e}
\gtrsim 1.4\,\Omega_c$, may be satisfied also by other
differential-rotation laws and EOSs. We plan to investigate this
conjecture in a forthcoming work.

Finally, in order to gain a better understanding of what are the actual
differences in the stellar properties induced by the various rotation
laws, we study in more detail the interior structure of three
differentially rotating models having the same gravitational mass of $M =
2.62 \,M_{\odot}$\footnote{Note that this mass is smaller than the
initial gravitational mass of the DD2 binary (\cf
Table~\ref{tab:merger_core}) because it does not include the mass
radiated in gravitational waves or the mass in the remnant disc.} and
relative to the uniform, the U8, the U9 and the CR rotation laws,
respectively; these models are marked by circles in
Fig.~\ref{Models_stability_lines}. The three panels in
Fig.~\ref{Models_mass262}, which show, respectively, the radial profiles
of the angular velocity, of the specific angular momentum and of the
rest-mass density, clearly indicate that when keeping the gravitational
mass constant, the three different rotation laws will produce three
significantly different internal structures. In particular, taking the
uniform-rotation law as the reference, it is clear from the left panel
that the U8 law -- for which the turning-point stability line has
negative slope -- leads to angular velocities that are almost always
{larger} than for the uniformly rotating model, and where the central
angular velocity is considerably larger than that at the equator
for this particular choice of parameters. By
contrast, the CR law, -- for which the turning-point stability line has
a positive slope -- has angular velocities that are systematically
{smaller} than the uniform-rotation model. In this latter case,
furthermore, the uniform-velocity core is much more extended and almost a
factor four larger than in the U8-law models. The data relative to the
additional U9 law can be considered as an intermediate case between the
U8 and CR laws and hence the corresponding stellar model exhibits a
behaviour that is in-between what discussed for the other two laws.

Finally, note that with the exception of the CR-law model, which is the
most compact of the four stellar models, the other three models have
rather similar radii, with the uniform-rotation model being the most
extended and being followed by the U8 and U9-law models. At the same
time, the density profiles are rather different. In particular, the
U8-law model has the core with the lowest rest-mass density of the four
models, while, by contrast, the CR-law model has the larger rest-mass
densities in the inner regions of the star. This is not surprising and
indeed what to be expected given that the angular velocity in the core of
the CR-law model is the smallest and hence it provides the smallest
centrifugal support. On the other hand, the angular velocity in the core
of the U8-law model is the largest and hence it provides the strongest
centrifugal support. A similar behaviour is found when comparing for
models with the $U9$ law and having the same gravitational mass but
different values values of the $\lambda_2$ parameter, \ie equal-mass
models in Fig.~\ref{U9_stability_clock}.

In summary, the panels in Fig.~\ref{Models_mass262} clarify that
differentially rotating stars with realistic differential-rotation laws
near the turning-point stability line tend to have extended, slowly but
uniformly rotating cores that have comparatively larger rest-mass
densities. The fact that the U8-law does not reproduce well such a
behaviour represents a very strong motivation towards the use of more
realistic laws such as U9+ or CR.

\section{Conclusions}
\label{sec:section6}

Modelling the properties of the remnants of core-collapse supernovae or
of BNS mergers has a long history and has reached a high degree of
sophistication, with a series of various differential-rotation laws that
have been proposed over the years. The most recent of such laws --
normally referred to as Uryu laws of differential rotation -- have been
inspired by simulations and have among the most important features: (i) a
slowly and uniformly rotating core; (ii) a rapid increase in the angular
velocity in the outer regions of the merger remnant; (iii) a radial
fall-off reflecting a Keplerian profile. While these laws have been used
extensively in the literature and studied to explore the space of
solutions of differentially rotating stars, they do not always represent
a good match to the simulations. To further improve the level of realism
in the construction of equilibrium models of differentially rotating
neutrons stars, we have here reconsidered this problem starting from the
actual data.

In particular, we have analysed the results of three, state-of-the-art
BNS merger simulations performed with different and realistic EOSs and
mass ratios, and have extracted from them the quasi-stationary rotational
properties of the remnants, both in terms of azimuthally- and
time-averaged profiles of the angular velocity and of the specific
angular momentum. This analysis has been useful to study the existence of
universal features in the angular-velocity profile and to point out a
simple criterion to recognise the transition between the ``core'' of the
remnant and the ``disc'', that is, the portion of the HMNS whose
angular-velocity profile follows a Keplerian behaviour. More
specifically, it was possible to realise that in the remnant disc
the angular velocity exhibits a robust and generic linear behaviour in
terms of the logarithm of the rest-mass density. Furthermore, the
departure from this linear behaviour also corresponds to the location in
the remnant where the Keplerian flow starts, so that it is simple to mark
the beginning of the remnant disc as the location where the angular
velocity is no longer linearly proportional to the logarithm of the
rest-mass density. Interestingly, this position does not correspond to
the maximum of the angular velocity, but is located at larger radii.

In addition to the analysis of the data from numerical simulations, we
have extensively tested and modified the \texttt{Hydro-RNS} code, whose
public version is unfortunately unable to reproduce the rather extreme
differential-rotation profiles that are present in realistic simulations,
in particular for equatorial angular velocities that are comparable or
larger than the central ones, \ie with $\Omega_e \gtrsim \Omega_c$. In
this way, after reviewing the basic properties of the traditional U8 and
U9 Uryu laws, we have provided evidence of the code's ability to
reproduce published results and to extend the ranges in which they have
been studied so far.

The use of the differential-rotation profiles extracted from the
simulations has allowed us to realise that while the U9 law can provide a
reasonable approximation of the data, the fidelity can be further
improved after a simple extension of the law in terms of an additional
exponent, leading to what we have referred to as the U9+ law. However,
both the U9 and the U9+ laws provide a rather inaccurate description in the
outer layers of the remnants, where the angular velocity and the
rest-mass density are low. To counter that, we have devised a novel law
of differential rotation, that we refer to as the CR law, which, thanks
to the addition of new parameters, allows to reproduce the
angular-velocity profiles also well beyond the nominal stellar surface
and with relative differences that are always below $5\%$.

A particularly interesting and important aspect of the research carried
out here is that related to the determination of the dynamical stability
line obtained from the turning-point criterion, namely, by the stability
line marking the maximum (turning point) of the gravitational mass along
sequences of constant angular momentum. Although this line reflects only
a sufficient condition for the dynamical collapse to rotating black
holes, its global properties are still largely unknown. Thanks to the
large number of differentially rotating models computed here, it has been
possible to carry out the first systematic study of the properties of
turning-point stability lines in realistic (and non-realistic) laws of
differential rotation. A particularly surprising result obtained in this
way is that the slope of the stability line, namely, the sign of the
derivative $\left.\partial M/\partial \rho_c\right|_{\rm tp}$ depends on
the properties of the law of differential rotation and is not necessarily
negative, as customary for the $j$-constant or the U8 laws. In
particular, we have shown that the U9+ and CR laws (but also the U9 law
with a suitable choice of parameters) -- which provide an accurate
representation of the properties of the post-merger remnant -- actually
lead to turning-point stability lines where the maximum mass grows with
the central rest-mass density, \ie with $\left.\partial M/\partial
\rho_c\right|_{\rm tp} > 0$.  We believe this reflects closely the
properties of the BNS-merger remnant, which has a larger amount of
angular momentum in the outer layers and a slowly rotating core, so that
the reduced centrifugal support leads also to larger central rest-mass
densities. On the basis of the EOSs and differential rotation laws
studied here, we conjecture that stellar models for which the equatorial
angular velocity is about 3/4 times larger than the maximum, \ie
$\Omega_{e} \gtrsim 3/4\,\Omega_{\rm max}$ will exhibit turning-point
stability lines with $\left.\partial M/\partial \rho_c\right|_{\rm tp} >
0$.

The systematic and comprehensive analysis presented here can be extended
and improved in a number of ways. First, by exploring the post-merger
properties of remnants produced with other EOSs so as to validate the
robustness of the proposed U9+ and CR laws. Particularly important in
this respect will be exploration of a broader range of mass ratios as
they seem to exhibit somewhat different differential-rotation
laws. Second, by improving the convergence properties of
\texttt{Hydro-RNS} and hence compute the Keplerian line of differentially
rotating models that so far in the literature has been computed only for
the simpler $j$-constant differential-rotation law. Third, compute the
location of the neutral-stability line~\citep{Takami:2011} and determine
its location with respect to the turning-point stability line. Finally,
use numerical simulations of dynamical spacetimes to explore the
equilibrium properties of the turning-point and neutral-stability
lines, to determine necessary and sufficient conditions for collapse to
rotating black holes, thus confirming also for more realistic
differentially rotating stars the analysis carried out for the
$j$-constant law~\citep{Weih2017, Szewczyk2023}. We plan to explore some
of these topics in future works.

\section*{Acknowledgements}

We thank K. Topolski, M. Chabanov and C. Ecker for providing the data of
the numerical simulations. We are also grateful to P. Iosif, C. Kr\"uger and
N. Stergioulas for help with the code and useful comments. This research
is supported by the European Research Council Advanced Grant ``JETSET:
Launching, propagation and emission of relativistic jets from binary
mergers and across mass scales'' (grant no. 884631). LR acknowledges the
Walter Greiner Gesellschaft zur F\"orderung der physikalischen
Grundlagenforschung e.V. through the Carl W. Fueck Laureatus Chair. The
calculations were performed on the local ITP Supercomputing Clusters
Iboga and Calea and on HPE Apollo HAWK at the High Performance Computing
Center Stuttgart (HLRS) under the grant BNSMIC.

\section*{Data Availability}

Data available on request. The data underlying this article will be
shared on reasonable request to the corresponding author.

\appendix
\label{sec:appendix}


\section{On the U8 models and the DD2 simulation}
\label{sec:appendixA}

\begin{figure*}
  \centering
  \includegraphics[width=0.8\textwidth]{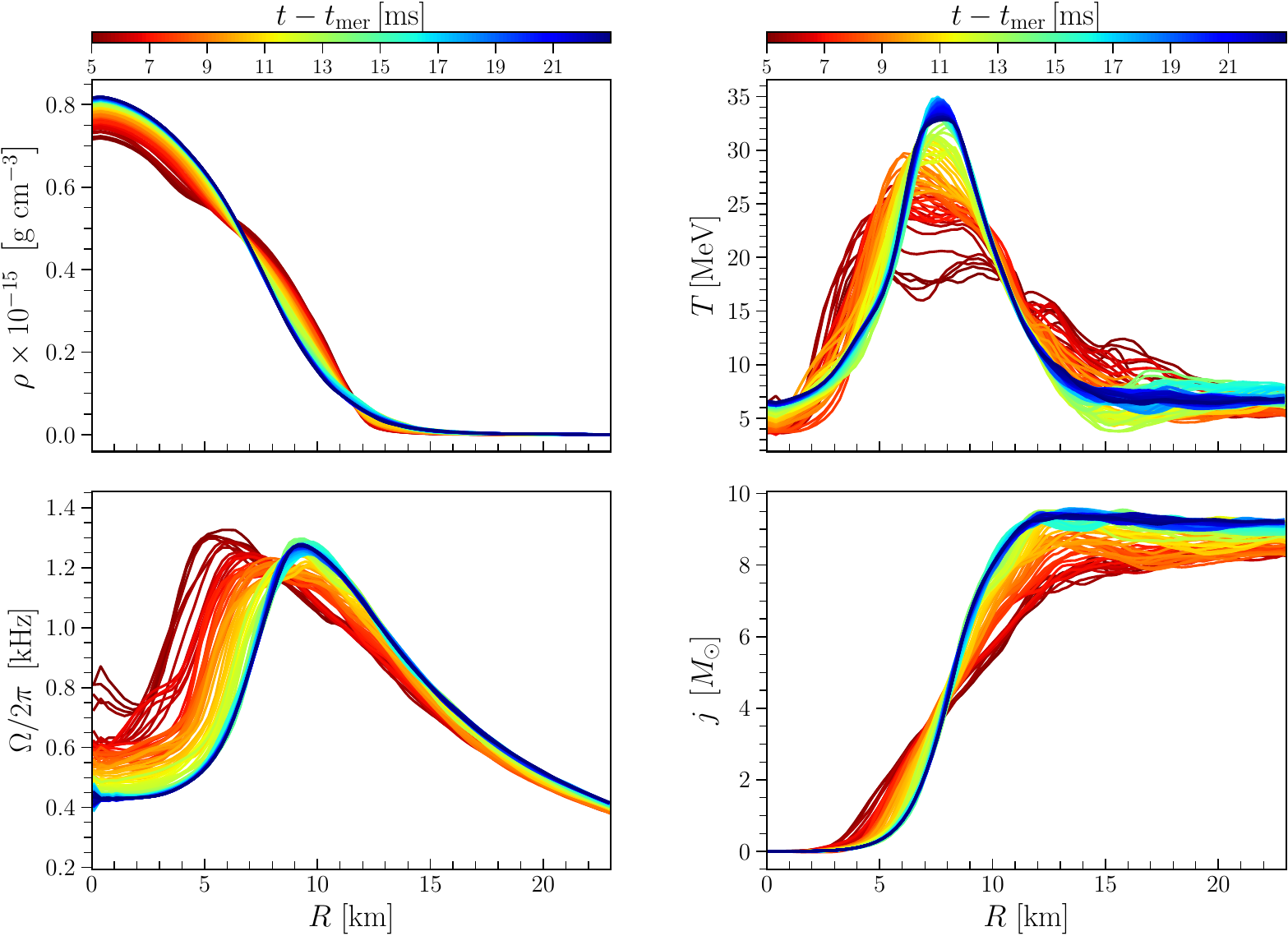}
  \caption{The same as in Fig.~\ref{fig:Data_1ms_Av} but for a higher
    cadence of $0.1\,{\rm ms}$. The data refers to the BNS merger
    simulation with the DD2 EOS and shows, in addition, the radial
    profiles of the rest-mass density (top-left panel) and temperature
    (top-right panel).}
  \label{DD2_outAll}
\end{figure*}

To facilitate the reproducibility of our results, in this Appendix we
provide additional information on the computed stellar models with the U8
and on the evolution of the remnant properties of the BNS simulation with
a DD2 EOS. More specifically, we present in
Table~\ref{tab:Uryu8_physical_quantities} the most important properties
of the equilibrium stellar models computed with \texttt{Hydro-RNS} and
the U8 rotational law when keeping the central rest-mass density at
$\rho_c=8.024 \times 10^{14}\, {\rm g~cm^{-3}}$ and the axis ratio at
$r_p/r_e=0.8$ (see Figs. \ref{U8_L1_L2_var} and \ref{U8_P_Q_var} in
the main text).

Similarly, we present in Fig.~\ref{DD2_outAll} a more refined in time
evolution of various quantities extracted from the BNS merger simulation
performed with the DD2 EOS. In particular, from the top right and in a
clock-wise sense we report the evolution of the radial profiles of
various quantities as the rest-mass density $\rho$, the temperature $T$,
the specific angular momentum $j$ and the angular velocity $\Omega$. All
quantities are evaluated on the equatorial plane ($z=0$) averaged in the
azimuthal direction and in time in an interval of $0.1\,{\rm ms}$ and
spanning a window in time $5 \, {\rm ms} < t-t_{\rm mer} < 23 \, {\rm
  ms}$ (solid lines from red to blue; some of this data was also shown in
Fig.~\ref{fig:Data_1ms_Av}, although with a lower cadence).

Note how the remnant is extremely dynamical up to $t-t_{\rm mer} \simeq
10\,{\rm ms}$, but also that soon after it reaches a quasi-stationary
equilibrium lasting from $t-t_{\rm mer} \simeq 12\,{\rm ms}$ till the end
of the simulation and in which all quantities settle to their typical
behaviour: a high-density, low angular velocity and cold core surrounded
by a low-density, high angular velocity and hot mantle. Interestingly,
due to the differential rotation and a higher rotation frequency off from
the center, the star heats up into a ring around the center, reaching temperatures 
up to $35 \, {\rm MeV}$. \footnote{Although the remnants have high
temperatures in the fast-rotating regions, for simplicity, we use cold
$\beta$-equilibrium slices of the EOSs to construct models with
\texttt{Hydro-RNS}.}

\begin{table*}
  \centering
  \caption{Physical quantities for stellar models computed with the U8
    rotation law, the DD2 EOS, a fixed central rest-mass density of
    $\rho_c=8.024 \times 10^{14}\, {\rm g~cm^{-3}}$ and fixed axis ratio
    $r_p/r_e=0.8$.}
  \label{tab:Uryu8_physical_quantities}
  \begin{tabular}{ccccccccccccccc}
    \hline
    Model & $\hat{A}$ & $\hat{B}$ & $p$ & $q$ & $M_0$ & $M$ & $J$ & $T/|W|$ & $\Omega_c/2\pi$ & $\Omega_\text{max}/2\pi$ & $\Omega_e/2\pi$ & $\Omega_K/2\pi$ & $R_e$ & $r_e$ \\
    & & & & &$[M_\odot]$ & $[M_\odot]$ & $[{M_{\odot}^2}]$ & $\times 10^{-2}$ & $[{\rm kHz}]$ & $[{\rm kHz}]$ & $[{\rm kHz}]$ & $[{\rm kHz}]$ & [km] & [km] \\
    \hline
    ${\mathrm A}$ & 0.808 & 0.467 & 1.00 & 3.00 & 2.653 & 2.287 & 2.074 & 5.198 & 0.915 & 2.290 & 0.458 & 1.711 &9.268 & 6.731 \\
    ${\mathrm B}$ & 0.930 & 0.865 & 1.00 & 3.00 & 2.666 & 2.302 & 2.385 & 6.444 & 1.067 & 1.600 & 0.533 & 1.677 &9.419 & 6.853 \\
    ${\mathrm C}$ & 1.427 & 0.988 & 1.00 & 3.00 & 2.667 & 2.310 & 2.676 & 7.215 & 0.640 & 1.279 & 0.639 & 1.627 &9.637 & 7.049 \\
    ${\mathrm D}$ & 2.042 & 1.414 & 1.00 & 3.00 & 2.630 & 2.283 & 2.639 & 7.200 & 0.535 & 1.071 & 0.803 & 1.564 &9.879 & 7.318 \\
    ${\mathrm X}$ & 0.854 & 0.591 & 1.00 & 3.00 & 2.659 & 2.295 & 2.214 & 5.663 & 0.977 & 1.955 & 0.489 & 1.697 &9.332 & 6.781 \\ 
    \hline
    ${\mathrm E}$ & 1.8 & 1.2 & 1.00 & 3.00 & 2.650 & 2.299 & 2.707 & 7.399 & 0.543 & 1.135 & 0.729 & 1.590 &9.785 & 7.206 \\               
    ${\mathrm F}$ & 1.8 & 1.6 & 1.00 & 3.00 & 2.641 & 2.291 & 2.680 & 7.183 & 0.674 & 1.050 & 0.774 & 1.578 &9.818 & 7.249 \\
    ${\mathrm G}$ & 1.3 & 1.0 & 1.00 & 3.00 & 2.669 & 2.310 & 2.641 & 7.070 & 0.734 & 1.310 & 0.621 & 1.636 &9.596 & 7.009 \\
    ${\mathrm H}$ & 2.3 & 1.8 & 1.00 & 3.00 & 2.603 & 2.260 & 2.488 & 6.626 & 0.586 & 1.028 & 0.904 & 1.538 &9.955 & 7.429 \\
    ${\mathrm I}$ & 1.8 & 1.4 & 1.25 & 3.00 & 2.639 & 2.290 & 2.676 & 7.272 & 0.606 & 1.095 & 0.768 & 1.574 &9.840 & 7.271 \\
    ${\mathrm J}$ & 1.8 & 1.4 & 1.50 & 3.00 & 2.635 & 2.287 & 2.655 & 7.292 & 0.595 & 1.114 & 0.777 & 1.569 &9.861 & 7.294 \\ 	
    ${\mathrm K}$ & 1.8 & 1.4 & 1.00 & 5.00 & 2.647 & 2.297 & 2.704 & 7.379 & 0.587 & 1.134 & 0.723 & 1.585 &9.805 & 7.227 \\ 
    ${\mathrm L}$ & 1.8 & 1.4 & 1.00 & 7.00 & 2.658 & 2.306 & 2.740 & 7.511 & 0.574 & 1.172 & 0.707 & 1.587 &9.807 & 7.221 \\ 
    ${\mathrm Y}$ & 1.8 & 1.4 & 1.00 & 3.00 & 2.645 & 2.294 & 2.698 & 7.279 & 0.614 & 1.083 & 0.756 & 1.582 &9.810 & 7.237 \\   
    \hline
  \end{tabular}
\end{table*}

\bibliographystyle{mnras}
\bibliography{Diff_rot_paper.bbl}

\begin{thebibliography}{}
\makeatletter
\relax
\def\mn@urlcharsother{\let\do\@makeother \do\$\do\&\do\#\do\^\do\_\do\%\do\~}
\def\mn@doi{\begingroup\mn@urlcharsother \@ifnextchar [ {\mn@doi@}
  {\mn@doi@[]}}
\def\mn@doi@[#1]#2{\def\@tempa{#1}\ifx\@tempa\@empty \href
  {http://dx.doi.org/#2} {doi:#2}\else \href {http://dx.doi.org/#2} {#1}\fi
  \endgroup}
\def\mn@eprint#1#2{\mn@eprint@#1:#2::\@nil}
\def\mn@eprint@arXiv#1{\href {http://arxiv.org/abs/#1} {{\tt arXiv:#1}}}
\def\mn@eprint@dblp#1{\href {http://dblp.uni-trier.de/rec/bibtex/#1.xml}
  {dblp:#1}}
\def\mn@eprint@#1:#2:#3:#4\@nil{\def\@tempa {#1}\def\@tempb {#2}\def\@tempc
  {#3}\ifx \@tempc \@empty \let \@tempc \@tempb \let \@tempb \@tempa \fi \ifx
  \@tempb \@empty \def\@tempb {arXiv}\fi \@ifundefined
  {mn@eprint@\@tempb}{\@tempb:\@tempc}{\expandafter \expandafter \csname
  mn@eprint@\@tempb\endcsname \expandafter{\@tempc}}}

\bibitem[\protect\citeauthoryear{{Abdikamalov}, {Ahmedov}  \&
  {Miller}}{{Abdikamalov} et~al.}{2009}]{Abdikamalov2009}
{Abdikamalov} E.~B.,  {Ahmedov} B.~J.,   {Miller} J.~C.,  2009, \mn@doi [Mon.
  Not. R. Astron. Soc.] {10.1111/j.1365-2966.2009.14540.x}, \href
  {http://adsabs.harvard.edu/abs/2009MNRAS.395..443A} {395, 443}

\bibitem[\protect\citeauthoryear{{Abdikamalov}, {Ott}, {Rezzolla}, {Dessart},
  {Dimmelmeier}, {Marek}  \& {Janka}}{{Abdikamalov}
  et~al.}{2010}]{Abdikamalov:2009aq}
{Abdikamalov} E.~B.,  {Ott} C.~D.,  {Rezzolla} L.,  {Dessart} L.,
  {Dimmelmeier} H.,  {Marek} A.,   {Janka} H.-T.,  2010, \mn@doi [Phys. Rev. D]
  {10.1103/PhysRevD.81.044012}, \href
  {http://adsabs.harvard.edu/abs/2010PhRvD..81d4012A} {81, 044012}

\bibitem[\protect\citeauthoryear{{Ansorg}, {Gondek-Rosi{\'n}ska}  \&
  {Villain}}{{Ansorg} et~al.}{2009}]{Ansorg2009}
{Ansorg} M.,  {Gondek-Rosi{\'n}ska} D.,   {Villain} L.,  2009, \mn@doi [Mon.
  Not. R. Astron. Soc.] {10.1111/j.1365-2966.2009.14904.x}, \href
  {http://adsabs.harvard.edu/abs/2009MNRAS.396.2359A} {396, 2359}

\bibitem[\protect\citeauthoryear{Baiotti \& Rezzolla}{Baiotti \&
  Rezzolla}{2017}]{Baiotti2016}
Baiotti L.,  Rezzolla L.,  2017, \mn@doi [Rept. Prog. Phys.]
  {10.1088/1361-6633/aa67bb}, 80, 096901

\bibitem[\protect\citeauthoryear{{Baiotti}, {De Pietri}, {Manca}  \&
  {Rezzolla}}{{Baiotti} et~al.}{2007}]{Baiotti06b}
{Baiotti} L.,  {De Pietri} R.,  {Manca} G.~M.,   {Rezzolla} L.,  2007, \mn@doi
  [Phys. Rev. D] {10.1103/PhysRevD.75.044023}, \href
  {http://adsabs.harvard.edu/abs/2007PhRvD..75d4023B} {75, 044023}

\bibitem[\protect\citeauthoryear{{Baiotti}, {Giacomazzo}  \&
  {Rezzolla}}{{Baiotti} et~al.}{2008}]{Baiotti08}
{Baiotti} L.,  {Giacomazzo} B.,   {Rezzolla} L.,  2008, \mn@doi [Phys. Rev. D]
  {10.1103/PhysRevD.78.084033}, \href
  {http://adsabs.harvard.edu/abs/2008PhRvD..78h4033B} {78, 084033}

\bibitem[\protect\citeauthoryear{{Baumgarte}, {Shapiro}  \&
  {Shibata}}{{Baumgarte} et~al.}{2000}]{Baumgarte00b}
{Baumgarte} T.~W.,  {Shapiro} S.~L.,   {Shibata} M.,  2000, \mn@doi [Astrophys.
  J. Lett.] {10.1086/312425}, \href
  {http://adsabs.harvard.edu/abs/2000ApJ...528L..29B} {528, L29}

\bibitem[\protect\citeauthoryear{{Bonazzola}, {Gourgoulhon}, {Salgado}  \&
  {Marck}}{{Bonazzola} et~al.}{1993}]{Bonazzola1993}
{Bonazzola} S.,  {Gourgoulhon} E.,  {Salgado} M.,   {Marck} J.~A.,  1993,
  Astron. and Astrophys., \href
  {http://adsabs.harvard.edu/abs/1993A%26A...278..421B} {278, 421}

\bibitem[\protect\citeauthoryear{{Bozzola}}{{Bozzola}}{2021}]{kuibit21}
{Bozzola} G.,  2021, \mn@doi [The Journal of Open Source Software]
  {10.21105/joss.03099}, \href
  {https://ui.adsabs.harvard.edu/abs/2021JOSS....6.3099B} {6, 3099}

\bibitem[\protect\citeauthoryear{{Bozzola}, {Stergioulas}  \&
  {Bauswein}}{{Bozzola} et~al.}{2018}]{Bozzola2017}
{Bozzola} G.,  {Stergioulas} N.,   {Bauswein} A.,  2018, \mn@doi [Mon. Not. R.
  Astron. Soc.] {10.1093/mnras/stx3002}, \href
  {https://ui.adsabs.harvard.edu/abs/2018MNRAS.474.3557B} {474, 3557}

\bibitem[\protect\citeauthoryear{{Bozzola}, {Espino}, {Lewin}  \&
  {Paschalidis}}{{Bozzola} et~al.}{2019}]{Bozzola2019}
{Bozzola} G.,  {Espino} P.~L.,  {Lewin} C.~D.,   {Paschalidis} V.,  2019,
  \mn@doi [European Physical Journal A] {10.1140/epja/i2019-12831-2}, \href
  {https://ui.adsabs.harvard.edu/abs/2019EPJA...55..149B} {55, 149}

\bibitem[\protect\citeauthoryear{{Breu} \& {Rezzolla}}{{Breu} \&
  {Rezzolla}}{2016}]{Breu2016}
{Breu} C.,  {Rezzolla} L.,  2016, \mn@doi [Mon. Not. R. Astron. Soc.]
  {10.1093/mnras/stw575}, \href
  {http://adsabs.harvard.edu/abs/2016MNRAS.459..646B} {459, 646}

\bibitem[\protect\citeauthoryear{{Burrows}, {Dessart}, {Livne}, {Ott}  \&
  {Murphy}}{{Burrows} et~al.}{2007}]{Burrows07a}
{Burrows} A.,  {Dessart} L.,  {Livne} E.,  {Ott} C.~D.,   {Murphy} J.,  2007,
  \mn@doi [Astrophys. J.] {10.1086/519161}, \href
  {http://adsabs.harvard.edu/abs/2007ApJ...664..416B} {664, 416}

\bibitem[\protect\citeauthoryear{Camelio, Dietrich, Rosswog  \&
  Haskell}{Camelio et~al.}{2021}]{Camelio2021}
Camelio G.,  Dietrich T.,  Rosswog S.,   Haskell B.,  2021, \mn@doi [Phys. Rev.
  D] {10.1103/PhysRevD.103.063014}, 103, 063014

\bibitem[\protect\citeauthoryear{{Chabanov}, {Tootle}, {Most}  \&
  {Rezzolla}}{{Chabanov} et~al.}{2023}]{Chabanov2022}
{Chabanov} M.,  {Tootle} S.~D.,  {Most} E.~R.,   {Rezzolla} L.,  2023, \mn@doi
  [Astrophys. J. Lett.] {10.3847/2041-8213/acbbc5}, \href
  {https://ui.adsabs.harvard.edu/abs/2023ApJ...945L..14C} {945, L14}

\bibitem[\protect\citeauthoryear{{Chi-Kit Cheong}, {Muhammed}, {Chawhan},
  {Duez}  \& {Foucart}}{{Chi-Kit Cheong} et~al.}{2024}]{Cheong2024}
{Chi-Kit Cheong} P.,  {Muhammed} N.,  {Chawhan} P.,  {Duez} M.~D.,   {Foucart}
  F.,  2024, \mn@doi [arXiv e-prints] {10.48550/arXiv.2402.18529}, \href
  {https://ui.adsabs.harvard.edu/abs/2024arXiv240218529C} {p. arXiv:2402.18529}

\bibitem[\protect\citeauthoryear{{Ciolfi}, {Kastaun}, {Giacomazzo}, {Endrizzi},
  {Siegel}  \& {Perna}}{{Ciolfi} et~al.}{2017}]{Ciolfi2017}
{Ciolfi} R.,  {Kastaun} W.,  {Giacomazzo} B.,  {Endrizzi} A.,  {Siegel} D.~M.,
   {Perna} R.,  2017, \mn@doi [Phys. Rev. D] {10.1103/PhysRevD.95.063016},
  \href {http://adsabs.harvard.edu/abs/2017PhRvD..95f3016C} {95, 063016}

\bibitem[\protect\citeauthoryear{{Cook}, {Shapiro}  \& {Teukolsky}}{{Cook}
  et~al.}{1992}]{Cook92b}
{Cook} G.~B.,  {Shapiro} S.~L.,   {Teukolsky} S.~A.,  1992, \mn@doi [Astrophys.
  J.] {10.1086/171849}, \href
  {https://ui.adsabs.harvard.edu/abs/1992ApJ...398..203C} {398, 203}

\bibitem[\protect\citeauthoryear{{De Pietri}, {Feo}, {Font}, {L{\"o}ffler},
  {Maione}, {Pasquali}  \& {Stergioulas}}{{De Pietri}
  et~al.}{2018}]{DePietri2018}
{De Pietri} R.,  {Feo} A.,  {Font} J.~A.,  {L{\"o}ffler} F.,  {Maione} F.,
  {Pasquali} M.,   {Stergioulas} N.,  2018, \mn@doi [Phys. Rev. Lett.]
  {10.1103/PhysRevLett.120.221101}, \href
  {https://ui.adsabs.harvard.edu/abs/2018PhRvL.120v1101D} {120, 221101}

\bibitem[\protect\citeauthoryear{{Demircik}, {Ecker}  \&
  {J{\"a}rvinen}}{{Demircik} et~al.}{2020}]{Demircik2020}
{Demircik} T.,  {Ecker} C.,   {J{\"a}rvinen} M.,  2020, arXiv e-prints, \href
  {https://ui.adsabs.harvard.edu/abs/2020arXiv200910731D} {p. arXiv:2009.10731}

\bibitem[\protect\citeauthoryear{Demircik, Ecker  \& J\"arvinen}{Demircik
  et~al.}{2022}]{Demircik:2021zll}
Demircik T.,  Ecker C.,   J\"arvinen M.,  2022, \mn@doi [Phys. Rev. X]
  {10.1103/PhysRevX.12.041012}, 12, 041012

\bibitem[\protect\citeauthoryear{{Dessart}, {Burrows}, {Ott}, {Livne}, {Yoon}
  \& {Langer}}{{Dessart} et~al.}{2006}]{Dessart2006}
{Dessart} L.,  {Burrows} A.,  {Ott} C.~D.,  {Livne} E.,  {Yoon} S.~C.,
  {Langer} N.,  2006, \mn@doi [\apj] {10.1086/503626}, \href
  {https://ui.adsabs.harvard.edu/abs/2006ApJ...644.1063D} {644, 1063}

\bibitem[\protect\citeauthoryear{{Dimmelmeier}, {Stergioulas}  \&
  {Font}}{{Dimmelmeier} et~al.}{2006}]{Dimmelmeier06}
{Dimmelmeier} H.,  {Stergioulas} N.,   {Font} J.~A.,  2006, \mn@doi [Mon. Not.
  R. Astron. Soc.] {10.1111/j.1365-2966.2006.10274.x}, \href
  {http://adsabs.harvard.edu/abs/2006MNRAS.368.1609D} {368, 1609}

\bibitem[\protect\citeauthoryear{Duez, Liu, Shapiro, Shibata  \& Stephens}{Duez
  et~al.}{2006}]{Duez:2006qe}
Duez M.~D.,  Liu Y.~T.,  Shapiro S.~L.,  Shibata M.,   Stephens B.~C.,  2006,
  \mn@doi [Phys. Rev. D] {10.1103/PhysRevD.73.104015}, 73, 104015

\bibitem[\protect\citeauthoryear{{East}, {Paschalidis}, {Pretorius}  \&
  {Tsokaros}}{{East} et~al.}{2019}]{East2019}
{East} W.~E.,  {Paschalidis} V.,  {Pretorius} F.,   {Tsokaros} A.,  2019,
  \mn@doi [Phys. Rev. D] {10.1103/PhysRevD.100.124042}, \href
  {https://ui.adsabs.harvard.edu/abs/2019PhRvD.100l4042E} {100, 124042}

\bibitem[\protect\citeauthoryear{{Ecker}, {J{\"a}rvinen}, {Nijs}  \& {van der
  Schee}}{{Ecker} et~al.}{2020}]{Ecker2019}
{Ecker} C.,  {J{\"a}rvinen} M.,  {Nijs} G.,   {van der Schee} W.,  2020,
  \mn@doi [Phys. Rev. D] {10.1103/PhysRevD.101.103006}, \href
  {https://ui.adsabs.harvard.edu/abs/2020PhRvD.101j3006E} {101, 103006}

\bibitem[\protect\citeauthoryear{Espino \& Paschalidis}{Espino \&
  Paschalidis}{2019}]{Espino2019}
Espino P.~L.,  Paschalidis V.,  2019, \mn@doi [Phys. Rev. D]
  {10.1103/PhysRevD.99.083017}, 99, 083017

\bibitem[\protect\citeauthoryear{Espino, Paschalidis, Baumgarte  \&
  Shapiro}{Espino et~al.}{2019}]{Espino2019b}
Espino P.~L.,  Paschalidis V.,  Baumgarte T.~W.,   Shapiro S.~L.,  2019,
  \mn@doi [Phys. Rev. D] {10.1103/PhysRevD.100.043014}, 100, 043014

\bibitem[\protect\citeauthoryear{{Font} et~al.,}{{Font} et~al.}{2002}]{Font02c}
{Font} J.~A.,  et~al., 2002, \mn@doi [Phys. Rev. D]
  {10.1103/PhysRevD.65.084024}, \href
  {http://adsabs.harvard.edu/abs/2002PhRvD..65h4024F} {65, 084024}

\bibitem[\protect\citeauthoryear{{Franceschetti} \& {Del
  Zanna}}{{Franceschetti} \& {Del Zanna}}{2020}]{Franceschetti2020}
{Franceschetti} K.,  {Del Zanna} L.,  2020, \mn@doi [Universe]
  {10.3390/universe6060083}, \href
  {https://ui.adsabs.harvard.edu/abs/2020Univ....6...83F} {6, 83}

\bibitem[\protect\citeauthoryear{{Franci}, {De Pietri}, {Dionysopoulou}  \&
  {Rezzolla}}{{Franci} et~al.}{2013}]{Franci2013}
{Franci} L.,  {De Pietri} R.,  {Dionysopoulou} K.,   {Rezzolla} L.,  2013,
  \mn@doi [Journal of Physics Conference Series]
  {10.1088/1742-6596/470/1/012008}, \href
  {http://adsabs.harvard.edu/abs/2013JPhCS.470a2008F} {470, 012008}

\bibitem[\protect\citeauthoryear{Friedman, Ipser  \& Sorkin}{Friedman
  et~al.}{1988}]{Friedman88}
Friedman J.~L.,  Ipser J.~R.,   Sorkin R.~D.,  1988, Astrophys. J., 325, 722

\bibitem[\protect\citeauthoryear{{Giacomazzo}, {Rezzolla}  \&
  {Stergioulas}}{{Giacomazzo} et~al.}{2011}]{Giacomazzo2011}
{Giacomazzo} B.,  {Rezzolla} L.,   {Stergioulas} N.,  2011, \mn@doi [Phys. Rev.
  D] {10.1103/PhysRevD.84.024022}, \href
  {http://adsabs.harvard.edu/abs/2011PhRvD..84b4022G} {84, 024022}

\bibitem[\protect\citeauthoryear{{Gondek-Rosi{\'n}ska}, {Kowalska}, {Villain},
  {Ansorg}  \& {Kucaba}}{{Gondek-Rosi{\'n}ska} et~al.}{2017}]{Gondek2016}
{Gondek-Rosi{\'n}ska} D.,  {Kowalska} I.,  {Villain} L.,  {Ansorg} M.,
  {Kucaba} M.,  2017, \mn@doi [Astrophys. J.] {10.3847/1538-4357/aa56c1}, \href
  {http://adsabs.harvard.edu/abs/2017ApJ...837...58G} {837, 58}

\bibitem[\protect\citeauthoryear{{Goussard}, {Haensel}  \& {Zdunik}}{{Goussard}
  et~al.}{1998}]{Goussard1998}
{Goussard} J.,  {Haensel} P.,   {Zdunik} J.~L.,  1998, Astron. and Astrophys.,
  \href {http://adsabs.harvard.edu/abs/1998A%26A...330.1005G} {330, 1005}

\bibitem[\protect\citeauthoryear{Haas \& \textit{et al.}}{Haas \& \textit{et
  al.}}{2020}]{EinsteinToolkit_etal:2020_11}
Haas R.,  \textit{et al.} 2020, {T}he {E}instein {T}oolkit,
  \mn@doi{10.5281/zenodo.4298887}, \url
  {https://doi.org/10.5281/zenodo.4298887}

\bibitem[\protect\citeauthoryear{{Hanauske}, {Takami}, {Bovard}, {Rezzolla},
  {Font}, {Galeazzi}  \& {St{\"o}cker}}{{Hanauske} et~al.}{2017}]{Hanauske2016}
{Hanauske} M.,  {Takami} K.,  {Bovard} L.,  {Rezzolla} L.,  {Font} J.~A.,
  {Galeazzi} F.,   {St{\"o}cker} H.,  2017, \mn@doi [Phys. Rev. D]
  {10.1103/PhysRevD.96.043004}, \href
  {http://adsabs.harvard.edu/abs/2017PhRvD..96d3004H} {96, 043004}

\bibitem[\protect\citeauthoryear{Hanauske, Weih, St\"ocker  \&
  Rezzolla}{Hanauske et~al.}{2021}]{Hanauske2021}
Hanauske M.,  Weih L.~R.,  St\"ocker H.,   Rezzolla L.,  2021, \mn@doi [Eur.
  Phys. J. ST] {10.1140/epjs/s11734-021-00003-5}, 230, 543

\bibitem[\protect\citeauthoryear{{Hempel} \& {Schaffner-Bielich}}{{Hempel} \&
  {Schaffner-Bielich}}{2010a}]{Hempel2010}
{Hempel} M.,  {Schaffner-Bielich} J.,  2010a, \mn@doi [Nuclear Physics A]
  {10.1016/j.nuclphysa.2010.02.010}, \href
  {http://adsabs.harvard.edu/abs/2010NuPhA.837..210H} {837, 210}

\bibitem[\protect\citeauthoryear{Hempel \& Schaffner-Bielich}{Hempel \&
  Schaffner-Bielich}{2010b}]{Hempel2009}
Hempel M.,  Schaffner-Bielich J.,  2010b, \mn@doi [Nucl. Phys.]
  {10.1016/j.nuclphysa.2010.02.010}, A837, 210

\bibitem[\protect\citeauthoryear{Iosif \& Stergioulas}{Iosif \&
  Stergioulas}{2021}]{Iosif2020}
Iosif P.,  Stergioulas N.,  2021, \mn@doi [Mon. Not. Roy. Astron. Soc.]
  {10.1093/mnras/stab392}, 503, 850

\bibitem[\protect\citeauthoryear{Iosif \& Stergioulas}{Iosif \&
  Stergioulas}{2022}]{Iosif2021}
Iosif P.,  Stergioulas N.,  2022, \mn@doi [Mon. Not. Roy. Astron. Soc.]
  {10.1093/mnras/stab3565}, 510, 2948

\bibitem[\protect\citeauthoryear{Ishii, J\"arvinen  \& Nijs}{Ishii
  et~al.}{2019}]{Ishii:2019gta}
Ishii T.,  J\"arvinen M.,   Nijs G.,  2019, \mn@doi [JHEP]
  {10.1007/JHEP07(2019)003}, 07, 003

\bibitem[\protect\citeauthoryear{Jarvinen \& Kiritsis}{Jarvinen \&
  Kiritsis}{2012}]{Jarvinen:2011qe}
Jarvinen M.,  Kiritsis E.,  2012, \mn@doi [JHEP] {10.1007/JHEP03(2012)002}, 03,
  002

\bibitem[\protect\citeauthoryear{{Kastaun} \& {Galeazzi}}{{Kastaun} \&
  {Galeazzi}}{2015}]{Kastaun2014}
{Kastaun} W.,  {Galeazzi} F.,  2015, \mn@doi [Phys. Rev. D]
  {10.1103/PhysRevD.91.064027}, \href
  {http://adsabs.harvard.edu/abs/2015PhRvD..91f4027K} {91, 064027}

\bibitem[\protect\citeauthoryear{{Komatsu}, {Eriguchi}  \& {Hachisu}}{{Komatsu}
  et~al.}{1989a}]{Komatsu89}
{Komatsu} H.,  {Eriguchi} Y.,   {Hachisu} I.,  1989a, \mn@doi [Mon. Not. R.
  Astron. Soc.] {10.1093/mnras/237.2.355}, \href
  {http://adsabs.harvard.edu/abs/1989MNRAS.237..355K} {237, 355}

\bibitem[\protect\citeauthoryear{{Komatsu}, {Eriguchi}  \& {Hachisu}}{{Komatsu}
  et~al.}{1989b}]{Komatsu89b}
{Komatsu} H.,  {Eriguchi} Y.,   {Hachisu} I.,  1989b, Mon. Not. R. Astron.
  Soc., \href
  {http://adsabs.harvard.edu/cgi-bin/nph-bib_query?bibcode=1989MNRAS.239..153K&db_key=AST}
  {239, 153}

\bibitem[\protect\citeauthoryear{{Kr{\"u}ger}, {Gaertig}  \&
  {Kokkotas}}{{Kr{\"u}ger} et~al.}{2010}]{Krueger2010}
{Kr{\"u}ger} C.,  {Gaertig} E.,   {Kokkotas} K.~D.,  2010, \mn@doi [Phys. Rev.
  D] {10.1103/PhysRevD.81.084019}, \href
  {https://ui.adsabs.harvard.edu/abs/2010PhRvD..81h4019K} {81, 084019}

\bibitem[\protect\citeauthoryear{{Lasota}, {Haensel}  \& {Abramowicz}}{{Lasota}
  et~al.}{1996}]{Lasota1996}
{Lasota} J.-P.,  {Haensel} P.,   {Abramowicz} M.~A.,  1996, \mn@doi [Astrophys.
  J.] {10.1086/176650}, \href
  {http://adsabs.harvard.edu/abs/1996ApJ...456..300L} {456, 300}

\bibitem[\protect\citeauthoryear{{Lin}, {Cheng}, {Chu}  \& {Suen}}{{Lin}
  et~al.}{2006}]{Lin2006}
{Lin} L.~M.,  {Cheng} K.~S.,  {Chu} M.~C.,   {Suen} W.~M.,  2006, \mn@doi
  [Astrophys. J.] {10.1086/499202}, \href
  {https://ui.adsabs.harvard.edu/abs/2006ApJ...639..382L} {639, 382}

\bibitem[\protect\citeauthoryear{{Loeffler} et~al.,}{{Loeffler}
  et~al.}{2012}]{loeffler_2011_et}
{Loeffler} F.,  et~al., 2012, \mn@doi [Class. Quantum Grav.]
  {10.1088/0264-9381/29/11/115001}, \href
  {http://adsabs.harvard.edu/abs/2012CQGra..29k5001L} {29, 115001}

\bibitem[\protect\citeauthoryear{{Lyford}, {Baumgarte}  \& {Shapiro}}{{Lyford}
  et~al.}{2003}]{Lyford2003}
{Lyford} N.~D.,  {Baumgarte} T.~W.,   {Shapiro} S.~L.,  2003, \mn@doi
  [Astrophys. J.] {10.1086/345350}, \href
  {http://adsabs.harvard.edu/abs/2003ApJ...583..410L} {583, 410}

\bibitem[\protect\citeauthoryear{{Manca}, {Baiotti}, {DePietri}  \&
  {Rezzolla}}{{Manca} et~al.}{2007}]{Manca07}
{Manca} G.~M.,  {Baiotti} L.,  {DePietri} R.,   {Rezzolla} L.,  2007, \mn@doi
  [Class. Quantum Grav.] {10.1088/0264-9381/24/12/S12}, \href
  {http://adsabs.harvard.edu/abs/2007CQGra..24S.171M} {24, S171}

\bibitem[\protect\citeauthoryear{{Muhammed} et~al.,}{{Muhammed}
  et~al.}{2024}]{Nishad2024}
{Muhammed} N.,  et~al., 2024, \mn@doi [arXiv e-prints]
  {10.48550/arXiv.2403.05642}, \href
  {https://ui.adsabs.harvard.edu/abs/2024arXiv240305642M} {p. arXiv:2403.05642}

\bibitem[\protect\citeauthoryear{{M{\"u}ller}}{{M{\"u}ller}}{2024}]{Mueller2024}
{M{\"u}ller} B.,  2024, \mn@doi [arXiv e-prints] {10.48550/arXiv.2403.18952},
  \href {https://ui.adsabs.harvard.edu/abs/2024arXiv240318952M} {p.
  arXiv:2403.18952}

\bibitem[\protect\citeauthoryear{{Musolino}, {Ecker}  \& {Rezzolla}}{{Musolino}
  et~al.}{2024}]{Musolino2023b}
{Musolino} C.,  {Ecker} C.,   {Rezzolla} L.,  2024, \mn@doi [Astrophys. J.]
  {10.3847/1538-4357/ad1758}, \href
  {https://ui.adsabs.harvard.edu/abs/2024ApJ...962...61M} {962, 61}

\bibitem[\protect\citeauthoryear{{Ng}, {Jiang}, {Musolino}, {Ecker}, {Tootle}
  \& {Rezzolla}}{{Ng} et~al.}{2024}]{Ng2023b}
{Ng} H. H.-Y.,  {Jiang} J.-L.,  {Musolino} C.,  {Ecker} C.,  {Tootle} S.~D.,
  {Rezzolla} L.,  2024, \mn@doi [Phys. Rev. D] {10.1103/PhysRevD.109.064061},
  \href {https://ui.adsabs.harvard.edu/abs/2024PhRvD.109f4061N} {109, 064061}

\bibitem[\protect\citeauthoryear{{Ott}, {Burrows}, {Thompson}, {Livne}  \&
  {Walder}}{{Ott} et~al.}{2006}]{Ott06c}
{Ott} C.~D.,  {Burrows} A.,  {Thompson} T.~A.,  {Livne} E.,   {Walder} R.,
  2006, \mn@doi [Astrophys. J. Suppl. Ser.] {10.1086/500832}, 164, 130

\bibitem[\protect\citeauthoryear{{Paschalidis}}{{Paschalidis}}{2017}]{Paschalidis2016}
{Paschalidis} V.,  2017, \mn@doi [Classical and Quantum Gravity]
  {10.1088/1361-6382/aa61ce}, \href
  {http://adsabs.harvard.edu/abs/2017CQGra..34h4002P} {34, 084002}

\bibitem[\protect\citeauthoryear{Passamonti \& Andersson}{Passamonti \&
  Andersson}{2020}]{Passamonti2020}
Passamonti A.,  Andersson N.,  2020, \mn@doi [Monthly Notices of the Royal
  Astronomical Society] {10.1093/mnras/staa2725}, 498, 5904

\bibitem[\protect\citeauthoryear{{Rezzolla} \& {Zanotti}}{{Rezzolla} \&
  {Zanotti}}{2013}]{Rezzolla_book:2013}
{Rezzolla} L.,  {Zanotti} O.,  2013, Relativistic Hydrodynamics.
Oxford University Press, Oxford, UK,
  \mn@doi{10.1093/acprof:oso/9780198528906.001.0001}

\bibitem[\protect\citeauthoryear{{Rezzolla}, {Lamb}  \& {Shapiro}}{{Rezzolla}
  et~al.}{2000}]{Rezzolla00}
{Rezzolla} L.,  {Lamb} F.~K.,   {Shapiro} S.~L.,  2000, \mn@doi [Astrophys. J.
  Lett.] {10.1086/312539}, \href
  {http://adsabs.harvard.edu/abs/2000ApJ...531L.139R} {531, L139}

\bibitem[\protect\citeauthoryear{{Rezzolla}, {Baiotti}, {Giacomazzo}, {Link}
  \& {Font}}{{Rezzolla} et~al.}{2010}]{Rezzolla:2010}
{Rezzolla} L.,  {Baiotti} L.,  {Giacomazzo} B.,  {Link} D.,   {Font} J.~A.,
  2010, \mn@doi [Class. Quantum Grav.] {10.1088/0264-9381/27/11/114105}, \href
  {http://adsabs.harvard.edu/abs/2010CQGra..27k4105R} {27, 114105}

\bibitem[\protect\citeauthoryear{{S{\'a}}}{{S{\'a}}}{2004}]{Sa2004}
{S{\'a}} P.~M.,  2004, \mn@doi [Phys. Rev. D] {10.1103/PhysRevD.69.084001},
  \href {http://adsabs.harvard.edu/abs/2004PhRvD..69h4001S} {69, 084001}

\bibitem[\protect\citeauthoryear{{Saijo}, {Baumgarte}  \& {Shapiro}}{{Saijo}
  et~al.}{2003}]{Saijo2003}
{Saijo} M.,  {Baumgarte} T.~W.,   {Shapiro} S.~L.,  2003, \mn@doi [Astrophys.
  J.] {10.1086/377334}, \href
  {http://adsabs.harvard.edu/abs/2003ApJ...595..352S} {595, 352}

\bibitem[\protect\citeauthoryear{Shibata, Karino  \& Eriguchi}{Shibata
  et~al.}{2002}]{Shibata:2002mr}
Shibata M.,  Karino S.,   Eriguchi Y.,  2002, Mon. Not. R. Astron. Soc., 334,
  L27

\bibitem[\protect\citeauthoryear{{Shibata}, {Taniguchi}  \&
  {Ury{\=u}}}{{Shibata} et~al.}{2003}]{Shibata:2003ga}
{Shibata} M.,  {Taniguchi} K.,   {Ury{\=u}} K.,  2003, \mn@doi [Phys. Rev. D]
  {10.1103/PhysRevD.68.084020}, \href
  {http://adsabs.harvard.edu/abs/2003PhRvD..68h4020S} {68, 084020}

\bibitem[\protect\citeauthoryear{{Staykov}, {Doneva}, {Heisenberg},
  {Stergioulas}  \& {Yazadjiev}}{{Staykov} et~al.}{2023}]{Staykov2023}
{Staykov} K.~V.,  {Doneva} D.~D.,  {Heisenberg} L.,  {Stergioulas} N.,
  {Yazadjiev} S.~S.,  2023, \mn@doi [Phys. Rev. D]
  {10.1103/PhysRevD.108.024058}, \href
  {https://ui.adsabs.harvard.edu/abs/2023PhRvD.108b4058S} {108, 024058}

\bibitem[\protect\citeauthoryear{Stergioulas}{Stergioulas}{1998}]{Stergioulas98}
Stergioulas N.,  1998, Living Rev. Relativ., 1, 8

\bibitem[\protect\citeauthoryear{{Stergioulas} \& {Friedman}}{{Stergioulas} \&
  {Friedman}}{1995}]{Stergioulas95}
{Stergioulas} N.,  {Friedman} J.~L.,  1995, \mn@doi [Astrophys. J.]
  {10.1086/175605}, \href {http://adsabs.harvard.edu/abs/1995ApJ...444..306S}
  {444, 306}

\bibitem[\protect\citeauthoryear{{Stergioulas}, {Apostolatos}  \&
  {Font}}{{Stergioulas} et~al.}{2004}]{Stergioulas04}
{Stergioulas} N.,  {Apostolatos} T.~A.,   {Font} J.~A.,  2004, \mn@doi [Mon.
  Not. R. Astron. Soc.] {10.1111/j.1365-2966.2004.07973.x}, \href
  {http://adsabs.harvard.edu/abs/2004MNRAS.352.1089S} {352, 1089}

\bibitem[\protect\citeauthoryear{{Studzi{\'n}ska}, {Kucaba},
  {Gondek-Rosi{\'n}ska}, {Villain}  \& {Ansorg}}{{Studzi{\'n}ska}
  et~al.}{2016}]{Studzinska2016}
{Studzi{\'n}ska} A.~M.,  {Kucaba} M.,  {Gondek-Rosi{\'n}ska} D.,  {Villain} L.,
    {Ansorg} M.,  2016, \mn@doi [Mon. Not. R. Astron. Soc.]
  {10.1093/mnras/stw2152}, \href
  {http://adsabs.harvard.edu/abs/2016MNRAS.463.2667S} {463, 2667}

\bibitem[\protect\citeauthoryear{Szewczyk, Gondek-Rosi\'nska  \&
  Cerd\'a-Dur\'an}{Szewczyk et~al.}{2023}]{Szewczyk2023}
Szewczyk P.,  Gondek-Rosi\'nska D.,   Cerd\'a-Dur\'an P.,  2023, \mn@doi [Acta
  Phys. Polon. Supp.] {10.5506/APhysPolBSupp.16.6-A8}, 16, 8

\bibitem[\protect\citeauthoryear{{Szkudlarek}, {Gondek-Rosi{\'n}ska}, {Villain}
   \& {Ansorg}}{{Szkudlarek} et~al.}{2019}]{Szkudlarek2019}
{Szkudlarek} M.,  {Gondek-Rosi{\'n}ska} D.,  {Villain} L.,   {Ansorg} M.,
  2019, \mn@doi [Astrophys. J.] {10.3847/1538-4357/ab1752}, \href
  {https://ui.adsabs.harvard.edu/abs/2019ApJ...879...44S} {879, 44}

\bibitem[\protect\citeauthoryear{{Takami}, {Rezzolla}  \& {Yoshida}}{{Takami}
  et~al.}{2011}]{Takami:2011}
{Takami} K.,  {Rezzolla} L.,   {Yoshida} S.,  2011, \mn@doi [Mon. Not. R.
  Astron. Soc.] {10.1111/j.1745-3933.2011.01085.x}, \href
  {http://adsabs.harvard.edu/abs/2011MNRAS.416L...1T} {416, L1}

\bibitem[\protect\citeauthoryear{{Takami}, {Rezzolla}  \& {Baiotti}}{{Takami}
  et~al.}{2014}]{Takami2014}
{Takami} K.,  {Rezzolla} L.,   {Baiotti} L.,  2014, \mn@doi [Phys. Rev. Lett.]
  {10.1103/PhysRevLett.113.091104}, \href
  {http://adsabs.harvard.edu/abs/2014PhRvL.113i1104T} {113, 091104}

\bibitem[\protect\citeauthoryear{{Takami}, {Rezzolla}  \& {Baiotti}}{{Takami}
  et~al.}{2015}]{Takami2015}
{Takami} K.,  {Rezzolla} L.,   {Baiotti} L.,  2015, \mn@doi [Phys. Rev. D]
  {10.1103/PhysRevD.91.064001}, \href
  {http://adsabs.harvard.edu/abs/2015PhRvD..91f4001T} {91, 064001}

\bibitem[\protect\citeauthoryear{Tassoul}{Tassoul}{1978}]{Tassoul-1978:theory-of-rotating-stars}
Tassoul J.-L.,  1978, Theory of Rotating Stars.
Princeton University Press

\bibitem[\protect\citeauthoryear{Togashi, Nakazato, Takehara, Yamamuro, Suzuki
  \& Takano}{Togashi et~al.}{2017}]{Togashi2017}
Togashi H.,  Nakazato K.,  Takehara Y.,  Yamamuro S.,  Suzuki H.,   Takano M.,
  2017, \mn@doi [Nucl. Phys.] {10.1016/j.nuclphysa.2017.02.010}, A961, 78

\bibitem[\protect\citeauthoryear{{Tsokaros}, {Ruiz}  \& {Shapiro}}{{Tsokaros}
  et~al.}{2020}]{Tsokaros2020c}
{Tsokaros} A.,  {Ruiz} M.,   {Shapiro} S.~L.,  2020, \mn@doi [Phys. Rev. D]
  {10.1103/PhysRevD.101.064069}, \href
  {https://ui.adsabs.harvard.edu/abs/2020PhRvD.101f4069T} {101, 064069}

\bibitem[\protect\citeauthoryear{Typel, Ropke, Klahn, Blaschke  \&
  Wolter}{Typel et~al.}{2010}]{Typel:2009sy}
Typel S.,  Ropke G.,  Klahn T.,  Blaschke D.,   Wolter H.~H.,  2010, \mn@doi
  [Phys. Rev. C] {10.1103/PhysRevC.81.015803}, 81, 015803

\bibitem[\protect\citeauthoryear{{Ury{\={u}}}, {Tsokaros}, {Baiotti},
  {Galeazzi}, {Taniguchi}  \& {Yoshida}}{{Ury{\={u}}} et~al.}{2017}]{Uryu2017}
{Ury{\={u}}} K.,  {Tsokaros} A.,  {Baiotti} L.,  {Galeazzi} F.,  {Taniguchi}
  K.,   {Yoshida} S.,  2017, \mn@doi [Phys. Rev. D]
  {10.1103/PhysRevD.96.103011}, \href
  {https://ui.adsabs.harvard.edu/abs/2017PhRvD..96j3011U} {96, 103011}

\bibitem[\protect\citeauthoryear{{Weih}, {Most}  \& {Rezzolla}}{{Weih}
  et~al.}{2018}]{Weih2017}
{Weih} L.~R.,  {Most} E.~R.,   {Rezzolla} L.,  2018, \mn@doi [Mon. Not. R.
  Astron. Soc.] {10.1093/mnrasl/slx178}, \href
  {http://adsabs.harvard.edu/abs/2018MNRAS.473L.126W} {473, L126}

\bibitem[\protect\citeauthoryear{Weih, Hanauske  \& Rezzolla}{Weih
  et~al.}{2020}]{Weih:2019xvw}
Weih L.~R.,  Hanauske M.,   Rezzolla L.,  2020, \mn@doi [Phys. Rev. Lett.]
  {10.1103/PhysRevLett.124.171103}, 124, 171103

\bibitem[\protect\citeauthoryear{Xie, Hawke, Passamonti  \& Andersson}{Xie
  et~al.}{2020}]{Xie2020}
Xie X.,  Hawke I.,  Passamonti A.,   Andersson N.,  2020, \mn@doi [Phys. Rev.
  D] {10.1103/PhysRevD.102.044040}, 102, 044040

\bibitem[\protect\citeauthoryear{Zhou, Tsokaros, Uryu, Xu  \& Shibata}{Zhou
  et~al.}{2019}]{Zhou2019}
Zhou E.,  Tsokaros A.,  Uryu K.,  Xu R.,   Shibata M.,  2019, \mn@doi [Phys.
  Rev. D] {10.1103/PhysRevD.100.043015}, 100, 043015

\bibitem[\protect\citeauthoryear{{Zink}, {Korobkin}, {Schnetter}  \&
  {Stergioulas}}{{Zink} et~al.}{2010}]{Zink:2010a}
{Zink} B.,  {Korobkin} O.,  {Schnetter} E.,   {Stergioulas} N.,  2010, Phys.
  Rev. D, 81, 084055

\makeatother
\end{thebibliography}

\bsp	
\label{lastpage}
\end{document}